\documentclass[preprint,review,12pt,3p]{elsarticle}

\usepackage{listings}
\usepackage{graphicx}
\usepackage{epsfig}
\usepackage{amssymb}
\usepackage{amsmath}
\usepackage{subfigure}
\usepackage{setspace}
\usepackage{booktabs}
\usepackage{multirow}

\usepackage{algorithm}
\usepackage{algpseudocode}

\usepackage{hyperref}
\hypersetup{colorlinks=false}

\biboptions{sort&compress} 

\journal{Journal of Computational Physics}

\begin{document}
\begin{frontmatter}
\title{Calibration of multi-physics computational models using Bayesian networks}
\author{You Ling, Joshua Mullins, and Sankaran Mahadevan\corref{corauthor}}
\address{Department of Civil and Environmental Engineering, Vanderbilt University, TN 37235}
\cortext[corauthor]{Corresponding author\\Phone: 1-615-322-3040. Fax: 1-615-322-3365. Email: sankaran.mahadevan@vanderbilt.edu. \\Postal address: Vanderbilt University, VU Station B 356077, Nashville, TN 37235-6077}
\date{May 3, 2012}

\begin{abstract}
This paper develops a Bayesian network-based method for the calibration of multi-physics models, integrating various sources of uncertainty with information from computational models and experimental data. We adopt the well-known Kennedy and O'Hagan (KOH) framework for model calibration under uncertainty, and develop extensions to multi-physics models and various scenarios of available data. Both aleatoric uncertainty (due to natural variability) and epistemic uncertainty (due to lack of information, including data uncertainty and model uncertainty) are accounted for in the calibration process. Challenging aspects of Bayesian calibration for multi-physics models are investigated, including: (1) calibration with different forms of experimental data (e.g., interval data and time series data), (2) determination of the identifiability of model parameters when the analytical expression of model is known or unknown, (3) calibration of multiple physics models sharing common parameters, and (4) efficient use of available data in a multi-model calibration problem especially when the experimental resources are limited. A first-order Taylor series expansion-based method is proposed to determine which model parameters are identifiable, i.e., to find the parameters that can be calibrated with the available data. Following the KOH framework, a probabilistic discrepancy function is estimated and added to the prediction of the calibrated model, attempting to account for model uncertainty. This discrepancy function is modeled as a Gaussian process when sufficient data are available for multiple model input combinations, and is modeled as a random variable when the available data set is small and limited. The overall approach is illustrated using two application examples related to microelectromechanical system (MEMS) devices: (1) calibration of a dielectric charging model with time-series data, and (2) calibration of two physics models (pull-in voltage and creep) using measurements of different physical quantities in different devices.
\end{abstract}


\end{frontmatter}
\section{Introduction}


Stochastic multi-physics simulation is a key component in the reliability analysis of engineering components/devices, which requires solving several computational models while accounting for various sources of uncertainty. Calibration of these multi-physics computational models can be challenging due to the complex structure of the system, existence of multiple uncertainty sources, and limited experimental data.

Model calibration can be viewed as the process of adjusting the value or the prior distribution of unknown model parameters in order to improve the agreement between the model output and observed data~\cite{Kennedy2001,Campbell2006,McFarland2008}. In comparison to other calibration methods (such as maximum likelihood and least squares) which return point/interval estimates, Bayesian inference returns the posterior probability density functions (PDF) of unknown parameters. These posterior PDFs account for various sources of uncertainty existing in the computer model and experimental observation, including natural variability in model inputs, data uncertainty (measurement uncertainty and epistemic uncertainty due to insufficient data), and model uncertainty~\cite{Sankararaman2011}.

Kennedy and O'Hagan~\cite{Kennedy2001} developed a Bayesian framework (commonly known as the Kennedy and O'Hagan (KOH) framework) for the calibration of computer models under various sources of uncertainty, and a discrepancy function was introduced in order to account for the discrepancy between observed data and the calibrated model prediction. In addition to the KOH framework, significant research efforts have been devoted to the development of Bayesian methods for scientific and engineering applications~\cite{Koutsourelakis2009,Bower2010,Sankararaman2010,Cheung2011,Wan2011,Moselhy2011}. However, several issues remain unclear in the implementation of Bayesian calibration for practical problems with complicated systems of models: (1) calibration with different types of available data, such as point data, interval data, and time series data, (2) identifiability of model parameters, i.e., how to find out which parameters can or cannot be calibrated using the Bayesian approach for a given computer model, (3) calibration of multiple models sharing common parameters, and (4) efficient use of experimental data in calibration, which may be useful for the cases that only a limited amount of data are available.

Aimed at providing potentially useful directions for solving the above issues, this paper develops a Bayesian network-based calibration approach for multi-physics computational models. The Bayesian network is a powerful tool to represent complicated systems with a set of nodes and the probabilistic relation between the nodes~\cite{Jensen2007,Torres-Toledano1998,Langseth2007}, and the observation data of some of the nodes can be conveniently incorporated into the network to facilitate the inference on other nodes. Based on the information contained in the Bayesian network and the observation data, model parameters can be calibrated accounting for different sources of uncertainty, and the posterior PDFs of the parameters can be obtained. Note that this paper focuses on model calibration with direct measurement data of the model output variable. In the case that the available information are the moments of the probability distributions of the model output variable, some recently developed methodologies based on optimization with constraints on the moments may be considered~\cite{Zabaras2008,Guan2009}. In addition, a Bayesian approach has been developed to include the information on the moments of unknown model parameters~\cite{Berry2012}.

In this paper, we first present the basic framework of model calibration using the Bayesian network in Section~\ref{section:CALbyBN}. In the subsequent sections, practical issues in the application of Bayesian calibration are discussed. In Section~\ref{section:diffTypeData}, two types of experimental data are considered, namely interval data and time series data, and the corresponding details of calibration are developed. For models with multiple parameters, it is possible that not all of the parameters can be calibrated due to the inner structure of the model or the amount of the available experimental data. Knowing which parameters are unidentifiable can save computational effort. A first-order Taylor series expansion-based method is developed for this purpose in Section~\ref{section:identifiability}. Some discussions on computing likelihood functions, which can be computationally expensive for complex systems, are provided in Section~\ref{section:compIssue}. A Bayesian network-based method is developed in Section~\ref{section:calMP} for multi-physics computational models, which efficiently uses the available experimental data in model calibration. Section~\ref{section:NumExp} illustrates the aforementioned methods for the calibration of (1) a dielectric charging model with time series data, and (2) a multi-physics modeling system for MEMS devices, which includes multiple interacting computational models (dynamic, electrostatic, damping, and creep models).

\section{Basics of model calibration using a Bayesian network}\label{section:CALbyBN}

\begin{figure}[h!]
\begin{center}
\subfigure[]{
\includegraphics[trim=51mm 92mm 112mm 42mm, clip, width=0.48\textwidth]{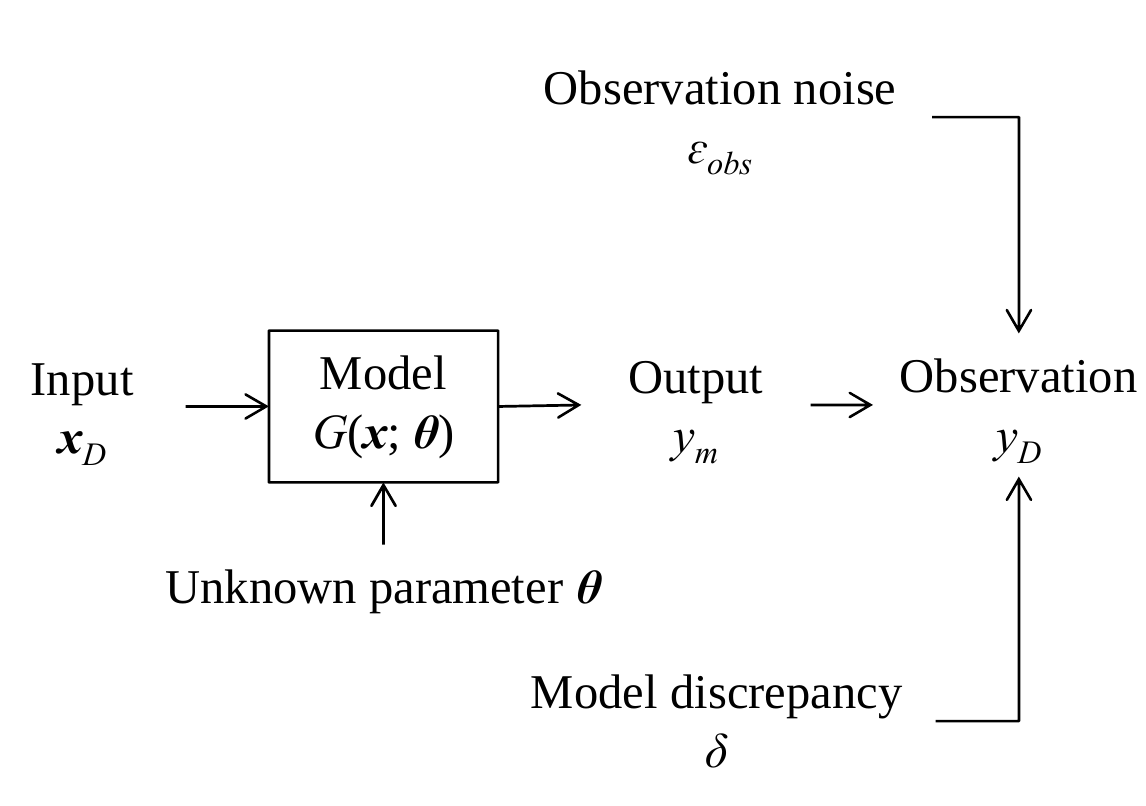}\label{fig:modelG}}
\subfigure[]{
\includegraphics[trim=55mm 55mm 115mm 70mm, clip, width=0.48\textwidth]{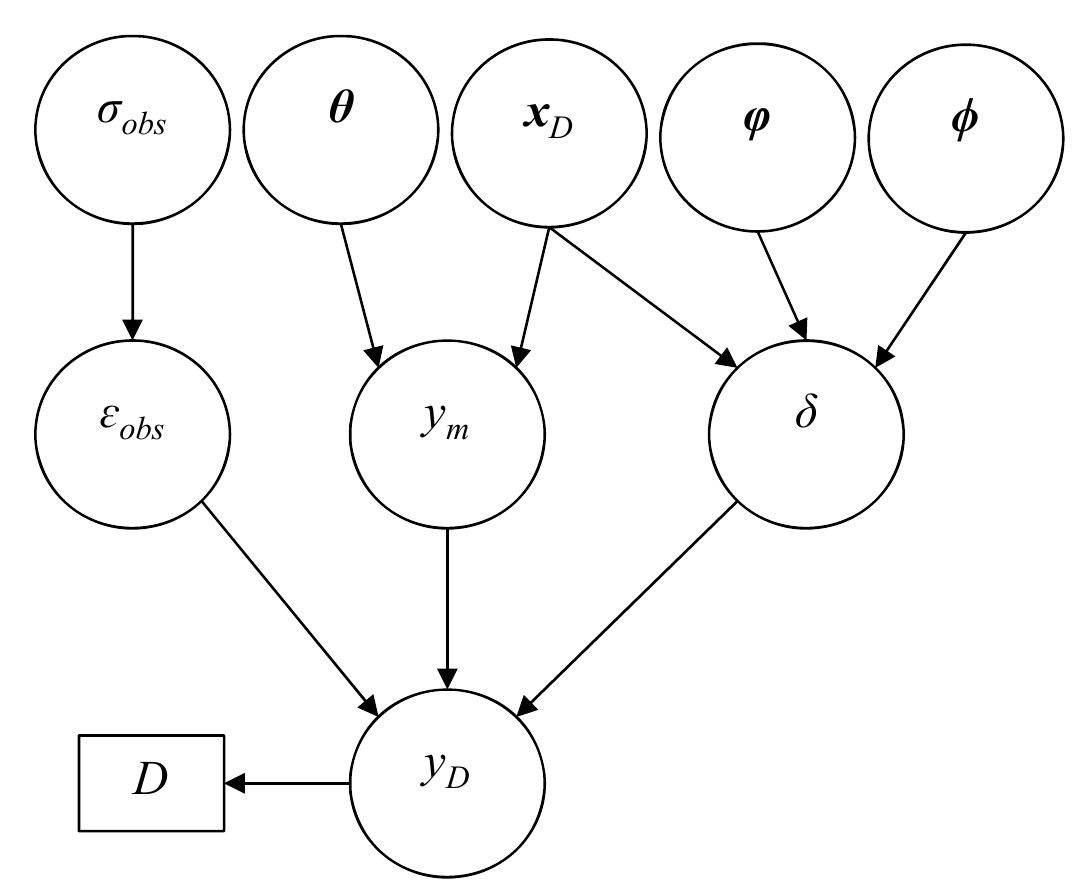}\label{fig:BN_1}}
\end{center}
\caption{(a) the illustrative diagram of a computer model, and (b) the corresponding Bayesian network}
\label{fig:example-1}
\end{figure}

Consider a computer model with inputs $\boldsymbol{x}$, parameters $\boldsymbol{\theta}$, and output $y_m$ (as shown in Fig.~\ref{fig:modelG})
\begin{equation}\label{eq:Model1}
y_m = G(\boldsymbol{x};\boldsymbol{\theta})
\end{equation}
This model is constructed to predict a physical quantity $y$, which is observable through experiments, i.e., the model output $y_m$ is a prediction of the actual quantity $y$. The distinction between the model inputs $\boldsymbol{x}$ and parameters $\boldsymbol{\theta}$ may not always be obvious. In this paper, we consider the model inputs as observable quantities, and are represented with specified deterministic values or probability distributions (if stochastic). In contrast, the model parameters are not observable, and the objective of model calibration is to estimate these parameters based on available information. Although both the terms "experimental inputs" and "model inputs" refer to the same physical quantities, it should be noted that the term "experimental inputs" (denoted as $\boldsymbol{x}_D$) is used to represent the measurement of these physical quantities in calibration experiments, whereas the term "model inputs" refers to the set of values of these quantities that go into the model. In the process of model calibration, the model inputs $\boldsymbol{x}$ are set to be $\boldsymbol{x}_D$, since the model will only be evaluated at $\boldsymbol{x}_D$.


Suppose the measurement uncertainty is represented using a zero-mean Guassian random variable $\varepsilon_{obs}$ with variance $\sigma_{obs}^2$. Following the KOH framework, the model uncertainty is represented by a model discrepancy function, which is approximated using a Gaussian process, i.e., $\delta \sim \mathcal{N} \big( m(\boldsymbol{x};\boldsymbol{\phi}),k(\boldsymbol{x},\boldsymbol{x'};\boldsymbol{\varphi}) \big)$; where $m(*)$ is the mean function of this Gaussian process $\delta$ and $\boldsymbol{\phi}$ is the set of coefficients of $m(*)$; $k(*)$ is the covariance function of $\delta$ and $\boldsymbol{\varphi}$ is the set of coefficients of $k(*)$. The choice of the form of the mean function can be rather subjective and is typically problem-specific. One such choice will be demonstrated in the numerical example in Section~\ref{subsection:NumExpDielectric}. The choice of covariance function may also be problem-specific, but a squared exponential function is commonly used for illustration, i.e.,
\begin{equation}\label{eq:sqExpCovFun}
k(\boldsymbol{x},\boldsymbol{x'};\boldsymbol{\varphi}) = \lambda \exp \Big( -\sum_{i=1}^q \frac{(x_i - x_i')^2}{2 l_i^2} \Big)
\end{equation}
where $\boldsymbol{\varphi}=[\lambda,l_1,l_2,...,l_q]$, and $q$ is the dimension of the inputs $\boldsymbol{x}$; $\lambda$ is the variance of this Gaussian process; $l_i$ is the length-scale parameter corresponding to the input $x_i$. Higher values of $l_i$ indicate higher statistical correlation between $x_i$ and $x_i'$. Note that if the KOH framework is strictly followed, the computer model $G(\boldsymbol{x};\boldsymbol{\theta})$ will also be approximated using a Gaussian process in order to reduce the computational effort. 

Since $\sigma_{obs}$, $\boldsymbol{\phi}$ and $\boldsymbol{\varphi}$ are usually unknown, they may also need to be calibrated. If some observation data (denoted as $D$) of $y$ are available, we can calibrate the unknown parameters $\boldsymbol{\theta}$, $\sigma_{obs}$, $\boldsymbol{\phi}$ and $\boldsymbol{\varphi}$ using Bayes' theorem as
\begin{eqnarray}\label{eq:Bayes}
\pi(\boldsymbol{\theta},\sigma_{obs},\boldsymbol{\phi},\boldsymbol{\varphi}| D) &=& \frac{\mathcal{L}(\boldsymbol{\theta},\sigma_{obs},\boldsymbol{\phi},\boldsymbol{\varphi}) \pi(\boldsymbol{\theta}) \pi(\sigma_{obs}) \pi(\boldsymbol{\phi}) \pi(\boldsymbol{\varphi})}{\int \mathcal{L}(\boldsymbol{\theta},\sigma_{obs},\boldsymbol{\phi},\boldsymbol{\varphi}) \pi(\boldsymbol{\theta}) \pi(\sigma_{obs}) \pi(\boldsymbol{\phi}) \pi(\boldsymbol{\varphi}) \mathrm{d} \boldsymbol{\theta} \mathrm{d} \sigma_{obs} \mathrm{d} \boldsymbol{\phi} \mathrm{d} \boldsymbol{\varphi}} \nonumber\\
\mathcal{L}(\boldsymbol{\theta},\sigma_{obs},\boldsymbol{\phi},\boldsymbol{\varphi}) &\propto& \pi( D|\boldsymbol{\theta},\sigma_{obs},\boldsymbol{\phi},\boldsymbol{\varphi})
\end{eqnarray}
where $\pi(\boldsymbol{\theta})$, $\pi(\sigma_{obs})$, $\pi(\boldsymbol{\phi})$ and $\pi(\boldsymbol{\varphi})$ are the prior PDFs of $\boldsymbol{\theta}$, $\sigma_{obs}$, $\boldsymbol{\phi}$ and $\boldsymbol{\varphi}$ respectively, representing prior knowledge of these parameters before calibration; $\pi(\boldsymbol{\theta},\sigma_{obs},\boldsymbol{\phi},\boldsymbol{\varphi}| D)$ is the joint posterior (or calibrated) PDF of $\boldsymbol{\theta}$, $\sigma_{obs}$, $\boldsymbol{\phi}$ and $\boldsymbol{\varphi}$; the joint likelihood function of $\boldsymbol{\theta}$, $\sigma_{obs}$, $\boldsymbol{\phi}$ and $\boldsymbol{\varphi}$, which is denoted as $\mathcal{L}(\boldsymbol{\theta},\sigma_{obs},\boldsymbol{\phi},\boldsymbol{\varphi})$, is proportional to  the conditional probability of observing the data $D$ given these parameters. Note that $\pi(*)$ denotes probability density function in this paper.

The likelihood function can be computed based on the construction of a Bayesian network. A Bayesian network is a directed acyclic graph formed by the variables (nodes) together with the directed edges, attached by a table of conditional probabilities of each variable on all its parents~\cite{Jensen2007}. Therefore, it is a graphical representation of uncertain quantities that explicitly incorporates the probabilistic causal dependence between the variables as well as the flow of information in the model~\cite{Mahadevan2001}. Fig.~\ref{fig:BN_1} shows the graph of the Bayesian network for the example model and experimental observation described in Eq.~\ref{eq:relationYandYD1} and Fig.~\ref{fig:modelG}. Herein we consider the experimental observation $y_D$ corresponding to a given experimental input setting $\boldsymbol{x}_D$ as a random variable, and the actual observation data point $D$ is a random realization of $y_D$. The relationship between $y_D$ and the model output $y_m$ can be written as
\begin{equation}\label{eq:relationYandYD1}
y_D = y_m+\delta+\varepsilon_{obs}
\end{equation}

Based on the constructed Bayesian network and the chain rule in probability theory, the joint likelihood function of $\boldsymbol{\theta}$, $\sigma_{obs}$, $\boldsymbol{\phi}$ and $\boldsymbol{\varphi}$ can be derived as
\begin{eqnarray}\label{eq:LK-ex1}
&& \mathcal{L}(\boldsymbol{\theta},\sigma_{obs},\boldsymbol{\phi},\boldsymbol{\varphi}) \propto \\
&& \int \pi(y_D=D|y_m,\varepsilon_{obs},\delta) \pi(y_m|\boldsymbol{x}_D,\boldsymbol{\theta}) \pi(\varepsilon_{obs}|\sigma_{obs}) \pi(\delta | \boldsymbol{x}_D, \boldsymbol{\phi},\boldsymbol{\varphi}) \pi(\boldsymbol{x}_D) \mathrm{d}y_m \mathrm{d}\varepsilon_{obs} \mathrm{d}\delta \mathrm{d}\boldsymbol{x}_D \nonumber
\end{eqnarray}


To compute $\mathcal{L}(\boldsymbol{\theta},\sigma_{obs},\boldsymbol{\phi},\boldsymbol{\varphi})$ in Eq.~\ref{eq:LK-ex1}, the conditional probabilities corresponding to the directed edges are needed. The PDF of the experimental inputs $\pi(\boldsymbol{x}_D)$ is assumed known as stated at the beginning of this subsection. Recall that the measurement error $\varepsilon_{obs}$ is assumed to be normally distributed, and thus $\pi(\varepsilon_{obs}|\sigma_{obs})$ is a normal probability density function with zero mean and variance $\sigma_{obs}^2$ which is evaluated at $\varepsilon_{obs}$. $\pi(\delta | \boldsymbol{x}_D, \boldsymbol{\phi},\boldsymbol{\varphi})$ is also a normal probability density function, since the model discrepancy function $\delta$ is approximated as a Gaussian process with index $\boldsymbol{x}_D$. Note that in this example, for given values of $\boldsymbol{x}_D$ and $\boldsymbol{\theta}$, the model output $y_m$ is deterministic; $y_D$ is also deterministic if the values of $y_m$, $\varepsilon_{obs}$, and $\delta$ are given. In general, inference in Bayesian networks with conditionally deterministic variables is not straightforward since conditional probability density functions such as $\pi(y_m|\boldsymbol{x}_D,\boldsymbol{\theta})$ and $\pi(y_D=D|y_m,\delta,\varepsilon_{obs})$ do not exist, and some related discussions and solutions can be found in~\cite{Cobb2006,Cinicioglu2009}. In this example, however, we can bypass the conditionally deterministic variables because of the assumptions that $\varepsilon_{obs}$ and $\delta$ are normally distributed given the corresponding parameters. That is to say, we can derive without computing the integration in Eq.~\ref{eq:LK-ex1} that $( y_D | \boldsymbol{x}_D, \boldsymbol{\theta},\sigma_{obs},\boldsymbol{\phi},\boldsymbol{\varphi} ) \sim \mathcal{N} \big( G(\boldsymbol{x}_D;\boldsymbol{\theta})+m(\boldsymbol{x}_D;\boldsymbol{\phi}), k(\boldsymbol{x}_D,\boldsymbol{x}_D;\boldsymbol{\varphi})+\sigma_{obs}^2 \big)$, where $m(\boldsymbol{x}_D;\boldsymbol{\phi})$ and $k(\boldsymbol{x}_D,\boldsymbol{x}_D;\boldsymbol{\varphi})$ are the mean and covariance function of $\delta$ respectively. Thus, $\mathcal{L}(\boldsymbol{\theta},\sigma_{obs},\boldsymbol{\phi},\boldsymbol{\varphi})$ can be obtained as
\begin{equation}\label{eq:LK-ex2}
\mathcal{L}(\boldsymbol{\theta},\sigma_{obs},\boldsymbol{\phi},\boldsymbol{\varphi}) \propto \int \pi(y_D=D|\boldsymbol{x}_D,\boldsymbol{\theta},\sigma_{obs},\boldsymbol{\phi},\boldsymbol{\varphi}) \ \pi(\boldsymbol{x}_D) \ \mathrm{d}\boldsymbol{x}_D
\end{equation}

The case discussed above considers calibration with a single data point. For the case that the experimental observations are taken for multiple input setting $\boldsymbol{X}_D = [ \boldsymbol{x}_{D1}, \boldsymbol{x}_{D2}, ..., \boldsymbol{x}_{Dn} ]$, let $\boldsymbol{y}_D =[ y_{D1}, y_{D2}, ..., y_{Dn} ]$ be the vector of experimental observations corresponding to $\boldsymbol{X}_D$. Similarly, we can derive that
\begin{equation}
(\boldsymbol{y}_D | \boldsymbol{X}_D, \boldsymbol{\theta},\sigma_{obs},\boldsymbol{\phi},\boldsymbol{\varphi} ) \sim \mathcal{N}(\boldsymbol{\mu_{y_D}},\Sigma+\sigma_{obs}^2 \boldsymbol{I})
\end{equation}
where
\begin{equation}\label{eq:observation1}
\boldsymbol{\mu} =
\begin{bmatrix}
G(\boldsymbol{x}_{D1};\boldsymbol{\theta})+m(\boldsymbol{x}_{D1};\boldsymbol{\phi}) \\
\vdots \\
G(\boldsymbol{x}_{Dn};\boldsymbol{\theta})+m(\boldsymbol{x}_{Dn};\boldsymbol{\phi}) 
\end{bmatrix}, \quad
\Sigma = 
\begin{bmatrix}
k(\boldsymbol{x}_{D1},\boldsymbol{x}_{D1};\boldsymbol{\varphi}) & \ldots & k(\boldsymbol{x}_{D1},\boldsymbol{x}_{Dn};\boldsymbol{\varphi}) \\
\vdots & \ddots &\vdots \\
k(\boldsymbol{x}_{Dn},\boldsymbol{x}_{D1};\boldsymbol{\varphi}) & \ldots & k(\boldsymbol{x}_{Dn},\boldsymbol{x}_{Dn};\boldsymbol{\varphi})
\end{bmatrix}
\end{equation}

Note that in Eq.~\ref{eq:observation1}, for given values of $\boldsymbol{X}_D$, $\boldsymbol{\theta}$, $\sigma_{obs}$, $\boldsymbol{\phi}$ and $\boldsymbol{\varphi}$, $\boldsymbol{y}_D$ is a Gaussian random vector that represents the experimental observation corresponding to multiple input settings. Further, let $\boldsymbol{D}= [ D_1,D_2,...,D_n ]$ represent the actual observation data which are random realizations of the random vector $\boldsymbol{y}_D$. In this case, $\mathcal{L}(\boldsymbol{\theta},\sigma_{obs},\boldsymbol{\phi},\boldsymbol{\varphi})$ can be obtained as
\begin{equation}\label{eq:LK-ex2}
\mathcal{L}(\boldsymbol{\theta},\sigma_{obs},\boldsymbol{\phi},\boldsymbol{\varphi}) \propto \int \pi(\boldsymbol{y}_D=\boldsymbol{D}|\boldsymbol{X}_D,\boldsymbol{\theta},\sigma_{obs},\boldsymbol{\phi},\boldsymbol{\varphi})  \ \pi(\boldsymbol{X}_D)  \ \mathrm{d}\boldsymbol{X}_D
\end{equation}

The calculation of the likelihood function is the key component in implementing Bayesian model calibration. In the cases discussed in this section, the likelihood function is constructed based on measurement data reported as point values. In practical problems, various types of experimental data may be available, including interval data and time series data. For these different types of data, the corresponding methods to calculate the likelihood function are presented in Section~\ref{section:BayesTech}.

\section{Practical issues in implementing Bayesian calibration}\label{section:BayesTech}
\subsection{Different types of experimental data}\label{section:diffTypeData}
\subsubsection{Bayesian calibration with interval data}\label{section:intervalData}
Due to the imprecision of measurement techniques and limited experimental resources, the measurement of many quantities is only available in the form of an interval, which brings in additional data uncertainty (i.e., the actual experimental value lies within an interval). Sankararaman and Mahadevan~\cite{Sankararaman2011a} developed a likelihood-based approach to quantify this type of uncertainty. In the example shown in Fig.~\ref{fig:example-1}, the experimental data $D$ may be reported as an interval, i.e., $D=[D^a, D^b]$,  and we can derive the corresponding expression for the likelihood function of unknown parameters based on the method developed in~\cite{Sankararaman2011a} as
\begin{eqnarray}\label{eq:LKintervalData}
\mathcal{L}(\boldsymbol{\theta},\sigma_{obs},\boldsymbol{\phi},\boldsymbol{\varphi}) &\propto& \Pr(D^a \le y_{D} \le D^b |\boldsymbol{\theta},\sigma_{obs},\boldsymbol{\phi},\boldsymbol{\varphi}) \nonumber\\
&& = \int_{D^{a}}^{D^{b}} \pi(y_D| \boldsymbol{x}_D,\boldsymbol{\theta},\sigma_{obs},\boldsymbol{\phi},\boldsymbol{\varphi}) \ \pi(\boldsymbol{x}_D) \ \mathrm{d} y_D \ \mathrm{d}\boldsymbol{x}_D
\end{eqnarray}
where $\pi(y_D|\boldsymbol{x}_D,\boldsymbol{\theta},\sigma_{obs},\boldsymbol{\phi},\boldsymbol{\varphi})$ is a normal PDF as discussed in Section~\ref{section:CALbyBN}. Note that the likelihood-based approach~\cite{Sankararaman2011a} is not limited to the case of normal distributions. If data are in the form of half intervals, i.e., $y_D \ge D^a$, the likelihood function can be obtained by letting $D^b = +\infty$ in Eq.~\ref{eq:LKintervalData}. Similarly, let $D^a = -\infty$ if $y_D \le D^b$.

In some problems, measurements may be available at multiple input settings $\boldsymbol{X}_D = [\boldsymbol{x}_{D1},\boldsymbol{x}_{D2},...\boldsymbol{x}_{Dn}]$, and the data may be in the form of multiple intervals or a mixture of intervals and point values. We can conveniently extend Eq.~\ref{eq:LKintervalData} to these two cases.  Suppose the available data are now a set of intervals, i.e., $\boldsymbol{D} = \{ [D_1^a, D_1^b], [D_2^a, D_2^b], ..., [D_n^a, D_n^b] \}$, which forms a $n$-dimensional hypercube $\Omega_n$. The probability of observing the data is thus equivalent to the probability of the $n$-dimensional random vector $\boldsymbol{y}_D = [y_{D1},y_{D2},...,y_{Dn}]$ lying inside the hypercube $\Omega_n$. Hence, the likelihood function of unknown parameters can be derived as
\begin{eqnarray}\label{eq:LKintervalMultiData}
\mathcal{L}(\boldsymbol{\theta},\sigma_{obs},\boldsymbol{\phi},\boldsymbol{\varphi}) &\propto& \Pr(\boldsymbol{y}_D \in \Omega_n | \boldsymbol{\theta},\sigma_{obs},\boldsymbol{\phi},\boldsymbol{\varphi}) \nonumber\\
&& = \int_{\Omega_n} \pi(\boldsymbol{y}_D|\boldsymbol{X}_D, \boldsymbol{\theta},\sigma_{obs},\boldsymbol{\phi},\boldsymbol{\varphi}) \ \pi(\boldsymbol{X}_D) \ \mathrm{d} \boldsymbol{y}_D \ \mathrm{d}\boldsymbol{X}_D
\end{eqnarray}

In the case that the available information is a mixture of $k$ intervals and $(n-k)$ point values, the $k$ intervals form a $k$-dimensional hypercube $\Omega_k$. Let $\boldsymbol{y}_{D,k}$ represent the elements of the random vector $\boldsymbol{y}_D$ corresponding to interval data, and $\boldsymbol{y}_{D,n-k}$ represent the rest of the elements corresponding to point data $\boldsymbol{D}_{point}=[D_{n-k+1},D_{n-k+2},...,D_{n}]$. The likelihood function $\mathcal{L}(\boldsymbol{\theta},\sigma_{obs},\boldsymbol{\phi},\boldsymbol{\varphi})$ can be derived as
\begin{eqnarray}
\mathcal{L}(\boldsymbol{\theta},\sigma_{obs},\boldsymbol{\phi},\boldsymbol{\varphi}) \propto \Pr \big( (\boldsymbol{y}_{D,k} \in \Omega_k) \cap (\boldsymbol{y}_{D,n-k}=\boldsymbol{D}_{point}) | \boldsymbol{\theta},\sigma_{obs},\boldsymbol{\phi},\boldsymbol{\varphi}) \nonumber\\
= \int_{\Omega_k} \pi(\boldsymbol{y}_{D,k}|\boldsymbol{y}_{D,n-k}=\boldsymbol{D}_{point},\boldsymbol{X}_D, \boldsymbol{\theta},\sigma_{obs},\boldsymbol{\phi},\boldsymbol{\varphi}) \ \pi(\boldsymbol{X}_D) \ \mathrm{d}\boldsymbol{y}_{D,k} \ \mathrm{d}\boldsymbol{X}_D
\end{eqnarray}
where $\pi(\boldsymbol{y}_{D,k}|\boldsymbol{y}_{D,n-k}=\boldsymbol{D}_{point},\boldsymbol{X}_D, \boldsymbol{\theta},\sigma_{obs},\boldsymbol{\phi},\boldsymbol{\varphi})$ is obtained by substituting $\boldsymbol{y}_{D,n-k}=\boldsymbol{D}_{point}$ into the joint PDF of the random vector $\boldsymbol{y}_D$.

\subsubsection{Time series data}\label{section:CALtsdata}

In dynamic systems, information is commonly available in the form of time series data. This type of information leads to several additional challenges; in particular, the model prediction and the corresponding measurement data are dependent on the states of the system in the previous time steps, and replicate time series observations may be taken with a large number of time points.  Both of these characteristics may complicate the computation of the likelihood function. To perform model calibration with time-series data, we again use the KOH framework discussed in Section~\ref{section:CALbyBN}. In general, a dynamic model can be written as $y_{m,t} = G(\boldsymbol{y}_{m,-t},\boldsymbol{x},t;\boldsymbol{\theta})$; $y_{m,t}$ represents the model prediction at time $t$; $\boldsymbol{y}_{m,-t}$ represents the model predictions for the previous time steps; $\boldsymbol{x}$ and $\boldsymbol{\theta}$ are the same as in Section~\ref{section:CALbyBN}. Note that $y_{m,t}$ is deterministic for given values of $\boldsymbol{y}_{m,-t}$, $\boldsymbol{x}$, $t$ and $\boldsymbol{\theta}$. The Gaussian process used to approximate the model discrepancy function also becomes time dependent, and thus $\delta_{t} \sim \mathcal{N} \big( m(\boldsymbol{x},t;\boldsymbol{\phi}), k(\boldsymbol{x},\boldsymbol{x}',t,t'; \boldsymbol{\varphi}) \big)$. The measurement uncertainty is still represented as $\epsilon_{obs} \sim \mathcal{N} (0, \sigma_{obs}^2)$, which is time-independent. Similar to Eq.~\ref{eq:relationYandYD1}, the relationship between experimental observation $y_{D,t}$  and the corresponding model prediction $y_{m,t}$ can be written as
\begin{equation}\label{eq:relationYandYD2}
y_{D,t} = y_{m,t}+\delta_{t}+\varepsilon_{obs}
\end{equation}

The fact that $y_{m,t}$ is dependent on $\boldsymbol{y}_{m,-t}$ renders the construction of the likelihood function $\mathcal{L}(\boldsymbol{\theta}, \sigma_{obs}, \boldsymbol{\phi}, \boldsymbol{\varphi})$ difficult. For example, suppose we have one set of time-series data available $\boldsymbol{D}_t=[D_{t1},D_{t2},...,D_{tn}]$, i.e., the measurements are taken at several time points $\boldsymbol{t}_D=t_1,t_2,...,t_n$ when the values of the experimental inputs $\boldsymbol{x}_D$ remain the same. Note that again we consider that the actual data $\boldsymbol{D}_t$ are random realizations of the random vector $\boldsymbol{y}_{D,t}=[y_{D,t1},y_{D,t2},...,y_{D,tn}]$. The corresponding likelihood function can be written as
\begin{eqnarray}\label{eq:LKtimeseries}
&&\mathcal{L}(\boldsymbol{\theta}, \sigma_{obs}, \boldsymbol{\phi}, \boldsymbol{\varphi}) \propto \pi(\boldsymbol{y}_{D,t}=\boldsymbol{D}_t  | \boldsymbol{\theta}, \sigma_{obs}, \boldsymbol{\phi}, \boldsymbol{\varphi}) \nonumber\\
&=& \int \pi(\boldsymbol{y}_{D,t} = \boldsymbol{D}_t | \boldsymbol{y}_{m,-t},\boldsymbol{x}_D,\boldsymbol{\theta}, \sigma_{obs}, \boldsymbol{\phi},\boldsymbol{\varphi}) \ \pi(\boldsymbol{y}_{m,-t} | \boldsymbol{x}_D,\boldsymbol{\theta}) \ \pi(\boldsymbol{x}_D) \ \mathrm{d}\boldsymbol{y}_{m,-t} \ \mathrm{d}\boldsymbol{x}_D 
\end{eqnarray}
Note that if strictly written, $\boldsymbol{y}_{m,-t}$ in Eq.~\ref{eq:LKtimeseries} should be different for different data points. We chose not to write out each "$\boldsymbol{y}_{m,-t}$" to avoid making the equation unnecessarily complex. It can be seen that the PDF of model predictions for the past time points $\pi(\boldsymbol{y}_{m,-t} | \boldsymbol{x}_D, \boldsymbol{\theta})$ and an integration over all the elements of $\boldsymbol{y}_{m,-t}$ are needed, which makes the evaluation of the likelihood function analytically intractable, and numerically expensive methods (such as Monte Carlo simulation) may be needed. 

The extension to the cases where multiple sets of time series data are available is straightforward in theory. The difference from the case of single time series data is that $\boldsymbol{y}_{D,t}$, $\boldsymbol{y}_{m,-t}$ and $\boldsymbol{x}_D$ in Eq.~\ref{eq:LKtimeseries} become matrices instead of being vectors. In the special case that the multiple series are replicates, i.e., we have repeated measurements at the same time points for the same set of inputs, the variation from one series to another can be attributed to the observation noise $\varepsilon_{obs}$, and thus we can directly compute $\varepsilon_{obs}$ based on these repeated time series. Assuming that $\varepsilon_{obs}$ is a zero-mean Gaussian random variable with variance $\sigma_{obs}^2$ independent of time $t$, the variance $\sigma_{obs}^2$ can be estimated as 
\begin{equation}\label{eq:obsVARfromReplica}
\sigma_{obs}^2 = \frac{1}{n_1 (n_2-1)} \sum_{j=1}^{n_1} \sum_{i=1}^{n_2} [y_{D,t_j}^{i} - \frac{1}{n_2} \sum_{i=1}^{n_2} (y_{D,t_j}^i) ]^2
\end{equation}
where $n_2$ is the number of repeated time series, and $n_1$ is the number of time points in each series. The estimated $\sigma_{obs}$ and the average time series $(1/n_2) \sum_{i=1}^{n_2} (y_{D,t_j}^i)$ can be further used to compute the likelihood as in the single time series case.

If the measurement uncertainty in the experimental inputs is negligible, $\boldsymbol{x}_D$ can be treated as constant. Note that $\boldsymbol{y}_{m,-t}$ is deterministic for given values of $\boldsymbol{x}_D$ and $\boldsymbol{\theta}$ (initial condition is also considered as input). Then, the calibration with time series data $\boldsymbol{D}_t$ becomes much simpler, as Eq.~\ref{eq:LKtimeseries} can be simplified as
\begin{equation}\label{eq:LKtimeseries_simple}
\mathcal{L}(\boldsymbol{\theta}, \sigma_{obs}, \boldsymbol{\phi}, \boldsymbol{\varphi}) \propto \pi(\boldsymbol{y}_{D,t} = \boldsymbol{D}_t | \boldsymbol{x}_D,\boldsymbol{\theta}, \sigma_{obs}, \boldsymbol{\phi},\boldsymbol{\varphi})
\end{equation}
where the calculation of $\pi(\boldsymbol{y}_{D,t} = \boldsymbol{D}_t | \boldsymbol{x}_D,\boldsymbol{\theta}, \sigma_{obs}, \boldsymbol{\phi},\boldsymbol{\varphi})$ is very similar to the cases discussed in Section~\ref{section:CALbyBN}, and the related information can be found in Eq.~\ref{eq:observation1} and the surrounding discussion.

In the case that the time series data set is becoming available in real time during the operation of a system, i.e., the observation is made in real time and the model is used to make predictions for future time points, we can continuously calibrate (or update) the model using algorithms such as Kalman filter (for linear models), extended Kalman filter (for non-linear models), particle filter (sampling implementation of sequential Bayesian calibration)~\cite{Thrun2005}, etc. 

\subsection{Identifiability of model parameters}\label{section:identifiability}

Before implementing the calibration of a model, it is often of interest to determine whether we can extract useful information from the calibration results. A model is non-identifiable if there are infinite "best" (depending on the criterion chosen) estimates for the model parameters. In the Bayesian model calibration framework, the typical sign of non-identifiabilty is that the posterior PDFs of some of the model parameters are close to the prior PDFs, which indicates that the marginal likelihoods of these parameters are nearly flat and there is an infinite number of maximum likelihood estimates of the model parameters. In general, model non-identifiability can be classified into two levels, namely structural non-identifiability and practical non-identifiability~\cite{Raue2009}. The first level of non-identifiability, structural non-identifiability is due to the redundant parameterization of the model structure. Even if the model is structurally identifiable, a second level of non-identifiability, practical non-identifiability, can still arise due to the insufficient amount and quality of observation data. The quality of data is related to the bias and noise in the data due to the imprecision of measurement techniques. Successful detection of structural non-identifiability may help reduce model redundancy. Also, by detecting the existence of practical non-identifiability, analysts may be able to overcome this issue by developing better design of experiments or improving data quality~\cite{Arendt2011}.

It is usually straightforward to detect structural non-identifiability if the analytical expression of a model is available; however, in many problems, the analytical expression of the model is not readily available. One example is a dynamic model without an explicit steady state solution. Another possible case occurs when we add a discrepancy function to the numerical solutions of some governing equations, which in fact forms a new model without any analytical expression to consider~\cite{Arendt2010}. Various analytical and numerical methods have been developed to detect the structural non-identifiability of dynamic models~\cite{Grewal1976,Walter1996,Jia-fan2011}, whereas the second possible case does not appear to have been studied in the literature. This section addresses this case. 

Since the second level of non-identifiability, practical non-identifiability, is related to both model structure and observation data, it is necessary to inspect the likelihood function in order to determine whether some parameters of a model are practically non-identifiable.  In fact, rigorous definitions of model non-identifiability can be constructed based on the analytical properties of likelihood functions~\cite{Gu1994,Paulino1994,Little2010}. In addition to the theoretical analysis of likelihood functions, Raue et al.~\cite{Raue2009,Raue2011} developed a numerical approach based on the concept of "profile likelihood"~\cite{Murphy2000}, which has been shown to be effective in detecting practical non-identifiability. When the analytical expression of the likelihood function is available, or its numerical evaluation is trivial, it may be preferable to apply the profile likelihood-based method and determine the practical non-identifiability directly. But this method becomes cumbersome when the construction of the likelihood function is computationally expensive, since repetitive evaluations of the likelihood function are required to compute the profile likelihood.

Given the above observations, we propose a first-order Taylor series expansion-based method, which can detect structural non-identifiability for models without analytical expressions, and can detect practical non-identifiability due to insufficient amount of data. This method does not involve computation of the likelihood function, and thus is simpler to implement and less computationally demanding. The limitations of this method are: (1) it uses a linear approximation of the model, and hence may fail to detect non-identifiability if the model is highly nonlinear; (2) it can only detect local non-identifiability as the Taylor series expansion is constructed based on the derivatives at a single point; (3) it does not apply to statistical models; and (4) it does not cover practical non-identifiability due to the \textit{quality} of data. 

Suppose the physics model to be calibrated is $y_m=G(\boldsymbol{x};\boldsymbol{\theta})$, and a Gaussian process discrepancy function $\delta \sim \mathcal{N} \big( m(\boldsymbol{x};\boldsymbol{\phi}), k(\boldsymbol{x},\boldsymbol{x}'; \boldsymbol{\varphi}) \big)$ is added to the model. Thus, a new model is formed as $G_{new}(\boldsymbol{x};\boldsymbol{\psi})=G(\boldsymbol{x};\boldsymbol{\theta})+m(\boldsymbol{x};\boldsymbol{\phi})$, where $\boldsymbol{\psi}=[\boldsymbol{\theta},\boldsymbol{\phi}]$ includes the parameters of the physics model ($\boldsymbol{\theta}$) and the parameters of the mean of the discrepancy function ($\boldsymbol{\phi}$). In the case that the measurement noise is a zero mean random variable, $G_{new}$ is the expectation of observation $y_D$ according to Eq.~\ref{eq:relationYandYD1}. Further assume that the analytical form of $G_{new}$ is not available. In such cases, the first-order Taylor series expansion of this model (as shown in Eq.~\ref{eq:1stTaylor}) can be used as an efficient approximation when the model is believed to be not highly nonlinear:
\begin{eqnarray}\label{eq:1stTaylor}
E[y_D] = G_{new}(\boldsymbol{x}_D;\boldsymbol{\psi}) \approx G_{new}(\boldsymbol{x}_D;\hat{\boldsymbol{\psi}}) + \sum_{i=1}^{p} \frac{\partial{G_{new}}}{\partial{\psi_i}} \bigg|_{\boldsymbol{\psi}=\hat{\boldsymbol{\psi}}} \psi_i
\end{eqnarray} 
where $\hat{\boldsymbol{\psi}}$ can be the mean value of the prior of $\boldsymbol{\psi}$, and $p$ is the number of model parameters.

Suppose there are $n$ data points available, i.e., experimentally observed values $\boldsymbol{D}=[D_1,D_2,...,D_n]$ corresponding to different input settings $\boldsymbol{X}_D = [\boldsymbol{x}_{D1},\boldsymbol{x}_{D2},...,\boldsymbol{x}_{Dn}]$. Without considering measurement uncertainty and the variance of the model discrepancy function, we can construct a linear system as
\begin{eqnarray}\label{eq:linearSYS1}
\boldsymbol{A} \boldsymbol{\psi}^T = \boldsymbol{D}, \quad\boldsymbol{A} = \begin{bmatrix}
\frac{\partial{G_{new}}}{\partial{\psi_1}} \big|_{\boldsymbol{x}_{D1},\hat{\boldsymbol{\psi}}} & ... & \frac{\partial{G_{new}}}{\partial{\psi_p}} \big|_{\boldsymbol{x}_{D1},\hat{\boldsymbol{\psi}}}\\
\vdots & \ddots & \vdots \\
\frac{\partial{G_{new}}}{\partial{\psi_1}} \big|_{\boldsymbol{x}_{Dn},\hat{\boldsymbol{\psi}}} & ... & \frac{\partial{G_{new}}}{\partial{\psi_p}} \big|_{\boldsymbol{x}_{Dn},\hat{\boldsymbol{\psi}}}
\end{bmatrix}
\end{eqnarray}

The linear system in Eq.~\ref{eq:linearSYS1} can be underdetermined or determined, depending on the rank of the matrix $\boldsymbol{A}$ (denoted as $r_{A}$). If $r_{A} < p$, the system is underdetermined and there will be an infinite number of $\boldsymbol{\psi}$ values satisfying Eq.~\ref{eq:linearSYS1}; if $r_{A} = p$, the system is determined and there will be a unique vector $\boldsymbol{\psi}$ satisfying Eq.~\ref{eq:linearSYS1}. The latter case suggests that the model is practically identifiable given the available data points (assuming the quality of the data does not cause non-identifiability). The former case suggests that the model is non-identifiable either due to the model structure or insufficient data. If the inequality $r_{A} < p$ continue to hold as we increase the number of observation data, then it can be inferred that the model is structurally non-identifiable. 

\doublespacing
In order to help researchers reduce model redundancy once a model is detected as structurally non-identifiable, it may be of interest to know which set of parameters can/cannot be identified. Using the formulation of the linear system in Eq.~\ref{eq:linearSYS1}, we can retrieve this information by checking the linear dependency between the column vectors of the matrix $\boldsymbol{A}$, since the $i$-th column of $\boldsymbol{A}$ corresponds to the parameter $\psi_i$. For example, if the $i$-th column vector $\boldsymbol{a}_i$ and the $j$-th column vector $\boldsymbol{a}_j$ are linearly dependent, it is apparent that the corresponding parameters $\psi_i$ and $\psi_j$ are non-identifiable using the linear model in Eq.~\ref{eq:1stTaylor}. We can also find one set of identifiable parameters using the simple algorithm below (note that there may be multiple sets of identifiable parameters)

\singlespacing
\begin{algorithm}                      
\caption{Find one set of identifiable parameters}          
\label{alg:Iden}                           
\begin{algorithmic}                    
    \Require The first-order derivative matrix $\boldsymbol{A}$
    \Ensure The index set of identifiable parameters $\boldsymbol{I}$
    \State $\boldsymbol{A}_{temp} = \boldsymbol{A}$
    \State $\boldsymbol{I}$ = empty set
    \For{$i=1$ to $p$}
    \State $r_1$ = the rank of $\boldsymbol{A}_{temp}$
    \State Remove the $i$-th column from $\boldsymbol{A}_{temp}$
    \State $r_2$ = the rank of $\boldsymbol{A}_{temp}$ (with the $i$-th column removed)
    \If{$r_1 > r_2$}
        \State Add the value of $i$ to the set $\boldsymbol{I}$ as an element
    \EndIf
    \EndFor
    \State \Return $\boldsymbol{I}$
\end{algorithmic}
\end{algorithm}
\doublespacing
To illustrate the proposed method, consider the following two test models:

Model 1: $y_m=(\psi_1+\psi_2)x_1+ 2\psi_1 \psi_3 x_2^2 + \psi_1$

Model 2: $y_m=(\psi_1+\psi_2)x_1+ 2\psi_1 \psi_3 x_2^2$

It is not difficult to see from the model expressions that the three parameters of model 1 are identifiable given no less than three observations of $y$ (the physical quantity to be predicted using model 1 and model 2) corresponding to different combinations of the inputs $x_1$ and $x_2$, whereas the three parameters of model 2 are not identifiable no matter how many data points are available. But in order to illustrate the proposed method, we assume that the analytical expressions of these two models are unavailable.

Suppose the measurement data of $y$ are taken at three points: $[x_1,x_2]^1_D=[5,2]$, $[x_1,x_2]^2_D=[6,3]$, $[x_1,x_2]^3_D=[7,1]$, we can calculate the derivatives $\partial y_m/\partial \psi_i$ numerically (e.g., forward difference or central difference) at these input points for given values of the parameters, and thus obtain the matrix $\boldsymbol{A}$. For example, for $[\psi_1,\psi_2,\psi_3]=[2,3,4]$, the matrix $\boldsymbol{A}$ for model 1 and model 2 are as follows
\begin{equation*}
\boldsymbol{A_1} = \begin{bmatrix}
38 & 5 & 16 \\
79 & 6 & 36 \\
16 & 7 & 4 \\
\end{bmatrix} \qquad
\boldsymbol{A_2} = \begin{bmatrix}
37 & 5 & 16 \\
78 & 6 & 36 \\
15 & 7 & 4 \\
\end{bmatrix}
\end{equation*}
The rank of $\boldsymbol{A_1}$ is equal to 3, whereas the rank of $\boldsymbol{A_2}$ is equal to 2. Thus we can infer that the parameters of model 1 are identifiable, but the parameters of model 2 are not identifiable. We can also use the program in Algorithm~\ref{alg:Iden} to infer that $\psi_2$ and $\psi_3$ of model 2 are identifiable if the value of $\psi_1$ is given.

\subsection{Computational issues}\label{section:compIssue}

Bayesian calibration in Eq.~\ref{eq:Bayes} can be computationally expensive due to two reasons: (1) the likelihood function may be expensive to compute numerically, and (2) the multivariate integration in the denominator of Eq.~\ref{eq:Bayes} can be time consuming if the number of parameters is large. 

The use of a Gaussian process to quantify the model discrepancy term as shown in Section~\ref{section:CALbyBN} can lead to a high dimensional-parameter space, as a set of parameters $\boldsymbol{\phi}$ and $\boldsymbol{\varphi}$ which characterize the Gaussian process also needs to be estimated in addition to the actual physics model parameters $\boldsymbol{\theta}$. For the case that the data points are sparse, it may not be feasible to calibrate the model along with the estimation of these parameters of the Gaussian process. A compromised solution is to use a simplified model discrepancy function with less flexibility, i.e., a smaller number of parameters. Another possible method is to estimate the model parameters $\boldsymbol{\theta}$ and the parameters of the Gaussian process $\boldsymbol{\phi}$ in two sequential steps. First, the model parameters $\boldsymbol{\theta}$ are calibrated without considering model discrepancy. Then, we can estimate $\boldsymbol{\phi}$ and $\boldsymbol{\varphi}$ based on the \textit{a posteriori} estimate of $\boldsymbol{\theta}$ (denoted as $\boldsymbol{\theta}^*$), i.e., we can obtain the posterior PDF of $\boldsymbol{\phi}$ and $\boldsymbol{\varphi}$, $\pi(\boldsymbol{\phi},\boldsymbol{\varphi}|\boldsymbol{\theta}^*)$, which is conditioned on $\boldsymbol{\theta}^*$. 


The likelihood function represents the probabilistic relationship between measured data and unknown parameters, and repeated runs of the computer model $G(\boldsymbol{x};\boldsymbol{\theta})$ are required to compute this relationship. Hence, previous studies have mostly focused on approximating the computational model with a surrogate model~\cite{Kennedy2001,Sankararaman2010}, i.e., replacing the physics-based model $G(\boldsymbol{x};\boldsymbol{\theta})$ with a faster model without losing much accuracy. Surrogate modeling techniques that have been developed in literature include Kriging or Gaussian Process (GP) interpolation~\cite{Rasmussen2006}, polynomial chaos expansion~\cite{Xiu2002,Ghanem2003,Marzouk2009}, support vector machine (SVM)~\cite{Vapnik1999}, relevance vector machine~\cite{Tipping2001}, adaptive sparse grid collocation~\cite{Ma2009}, etc. Then, the likelihood function of the parameters can be evaluated based on executing the surrogate model a number of times. 

If the measurement uncertainty is the only source of uncertainty considered and can be represented using a Gaussian random variable, the likelihood function can be calculated analytically based on the model predictions. However, in the case that various sources of uncertainty exist (e.g., natural variability in the input $\boldsymbol{x}$, data uncertainty in input and output measurement, and model uncertainty), the likelihood function is no longer simple to compute (e.g., Eq.~\ref{eq:LK-ex1}). In that case, sampling methods like Monte Carlo simulation are needed to compute the function for given parameter values. If the number of calibration parameters is relatively large, the evaluation of the likelihood function can become expensive even with a fast surrogate model for $G(\boldsymbol{x};\boldsymbol{\theta})$. In such cases, another surrogate model can be built to directly approximate the joint likelihood function of all the parameters, based on actual evaluations of the likelihood function for selected values of the parameters. Thereafter, we can evaluate this surrogate model, instead of the actual likelihood function, in the calculation of the posterior PDFs, which can speed up Bayesian calibration under multiple sources of uncertainty. For example, Bliznyuk et al.~\cite{Bliznyuk2008} approximated the unnormalized posterior density (the product of likelihood function and prior density)  using radial basis functions.

If the number of parameters is relatively small, the integration of the product of the likelihood function and the prior PDFs of parameters can be conducted accurately and efficiently using numerical integration methods, such as Gaussian quadrature or the trapezoidal rule. When the number of parameters becomes large, Markov Chain Monte Carlo (MCMC) methods are widely used due to the relative insensitivity of the computational effort to the number of parameters. MCMC methods do not conduct the integration explicitly, but instead directly generate the random samples from the unnormalized posterior density of the parameters, upon convergence. Several algorithms are available for MCMC sampling, including Metropolis-Hastings~\cite{Chib1995,Green1995,Gelman1996,Haario2006,Zuev2012}, Gibbs~\cite{Casella1992}, slice sampling~\cite{Neal2003}, etc.

\section{Calibration of multi-physics computational models}\label{section:calMP}

Multi-physics modeling usually involves the combination of several models from different individual physics analyses. Ideally, these models would be calibrated separately with input-output experimental data corresponding to individual models. But in practice the experimental data may not be sufficient or available for some of the models. To calibrate all the models with limited information, a Bayesian network-based method is proposed below. The techniques discussed in Section~\ref{section:BayesTech} will be employed. 

\subsection{Integration of multi-physics models and experimental data via Bayesian network}\label{section:intModel&BN}

\begin{figure}[h!]
\begin{center}
\includegraphics[trim=55mm 94mm 55mm 52mm, clip=true, width=0.9\textwidth]{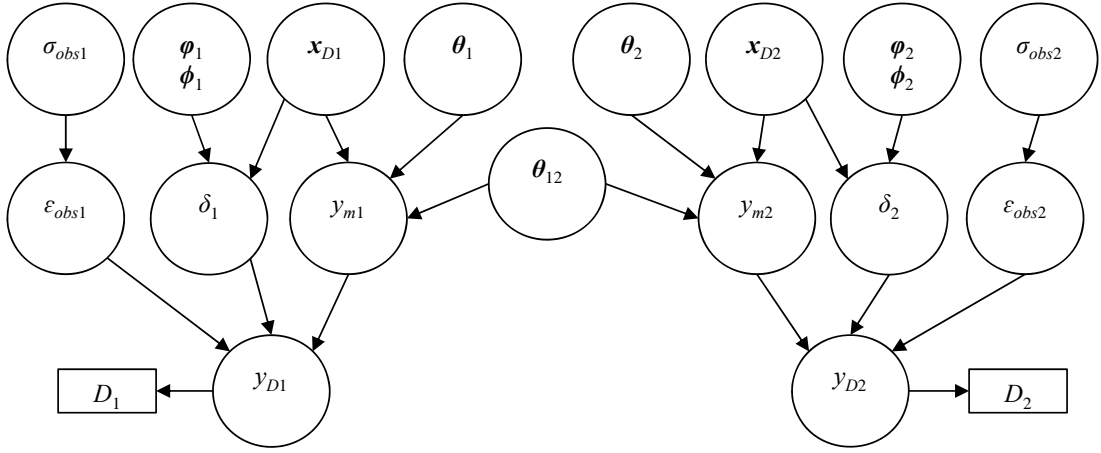}
\end{center}
\caption{Bayesian network for two physics models}
\label{fig:BN_2}
\end{figure}

Suppose we have two physics models $y_{m1}=G_1(\boldsymbol{x}_1;\boldsymbol{\theta}_1,\boldsymbol{\theta}_{12})$ and $y_{m2}=G_2(\boldsymbol{x}_2;\boldsymbol{\theta}_2,\boldsymbol{\theta}_{12})$. Note that these two models have different input variables ($\boldsymbol{x}_1$ versus $\boldsymbol{x}_2$) and parameters ($\boldsymbol{\theta}_1$ versus $\boldsymbol{\theta}_2$), but they also share some common parameters $\boldsymbol{\theta}_{12}$. Based on the discussion in Section~\ref{section:CALbyBN}, two Bayesian networks can be constructed for these two models individually. Further, due to the existence of the common parameters, these two networks can be connected to form a full network as shown in Fig.~\ref{fig:BN_2}, which enables information flow from one network to the other.

\subsection{Strategy of Bayesian calibration for multi-physics models}\label{section:stratBayesMP}

If both the observation data $D_{1}$ and $D_{2}$ are available, we have three options for model calibration based on the Bayesian networks in Fig.~\ref{fig:BN_2}, as presented below. 

The first option is to calibrate the two models simultaneously. Let $\boldsymbol{\Phi}_1$ and $\boldsymbol{\Phi}_2$ represents the calibration parameters of the two networks respectively except for the common parameters $\boldsymbol{\theta}_{12}$, i.e., $\boldsymbol{\Phi}_1 = [\boldsymbol{\theta}_1, \sigma_{obs1},\boldsymbol{\phi}_1,\boldsymbol{\varphi}_1]$, $\boldsymbol{\Phi}_2=[\boldsymbol{\theta}_2,\sigma_{obs2},\boldsymbol{\phi}_2,\boldsymbol{\varphi}_2]$.
\begin{eqnarray}\label{eq:CALoption1}
&& \pi(\boldsymbol{\Phi}_1,\boldsymbol{\Phi}_2,\boldsymbol{\theta}_{12}|y_{D1}=D_1,y_{D2}=D_2) \\
&& = \frac{\pi(y_{D1}=D_1,y_{D2}=D_2|\boldsymbol{\Phi}_1,\boldsymbol{\Phi}_2,\boldsymbol{\theta}_{12}) \ \pi(\boldsymbol{\Phi}_1) \ \pi(\boldsymbol{\Phi}_2) \ \pi(\boldsymbol{\theta}_{12})}{ \int \pi(y_{D1}=D_1,y_{D2}=D_2|\boldsymbol{\Phi}_1,\boldsymbol{\Phi}_2,\boldsymbol{\theta}_{12}) \ \pi(\boldsymbol{\Phi}_1) \ \pi(\boldsymbol{\Phi}_2) \ \pi(\boldsymbol{\theta}_{12}) \ \mathrm{d}\boldsymbol{\Phi}_1 \ \mathrm{d}\boldsymbol{\Phi}_2 \ \mathrm{d}\boldsymbol{\theta}_{12}} \nonumber
\end{eqnarray}

From the first option, we can obtain the posterior PDFs of all the parameters using both $D_{1}$ and $D_{2}$. However, this option can be computationally expensive because of the high dimensional parameter space.

The second option is to let information flow from left to right, conducting a two-step calibration. Following the procedure of Bayesian calibration for a single model presented in Section~\ref{section:CALbyBN}, $\boldsymbol{\Phi}_1$ and $\boldsymbol{\theta}_{12}$ are first calibrated using the observation data $D_{1}$. Then, the marginal posterior PDF of $\boldsymbol{\theta}_{12}$, $\pi(\boldsymbol{\theta}_{12}|y_{D1}=D_1)$, is used as the prior when we calibrate the parameters of the other network ($\boldsymbol{\Phi}_2$ and $\boldsymbol{\theta}_{12}$). Applying Bayes' theorem, we have
\begin{eqnarray}\label{eq:CALoption2}
&& \pi(\boldsymbol{\theta}_{12}|y_{D1}=D_1) = \frac{\int \pi(y_{D1}=D_1|\boldsymbol{\Phi}_1,\boldsymbol{\theta}_{12}) \ \pi(\boldsymbol{\Phi}_1) \ \pi(\boldsymbol{\theta}_{12}) \ \mathrm{d}\boldsymbol{\Phi}_1 } {\int \pi(y_{D1}=D_1|\boldsymbol{\Phi}_1,\boldsymbol{\theta}_{12}) \ \pi(\boldsymbol{\Phi}_1) \ \pi(\boldsymbol{\theta}_{12}) \ \mathrm{d}\boldsymbol{\Phi}_1 \ \mathrm{d}\boldsymbol{\theta}_{12}} \nonumber\\
&& \pi'(\boldsymbol{\Phi}_2,\boldsymbol{\theta}_{12} | y_{D1}=D_1, y_{D2}=D_2) \nonumber\\
&& = \frac{\pi(y_{D2}=D_2|\boldsymbol{\Phi}_2,\boldsymbol{\theta}_{12}) \ \pi(\boldsymbol{\Phi}_2) \ \pi(\boldsymbol{\theta}_{12}|y_{D1}=D_1)}{\int \pi(y_{D2}=D_2|\boldsymbol{\Phi}_2,\boldsymbol{\theta}_{12}) \ \pi(\boldsymbol{\Phi}_2) \ \pi(\boldsymbol{\theta}_{12}|y_{D1}=D_1) \ \mathrm{d}\boldsymbol{\Phi}_2 \ \mathrm{d}\boldsymbol{\theta}_{12}}
\end{eqnarray}

We can prove that Eq.~\ref{eq:CALoption2} gives the same joint posterior PDF of $\boldsymbol{\Phi}_2$ and $\boldsymbol{\theta}_{12}$ as from Eq.~\ref{eq:CALoption1}. By combining the two expressions in Eq.~\ref{eq:CALoption2}, we have
\begin{eqnarray}\label{eq:CALoption2proof}
&& \pi'(\boldsymbol{\Phi}_2,\boldsymbol{\theta}_{12} | y_{D1}=D_1, y_{D2}=D_2) \nonumber\\
&& = \frac{\pi(y_{D2}=D_2|\boldsymbol{\Phi}_2,\boldsymbol{\theta}_{12}) \pi(\boldsymbol{\Phi}_2) \int \pi(y_{D1}=D_1|\boldsymbol{\Phi}_1,\boldsymbol{\theta}_{12}) \pi(\boldsymbol{\Phi}_1) \pi(\boldsymbol{\theta}_{12}) \mathrm{d}\boldsymbol{\Phi}_1 } {\Big( \int \pi(y_{D2}=D_2,\boldsymbol{\Phi}_2,\boldsymbol{\theta}_{12} | y_{D1}=D_1) \mathrm{d}\boldsymbol{\Phi}_2 \mathrm{d}\boldsymbol{\theta}_{12} \Big) \Big( \int \pi(y_{D1}=D_1|\boldsymbol{\Phi}_1,\boldsymbol{\theta}_{12}) \pi(\boldsymbol{\Phi}_1) \pi(\boldsymbol{\theta}_{12}) \mathrm{d}\boldsymbol{\Phi}_1 \mathrm{d}\boldsymbol{\theta}_{12} \Big) } \nonumber\\
&& = \int \frac{\pi(y_{D1}=D_1|\boldsymbol{\Phi}_1,\boldsymbol{\theta}_{12}) \ \pi(y_{D2}=D_2|\boldsymbol{\Phi}_2,\boldsymbol{\theta}_{12}) \ \pi(\boldsymbol{\Phi}_1) \ \pi(\boldsymbol{\Phi}_2) \ \pi(\boldsymbol{\theta}_{12}) }       {\pi(y_{D2}=D_2|y_{D1}=D_1) \ \pi(y_{D1}=D_1)} \ \mathrm{d}\boldsymbol{\Phi}_1 \nonumber\\
&& = \int \frac{\pi(y_{D1}=D_1,y_{D2}=D_2|\boldsymbol{\Phi}_1,\boldsymbol{\Phi}_2,\boldsymbol{\theta}_{12}) \ \pi(\boldsymbol{\Phi}_1) \ \pi(\boldsymbol{\Phi}_2) \ \pi(\boldsymbol{\theta}_{12}) }       {\pi(y_{D2}=D_2|y_{D1}=D_1) \ \pi(y_{D1}=D_1)} \ \mathrm{d}\boldsymbol{\Phi}_1 \nonumber\\
&& = \int \pi(\boldsymbol{\Phi}_1,\boldsymbol{\Phi}_2,\boldsymbol{\theta}_{12}|y_{D1}=D_1,y_{D2}=D_2) \ \mathrm{d}\boldsymbol{\Phi}_1
\end{eqnarray}

Therefore, the second option provides us the posterior PDF of $\boldsymbol{\Phi}_1$ based on $D_{1}$, and the posterior PDFs of $\boldsymbol{\Phi}_2$ and $\boldsymbol{\theta}_{12}$ based on both $D_{1}$ and $D_{2}$. Note that the computational effort in the second option can be much smaller than in the first option, due to the reduced number of parameters in each step of the calibration.

The third option is similar to the second one, except that the information flows from right to left, i.e., $\boldsymbol{\Phi}_2$ and $\boldsymbol{\theta}_{12}$ are first calibrated using the observation data $D_{2}$, and then the marginal posterior PDF of $\boldsymbol{\theta}_{12}$, $\pi(\boldsymbol{\theta}_{12}|y_{D2}=D_2)$, is used as prior in the calibration of $\boldsymbol{\Phi}_2$ and $\boldsymbol{\theta}_{12}$ using the data $D_1$. Hence, the posterior PDFs of $\boldsymbol{\Phi}_1$ and $\boldsymbol{\theta}_{12}$ are obtained using both $D_1$ and $D_{2}$, whereas the posterior PDF of $\boldsymbol{\Phi}_2$ is only based on $D_2$.

Note that although the above Bayesian network-based method is presented using a two-model problem, it can be extended to the cases that $N (N \ge 2)$ physics models are to be calibrated with limited information, and there will be up to $N!+1$ options of calibration available, depending on the existence of common parameters between different models.
\subsection{Summary of contributions}
In Section~\ref{section:CALbyBN}, we presented the basic theory of model calibration using Bayesian network following the KOH framework. In Section~\ref{section:BayesTech}, methods were developed to address various issues in the practical applications of Bayesian model calibration: (1) calibration using interval data or time series data (Section~\ref{section:diffTypeData}); (2) identifiability of model parameters (Section~\ref{section:identifiability}; and (3) computational difficulty and possible solutions (Section~\ref{section:compIssue}). In Section~\ref{section:calMP}, a Bayesian network-based approach was developed for the calibration of multiple physics models. It is shown that the available data can be used more efficiently, since the data of different physical quantities can be exchanged through the Bayesian network and thus we can calibrate the parameters of one model using information from experiments that are related to other physics models.
\section{Numerical examples}\label{section:NumExp}
We illustrate the methods presented in the previous sections using two numerical examples. In Section~\ref{subsection:NumExpDielectric}, we calibrate a transient dielectric charging model~\cite{Palit2012} with time series data to illustrate the Bayesian approach discussed in Section~\ref{section:CALtsdata}. In Section~\ref{subsection:NumExp-2}, we implement Bayesian calibration of multi-physics models, based on the methods presented in Section~\ref{section:intervalData} and~\ref{section:calMP}. Two types of radio-frequency (RF) microelectromechanical system (MEMS) devices are used in this example. Examination of model parameter identifiability is performed in both examples using the first-order Taylor series expansion-based method proposed in Section~\ref{section:identifiability}.
\subsection{Calibration of a dielectric charging model with time series data}\label{subsection:NumExpDielectric}
Dielectric charging has been identified as an important failure mechanism of RF MEMS switches, causing the switches to either remain stuck or fail to actuate~\cite{Jain2011,Hartzell2011}. Since our focus is on the calibration method instead of the physics aspects of dielectric charging, we only describe what is related to calibration (model inputs, unknown model parameters, model output, and observation data). Details of the mechanism and model development are provided in~\cite{Jain2011,Palit2012}. The model has two input variables (voltage $V$, temperature $T$), six unknown parameters (trap density $N_T$, barrier height $\Phi_B$, capture cross section $\sigma$, Frenkel-Poole (FP) attempt frequency $\gamma$, high frequency dielectric constant $\varepsilon_{INF}$, and effective mass $m^*$), and a single output variable (transient current density $J_t$ at time $t$). Experiments were conducted on a 200-nm silicon nitride ($\text{Si}_3\text{N}_4$) dielectric with 2 mm*2 mm area for 12 different combinations of $V$ and $T$, and these experiments were repeated for four times. The transient current density was measured at about 190 discrete time points between 0 and 100 seconds. Therefore, a large data set with size $n \approx 12*4*190=9120$ is available, which is typical for time-dependent problems. Since we have four repeated time series observations for each combination of inputs ($V$ and $T$), and each series contains measurements at the same time points, the replicates can be utilized directly to compute the measurement noise statistics as shown in Eq.~\ref{eq:obsVARfromReplica}. In this example, the number of time points in one series $n_1 \approx 190$, and the number of repeated time series $n_2=4$.
 
As presented in Section~\ref{section:CALtsdata}, the model discrepancy term is approximated by a time-dependent Gaussian process (GP) discrepancy function $\delta_{t} \sim \mathcal{N} \big( m(\boldsymbol{x},t;\boldsymbol{\phi}), k(\boldsymbol{x},\boldsymbol{x'},t,t';\boldsymbol{\varphi}) \big)$. In this example, we select the form of the mean function and covariance function based on a heuristic approach. First, the parameters of the dielectric charging model are estimated using the method of least squares without considering any uncertainty source. Then, we compute the difference between model predictions (based on least squares parameter estimation) and the corresponding experimental data. By plotting the difference versus the input variables, we obtain a rough idea about the trend of the model discrepancy function. In this example, we observe three issues: (1) the discrepancy between model prediction and observation appears to vary exponentially with respect to time; (2) there is significant statistical correlation in time; and (3) the variance of the discrepancy varies with time. Therefore, we adopt the combination of a linear function of model inputs and an exponential function of time to model the mean of the Gaussian process discrepancy function. In addition to the squared exponential covariance function shown in Eq.~\ref{eq:sqExpCovFun} (denoted as $k_1$), a time-variant term $k_2(t;\boldsymbol{w})=w_1 \exp(-w_2 t)$ is also used in order to account for the non-stationary trend of variance with respect to time. Thus the selected mean and covariance functions are:
\begin{eqnarray}\label{eq:meanAndCovGPnumExp}
&&m(\boldsymbol{x},t;\boldsymbol{\phi}) = \phi_1 V + \phi_2 T + \phi_3 t + \phi_4 \exp(\phi_5 t) \nonumber\\
&&k(\boldsymbol{x},\boldsymbol{x'},t,t'; \boldsymbol{\varphi}) = 
\begin{cases}
k_1( \boldsymbol{x},\boldsymbol{x'},t,t';\boldsymbol{\varphi}_1), & t \ne t'\\
k_1( \boldsymbol{x},\boldsymbol{x'},t,t';\boldsymbol{\varphi}_1)+ k_2(t;\boldsymbol{w}), & t=t'
\end{cases}
\end{eqnarray}
where the parameters of the covariance function are $\boldsymbol{\varphi} = [\boldsymbol{\varphi}_1,\boldsymbol{w}]$, and $\boldsymbol{\varphi}_1=[\lambda,l_1,l_2,l_3]$. 

Note that a check of identifiability is needed after selecting the mean function $m(\boldsymbol{x},t;\boldsymbol{\phi})$, since the addition of the discrepancy function to the original model may cause non-identifiability. In this example, there are 17 unknown parameters to calibrate, i.e., $p=17$, and the corresponding matrix $\boldsymbol{A}$ is full rank, which suggests that the combination of the dielectric model and the discrepancy function is identifiable. In fact, the reason that there is no constant term in the linear mean function is because the constant term is not identifiable according to the first-order Taylor series expansion-based method developed in Section~\ref{section:identifiability}. 

Once the forms of the above functions are selected, the joint likelihood function of the dielectric charging model parameters and the parameters of $\delta_{t}$ is then formulated as in Eq.~\ref{eq:LKtimeseries_simple}.  Note that the likelihood is proportional to the joint probability of all observations conditioned on model inputs and parameters, and thus the construction of likelihood requires computing the determinant and inverse of the covariance matrix of these data points. If all the data points are used, the size of the covariance matrix will rise to around $2280*2280$, which can cause several numerical difficulties, including matrix singularity and expensive computation of matrix determinant. We bypass such numerical difficulties with the large set of time series data by including only a subset of the time points in the likelihood construction. These points are selected in a manner which reflects all the features of the dynamic response as closely as possible; however, the precise number of points is largely a matter of computational convenience. In this case, we select measurements at $8$ time points from each data series, and thus the size of the covariance matrix reduces to $96*96$. Note that more advanced methods, which approximate the original covariance matrix with a sparse matrix while taking into account the whole data set, can be found in~\cite{Quinonero-Candela2005}.

\begin{figure}[h!]
\begin{center}
\subfigure[$m^*$]{
\includegraphics[width=0.15\textwidth]{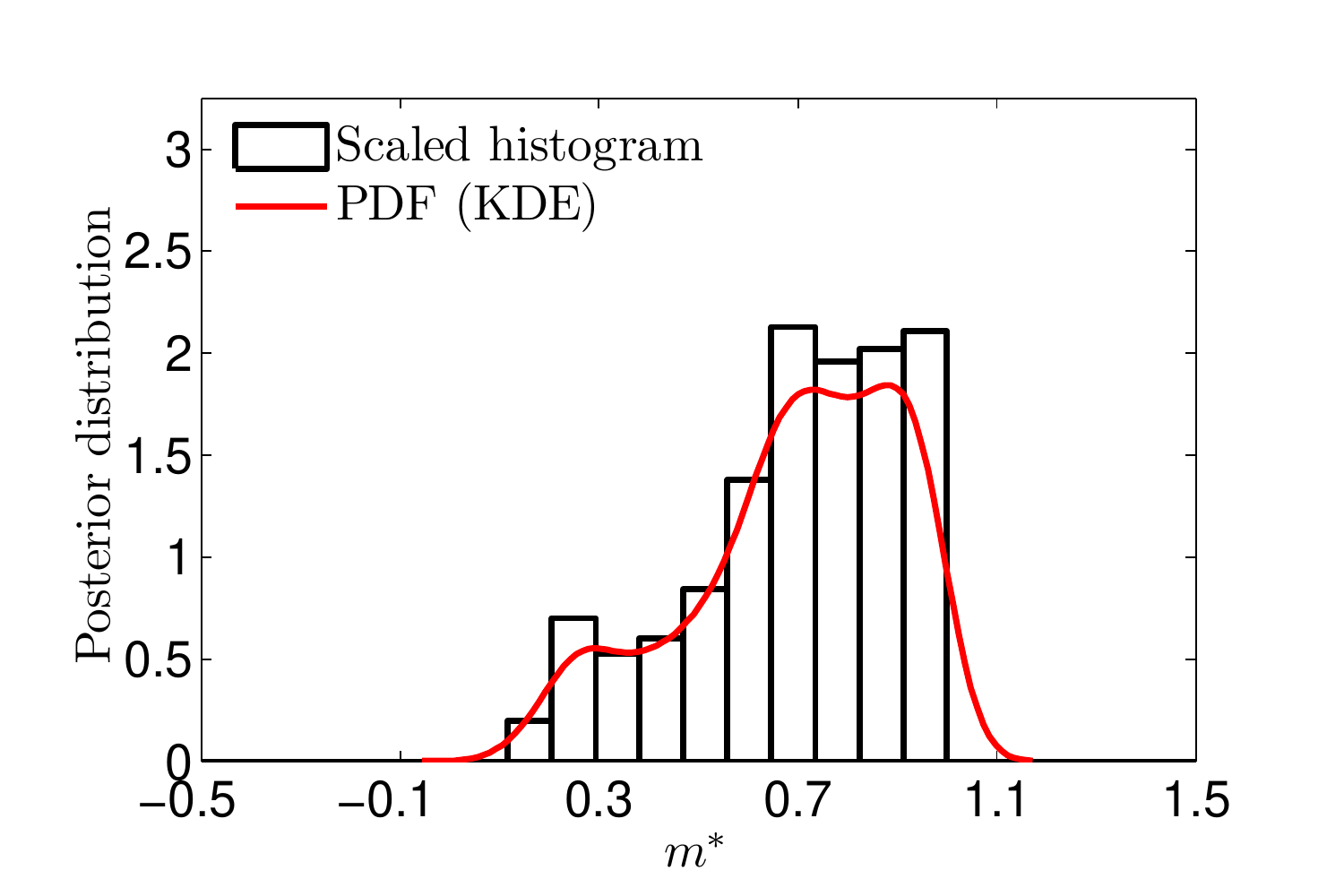}}
\subfigure[$\Phi_b$]{
\includegraphics[width=0.15\textwidth]{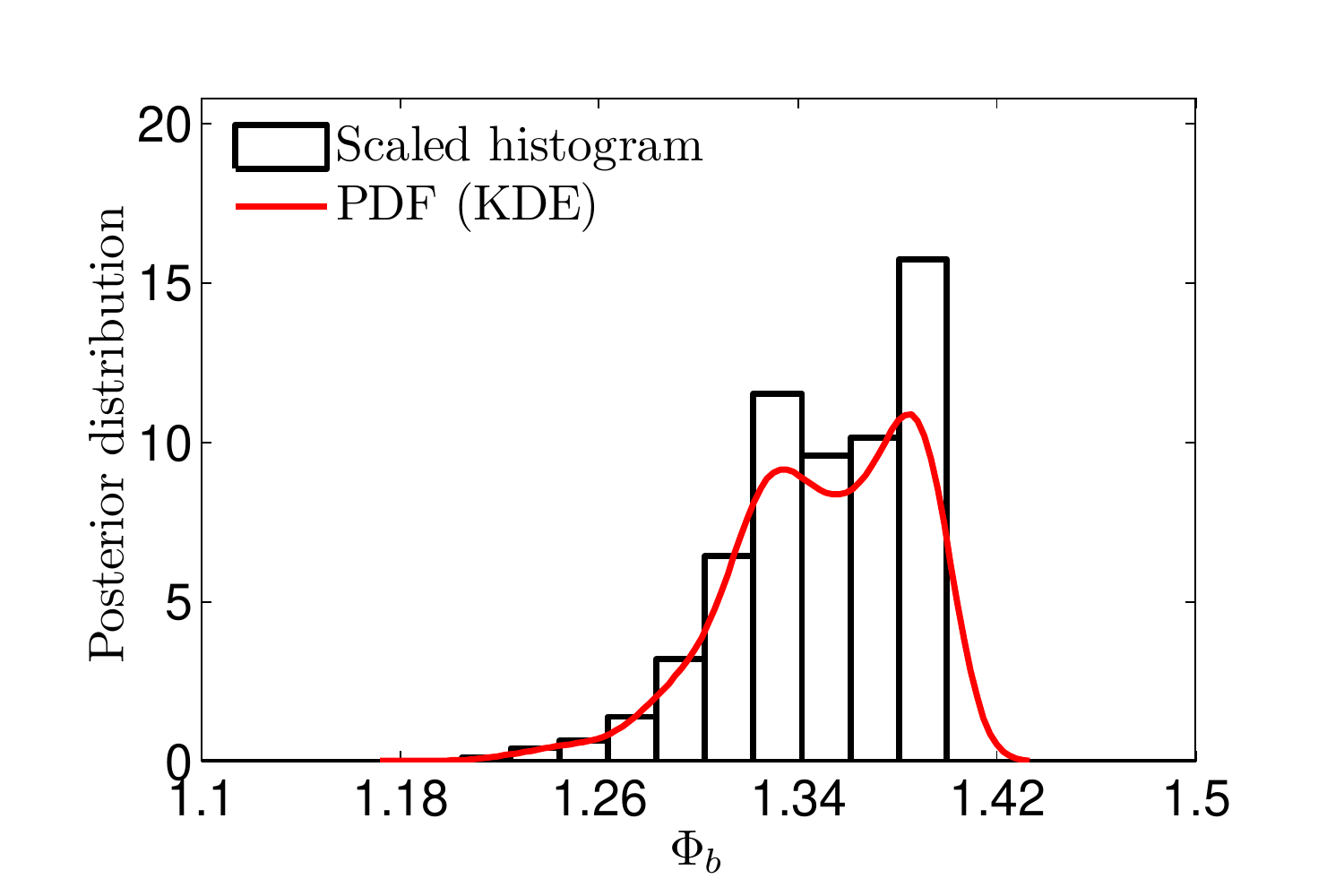}}
\subfigure[$\gamma$]{
\includegraphics[width=0.15\textwidth]{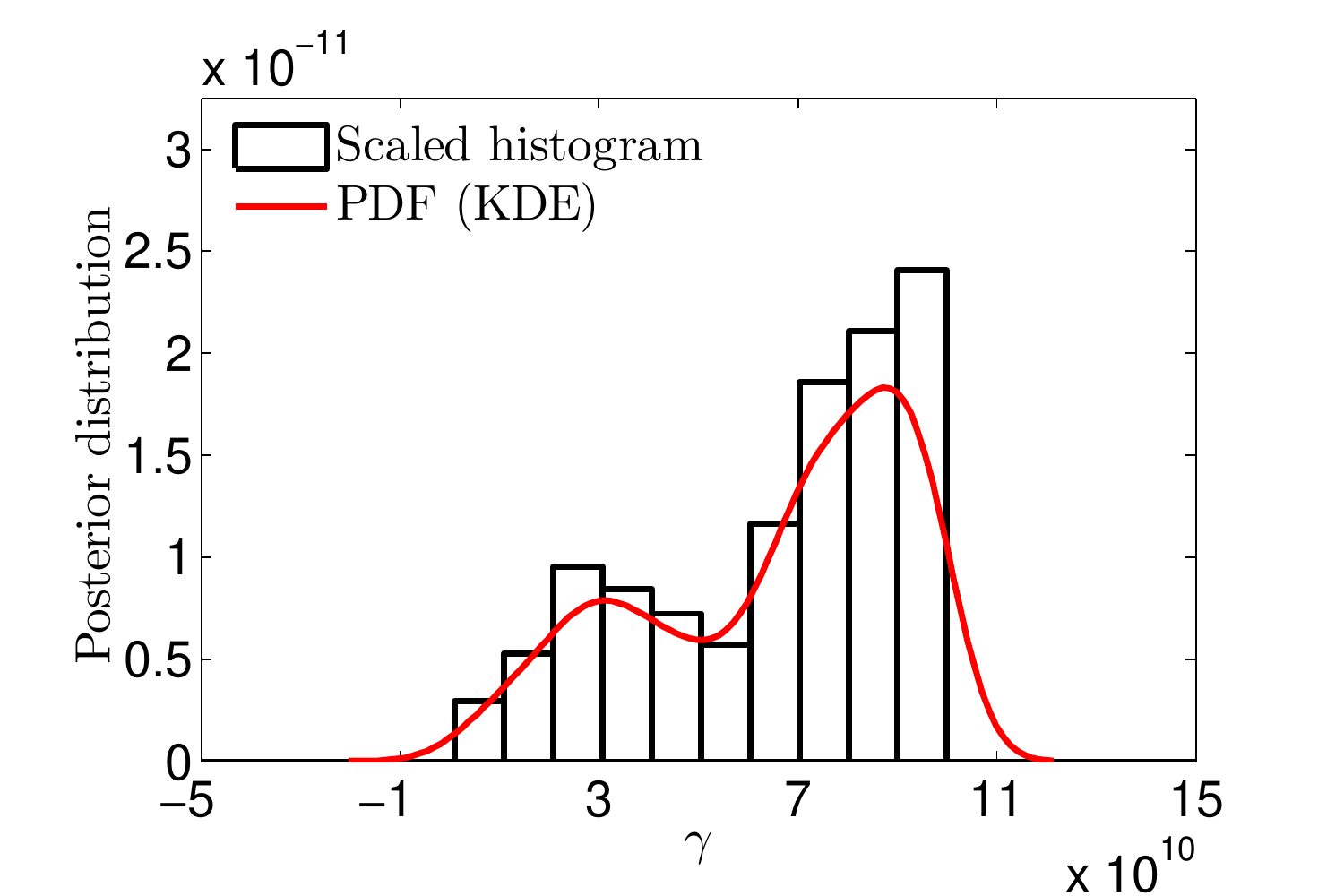}}
\subfigure[$\sigma$]{
\includegraphics[width=0.15\textwidth]{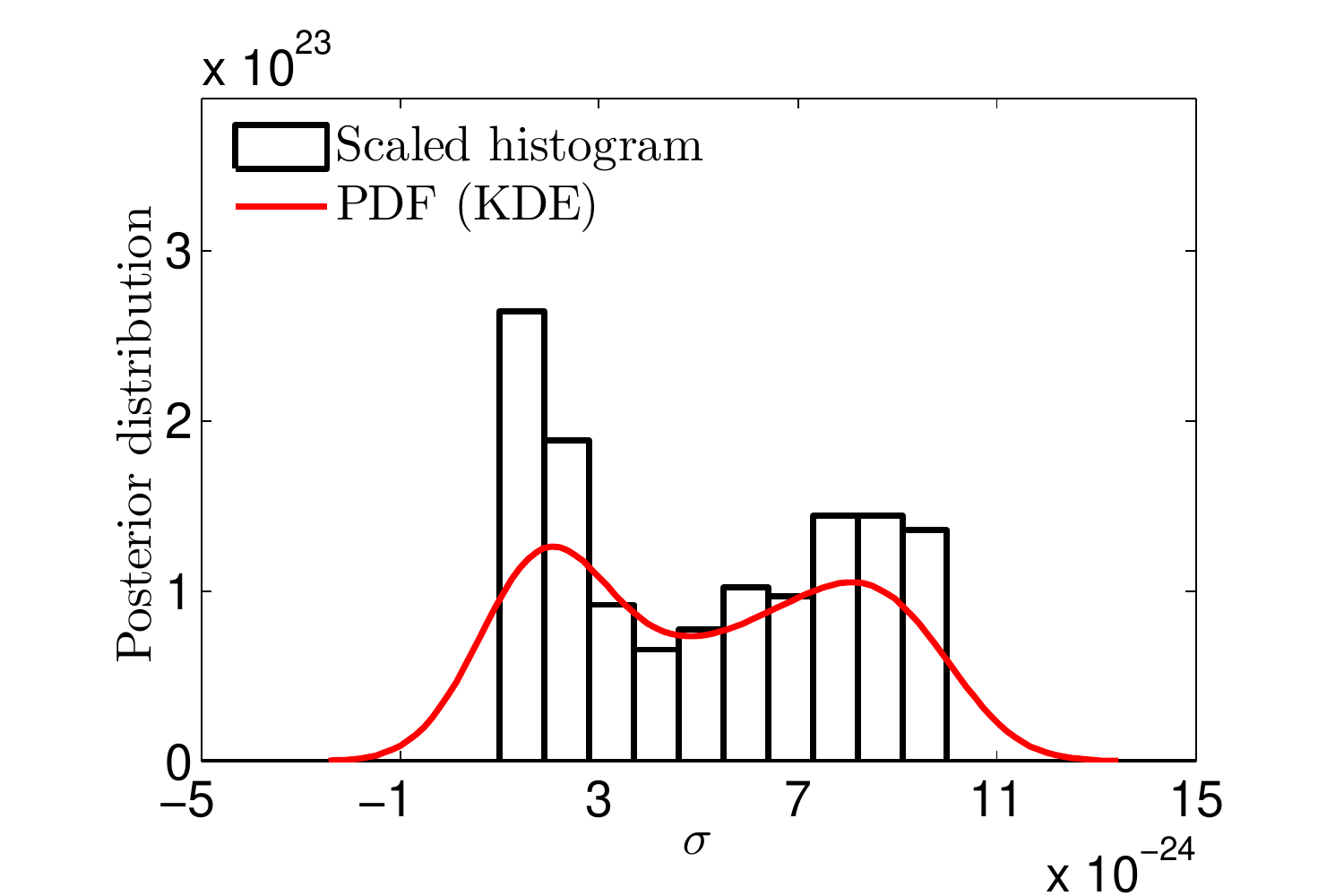}}
\subfigure[$\varepsilon_{inf}$]{
\includegraphics[width=0.15\textwidth]{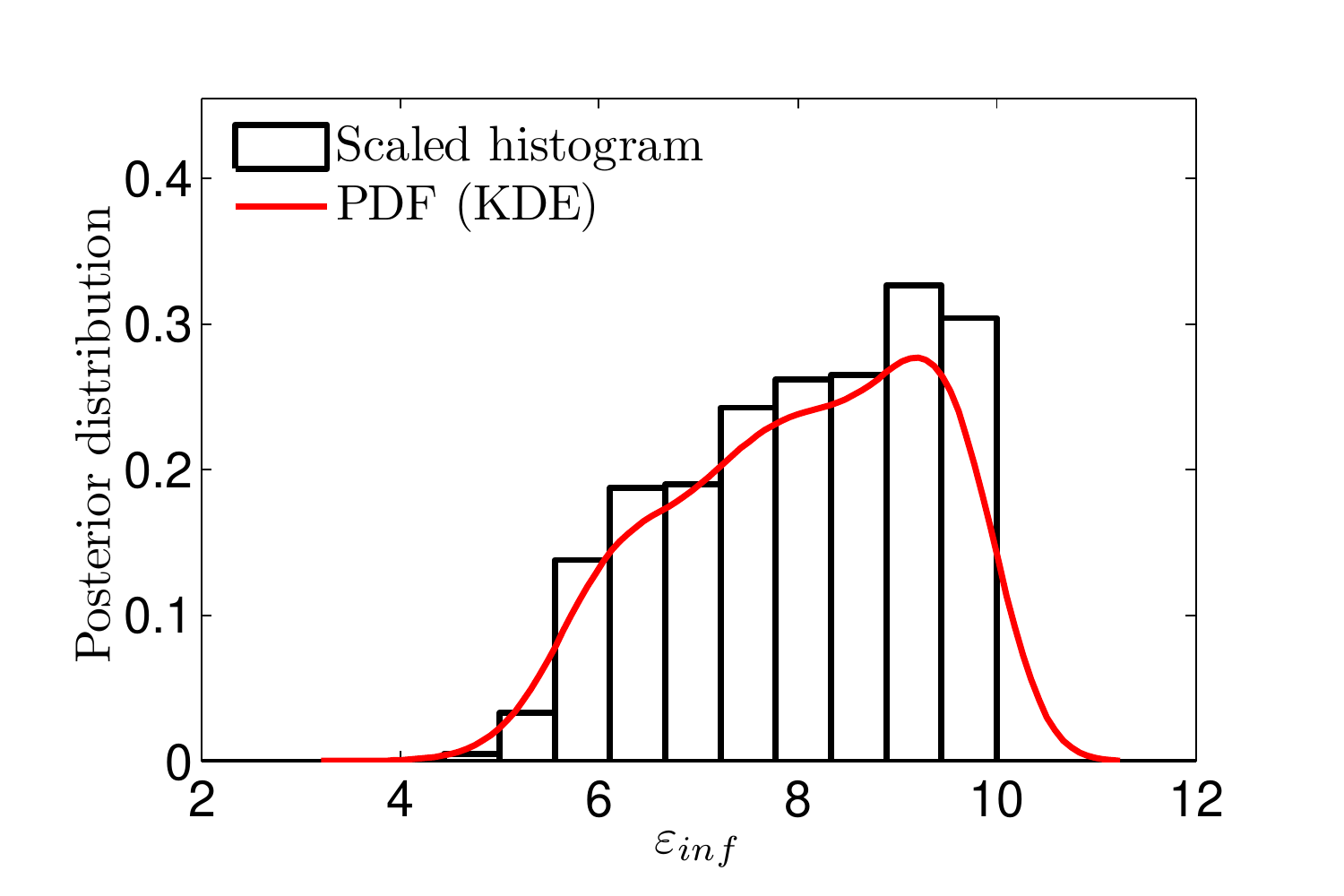}}
\subfigure[$N_T$]{
\includegraphics[width=0.15\textwidth]{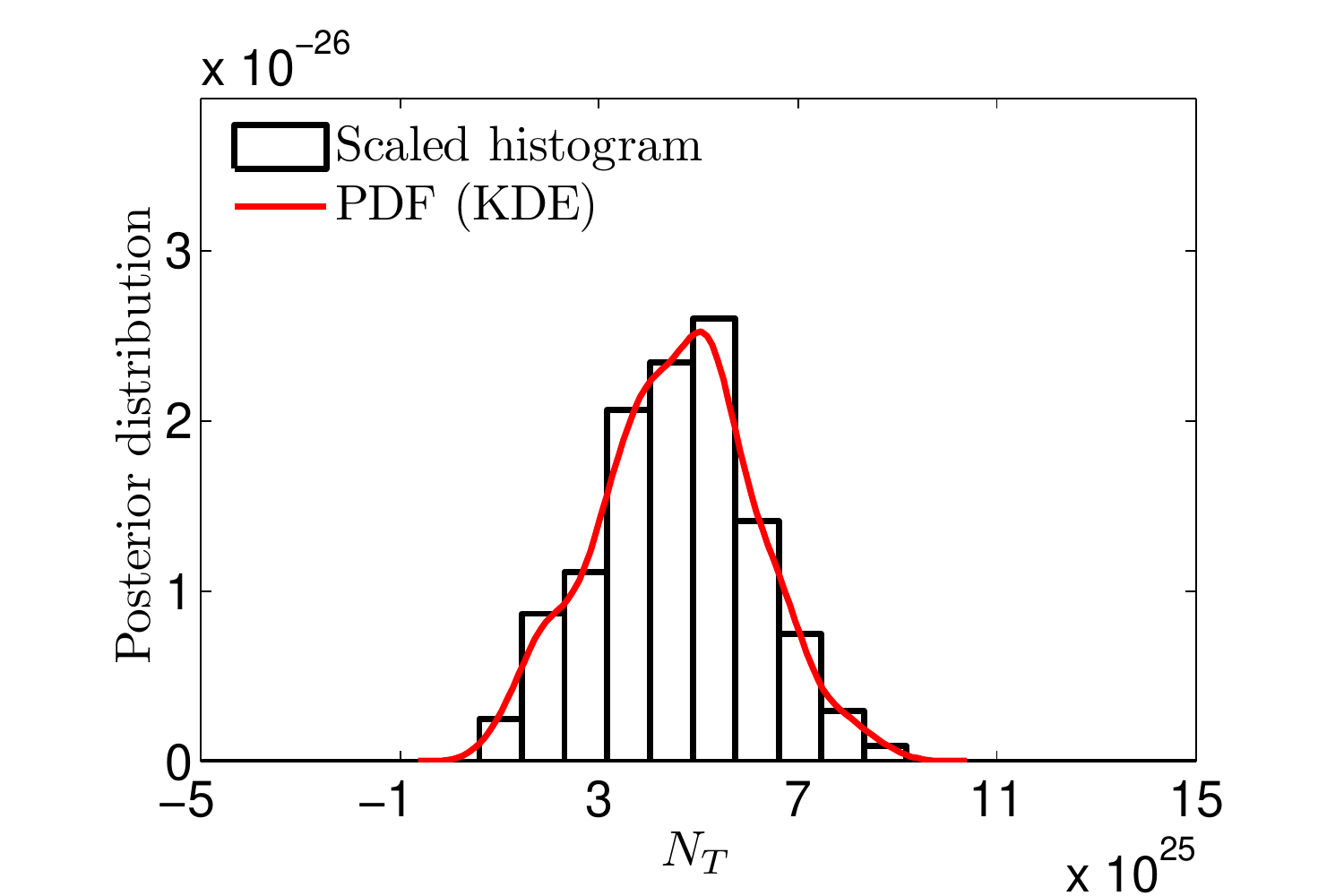}}\\
\subfigure[$\phi_1$]{
\includegraphics[width=0.15\textwidth]{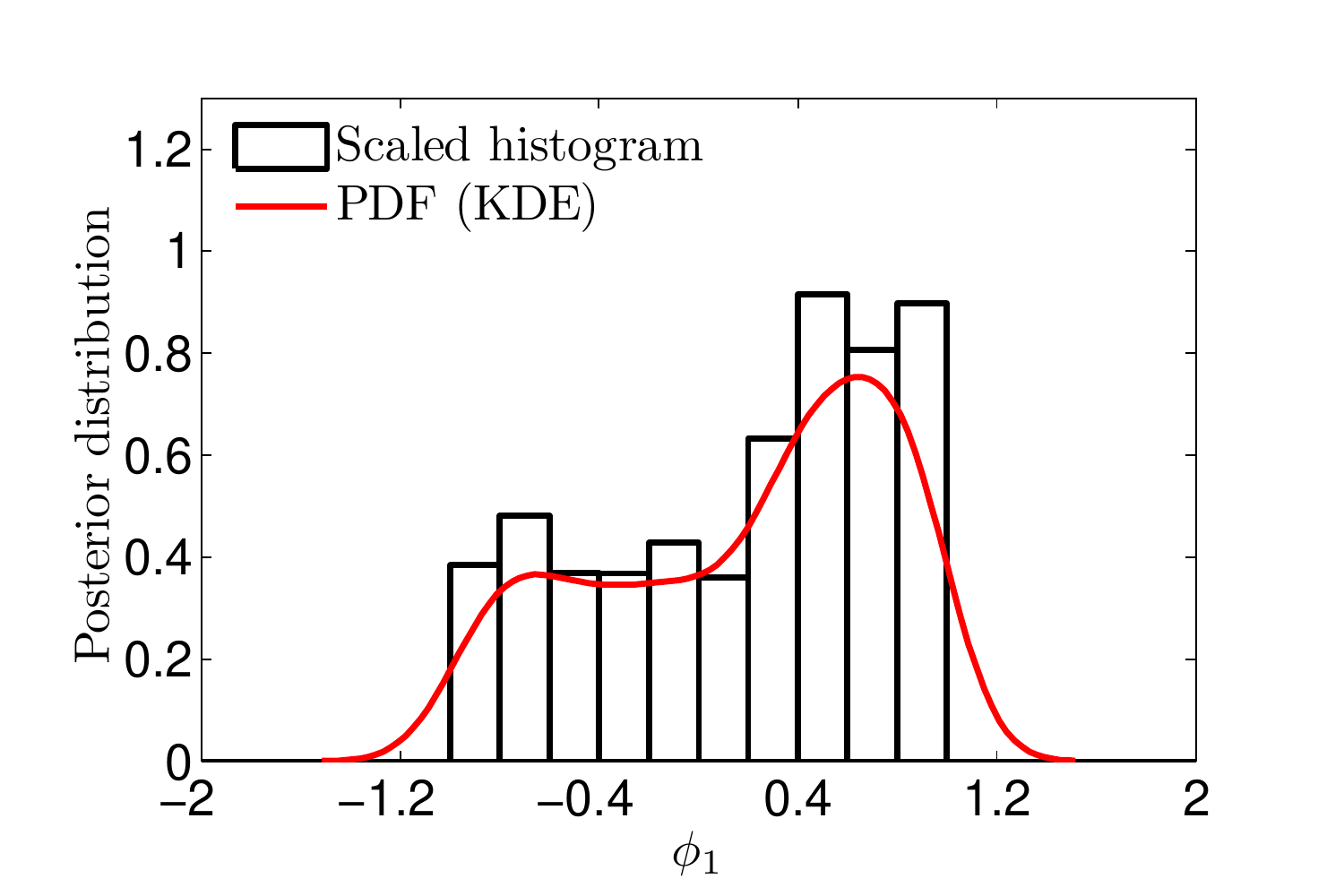}}
\subfigure[$\phi_2$]{
\includegraphics[width=0.15\textwidth]{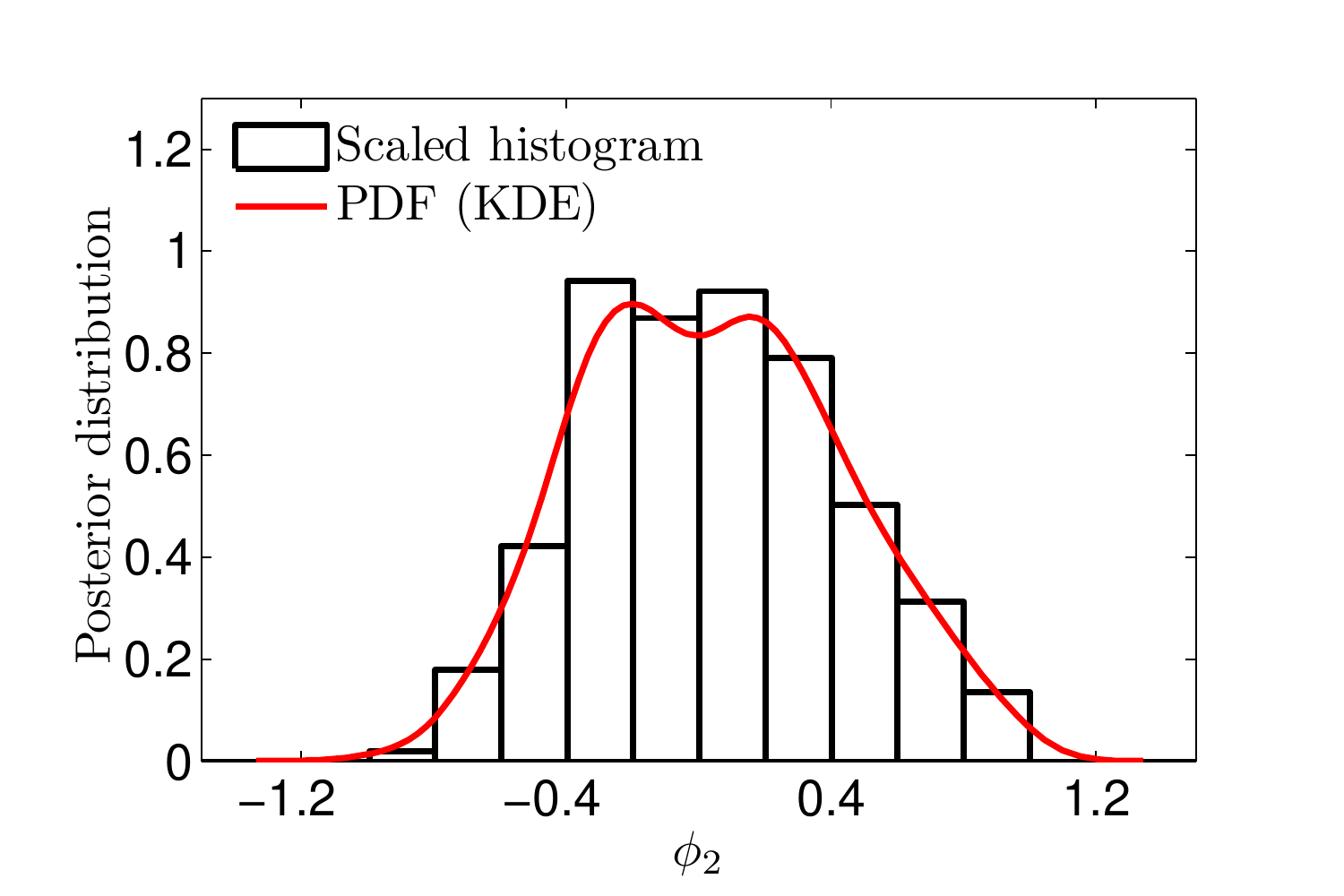}} 
\subfigure[$\phi_3$]{
\includegraphics[width=0.15\textwidth]{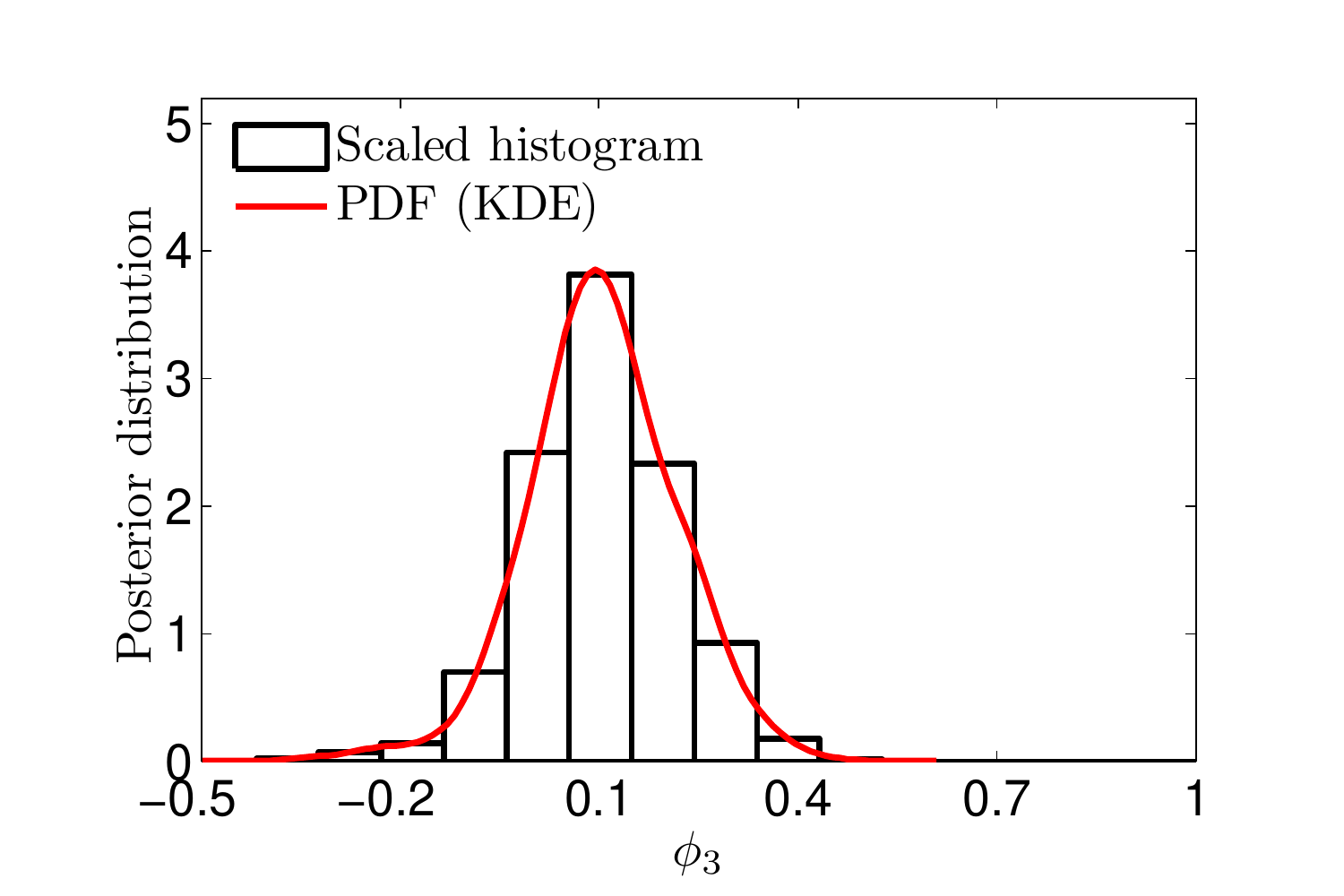}}
\subfigure[$\phi_4$]{
\includegraphics[width=0.15\textwidth]{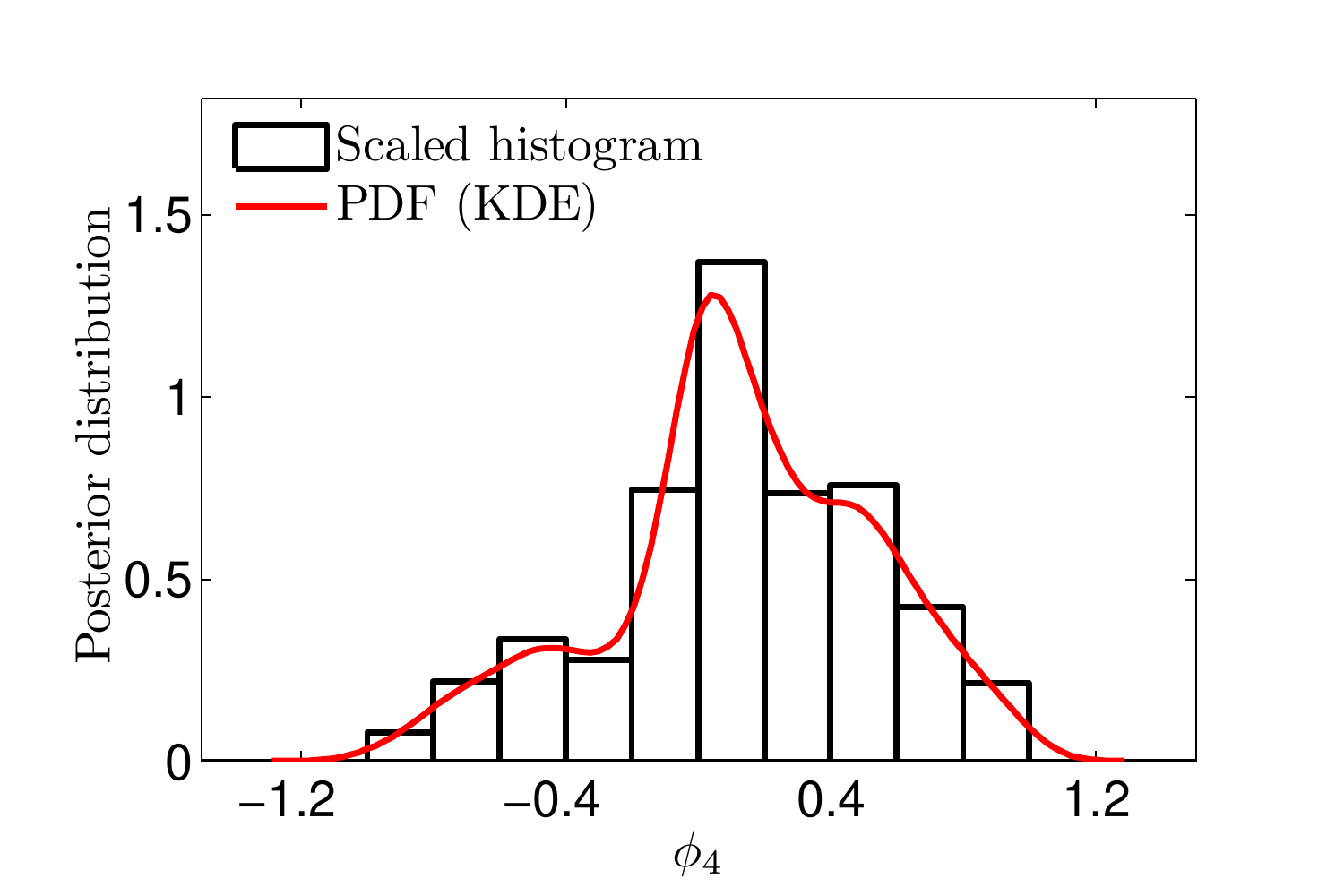}}
\subfigure[$\phi_5$]{
\includegraphics[width=0.15\textwidth]{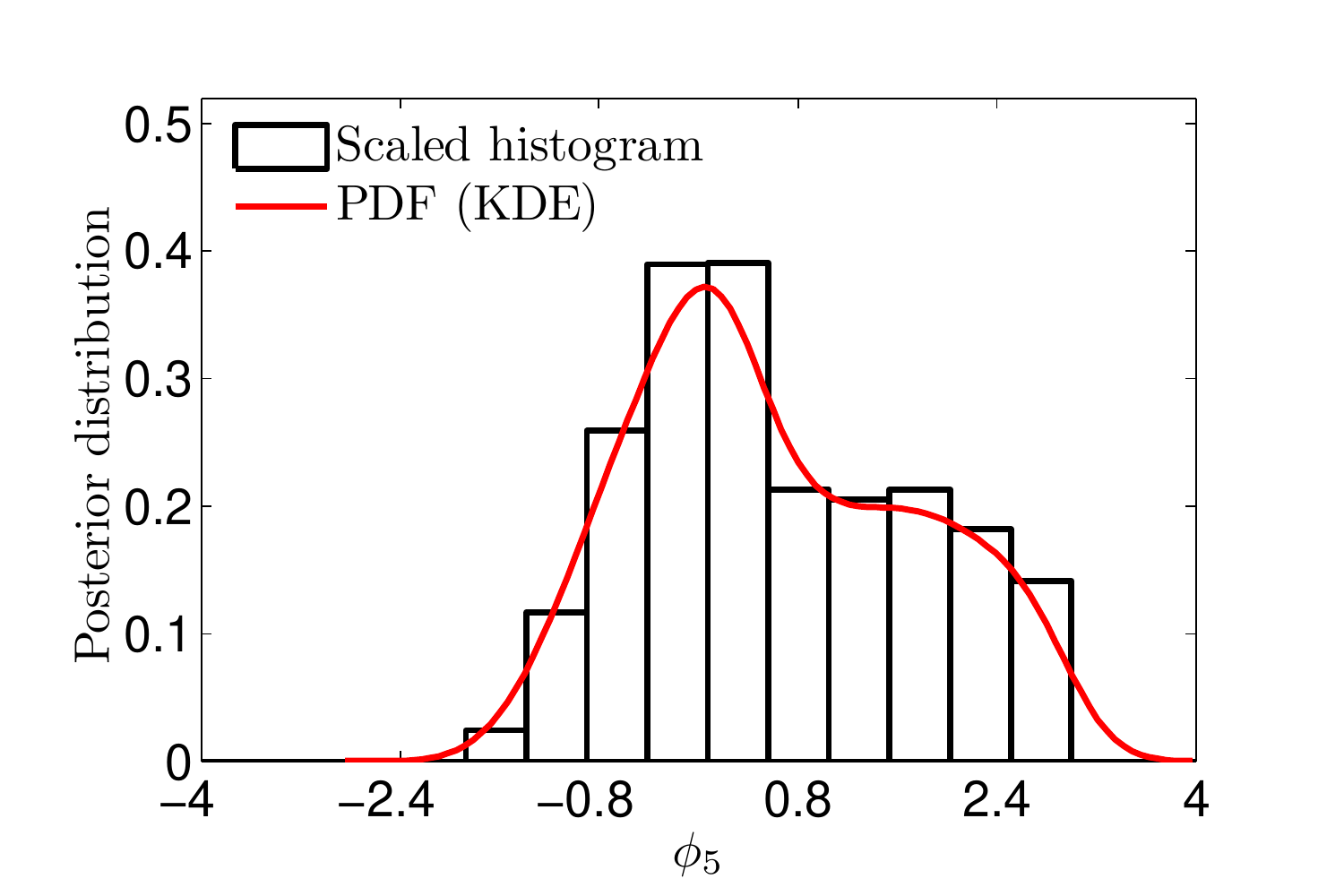}}
\subfigure[$\lambda$]{
\includegraphics[width=0.15\textwidth]{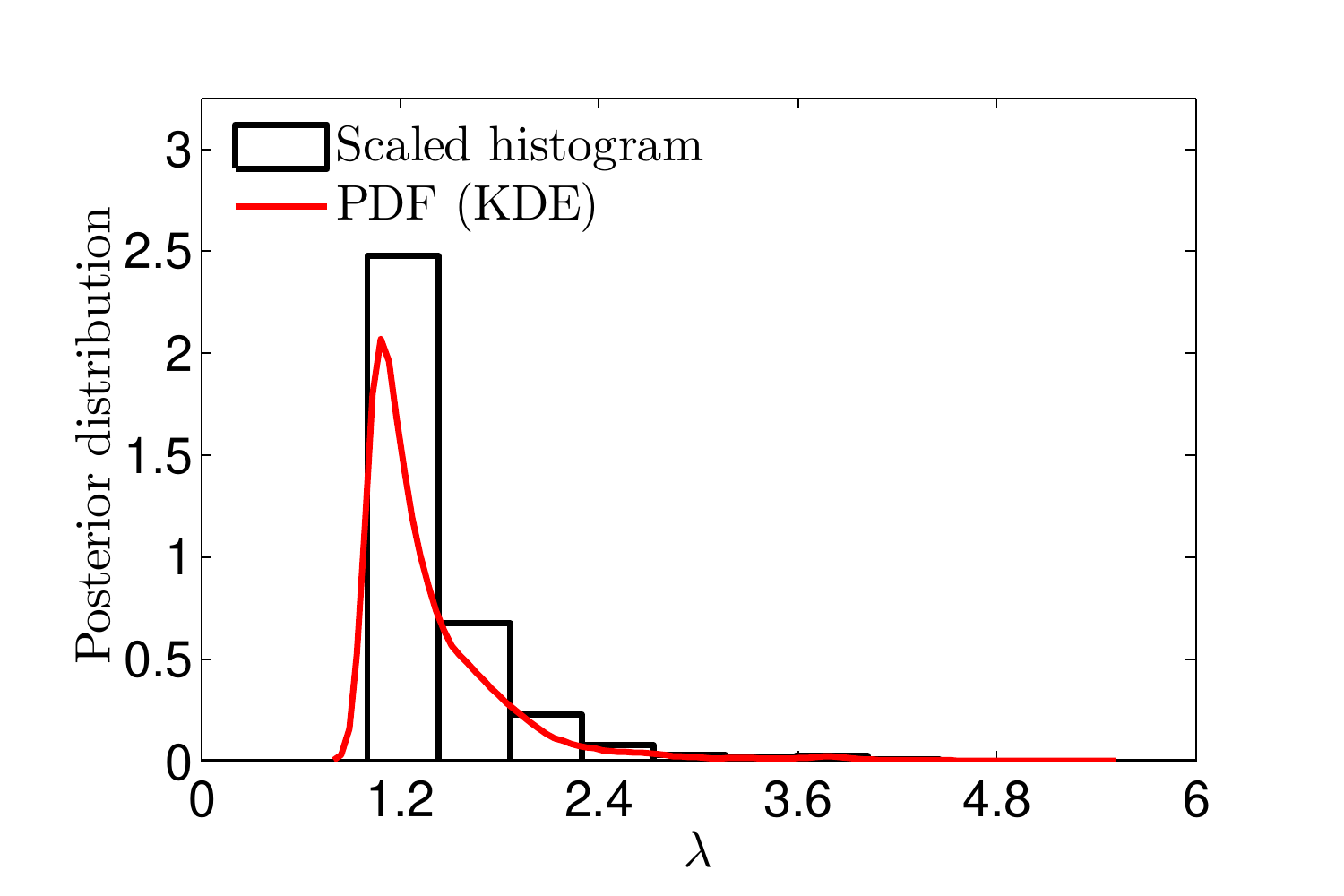}}\\
\subfigure[$l_1$]{
\includegraphics[width=0.15\textwidth]{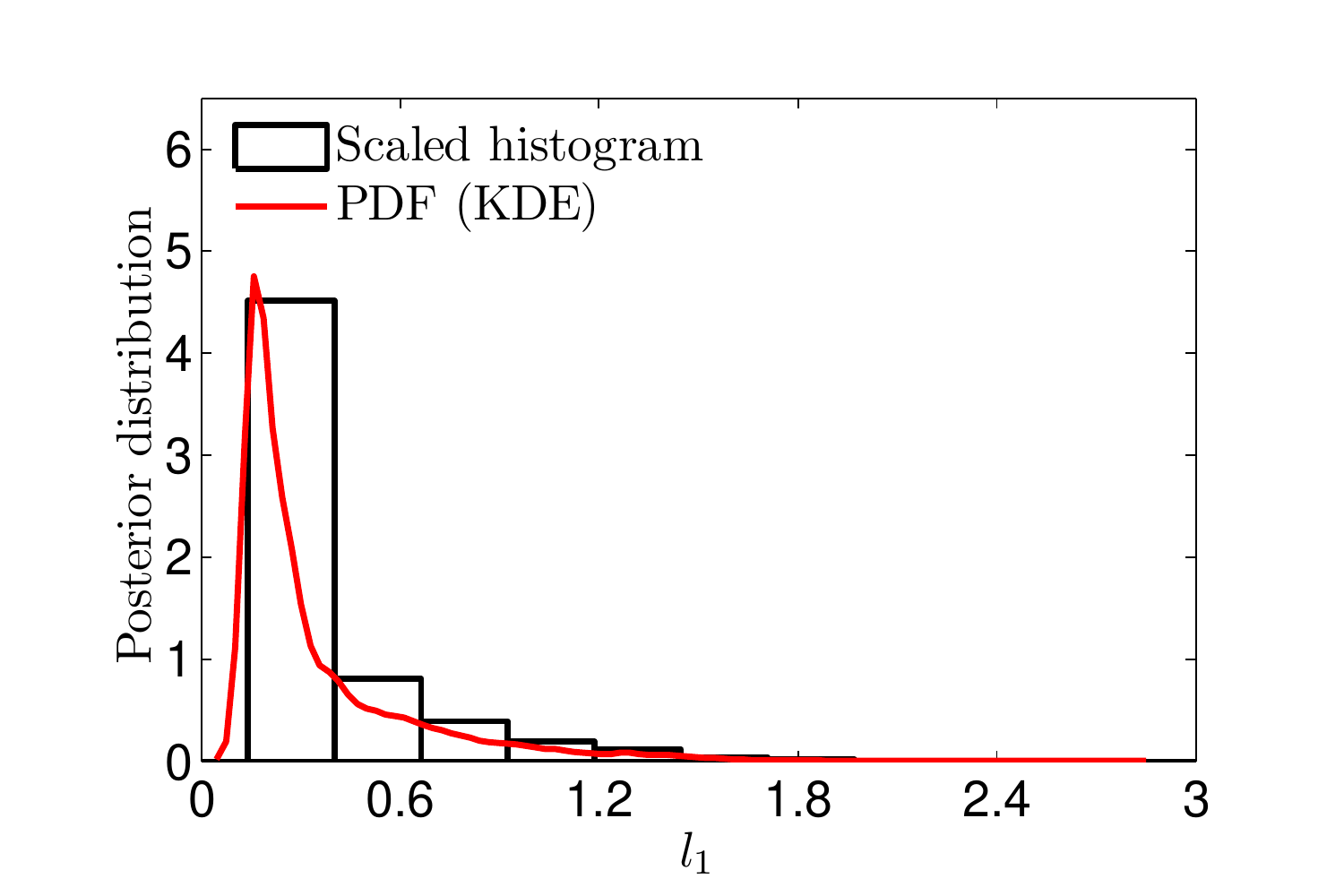}}
\subfigure[$l_2$]{
\includegraphics[width=0.15\textwidth]{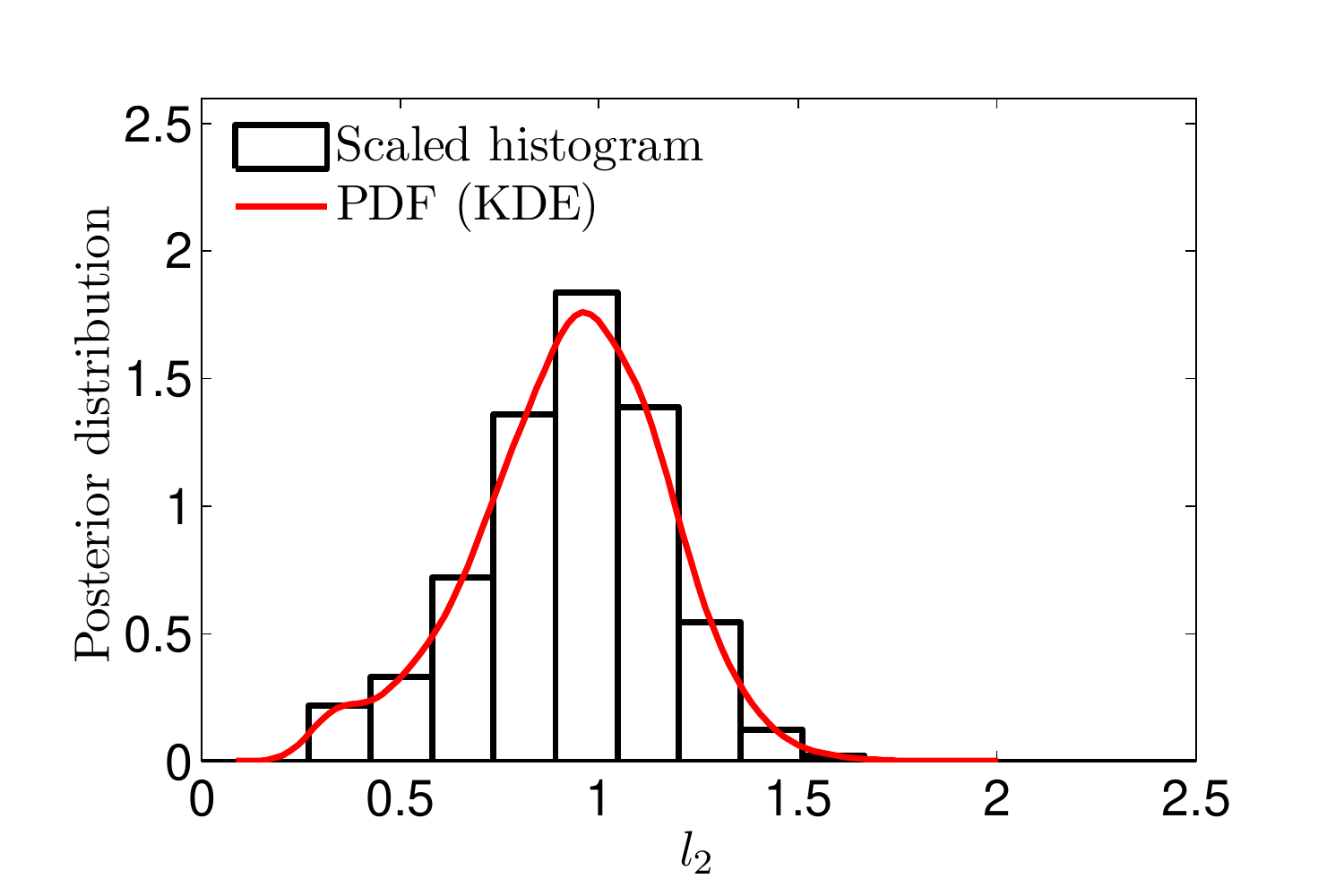}}
\subfigure[$l_3$]{
\includegraphics[width=0.15\textwidth]{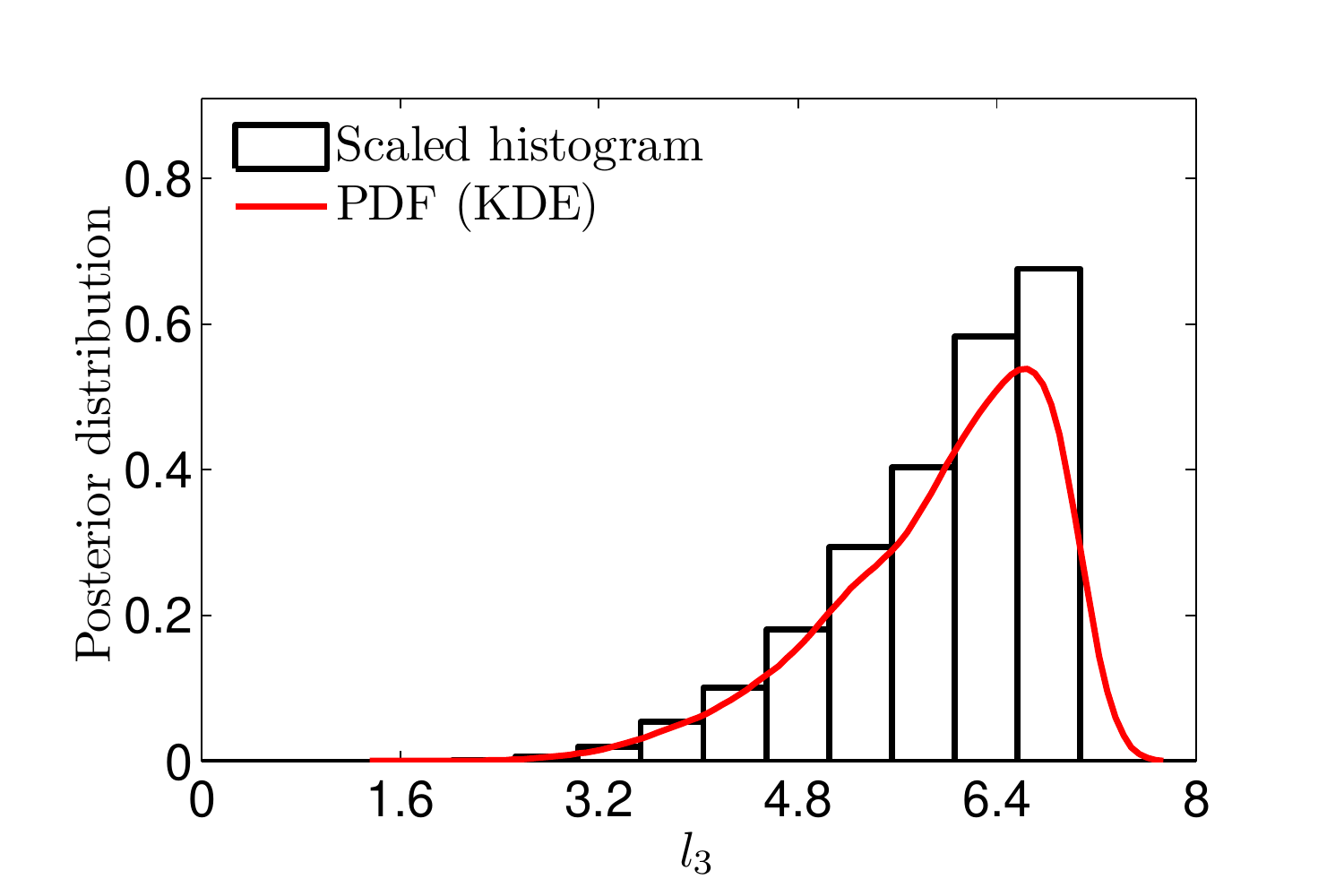}}
\subfigure[$w_1$]{
\includegraphics[width=0.15\textwidth]{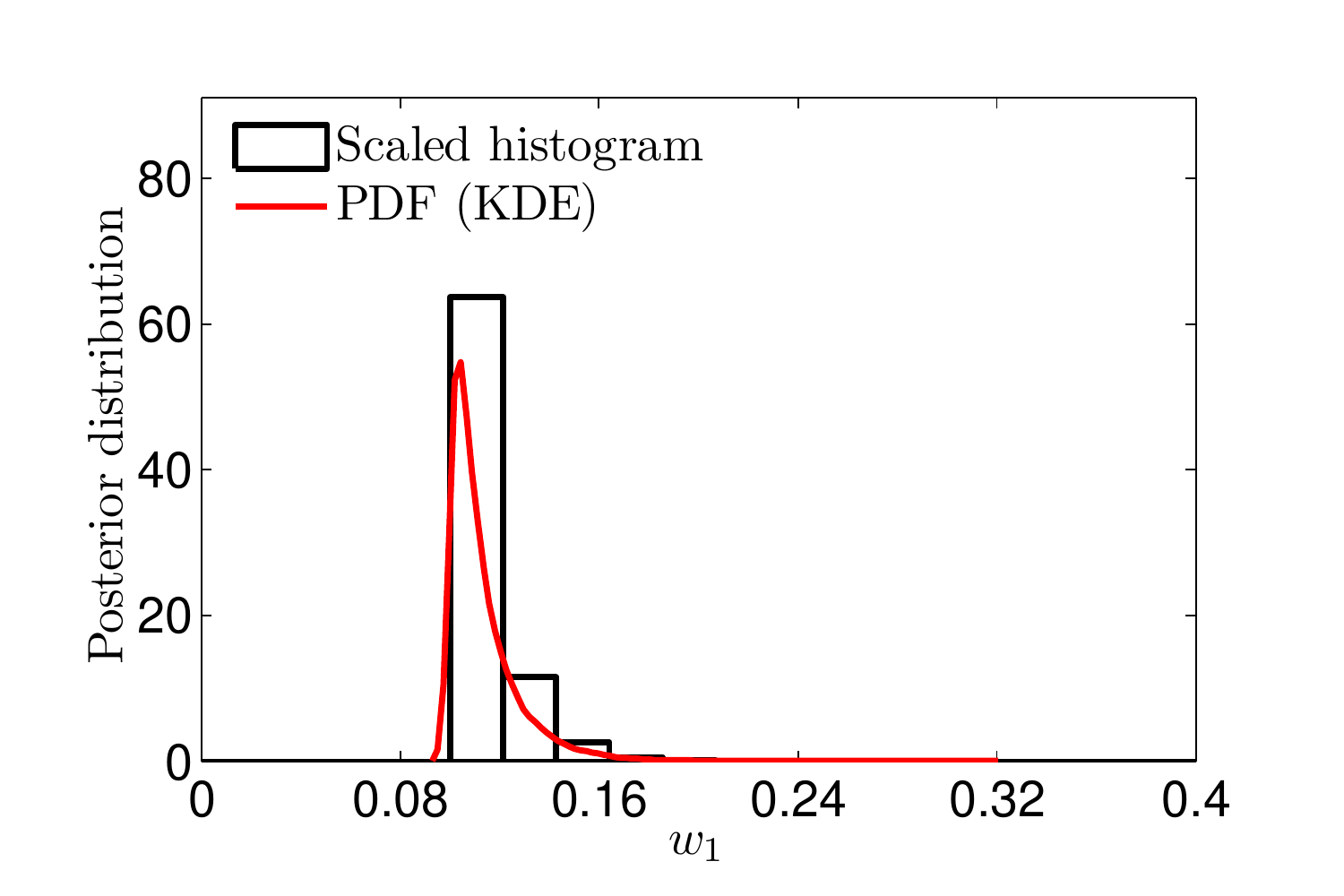}}
\subfigure[$w_2$]{
\includegraphics[width=0.15\textwidth]{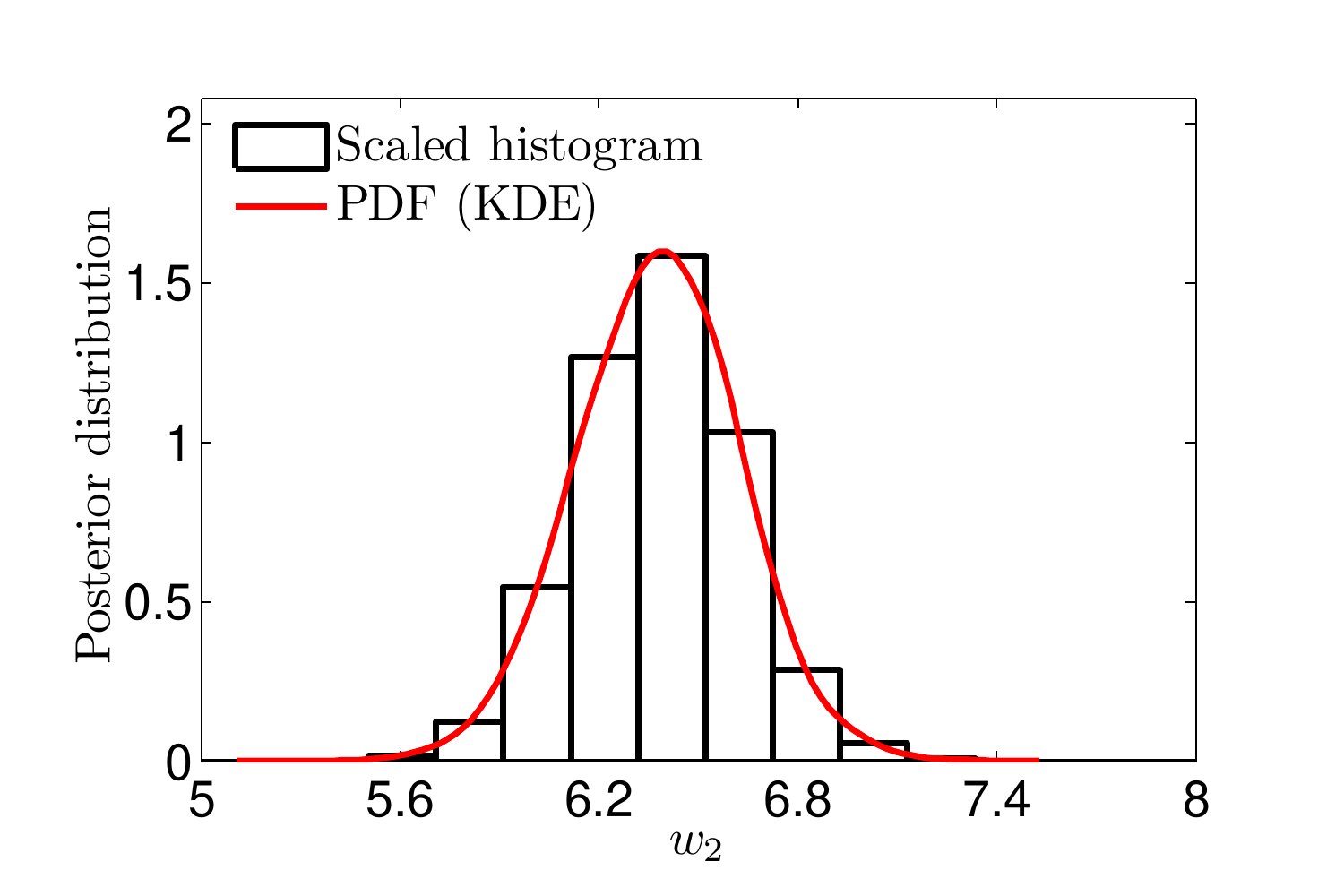}}
\end{center}
\caption{Plots of the scaled histogram and posterior PDF of parameters}
\label{fig:histAndPostpdfDielectric}
\end{figure}

Due to the high number of unknown parameters ($=17$), we use the Metropolis-Hastings MCMC algorithm to sample these 17 parameters from their posterior probability distribution. Note that we use uniform priors for all the parameters, since no information on the prior distributions is available except for the possible ranges of these parameters. The scaled histograms and the kernel density estimation (KDE) of posterior PDFs  based on $10^6$ samples are shown in Fig.~\ref{fig:histAndPostpdfDielectric}. Note that $l_3$ is the length-scale parameter corresponding to time $t$, and the posterior PDF of $l_3$ indicates significant statistical correlation in time, which is what we expected.

\begin{figure}[h!]
\begin{center}
\subfigure[]{
\includegraphics[width=0.48\textwidth]{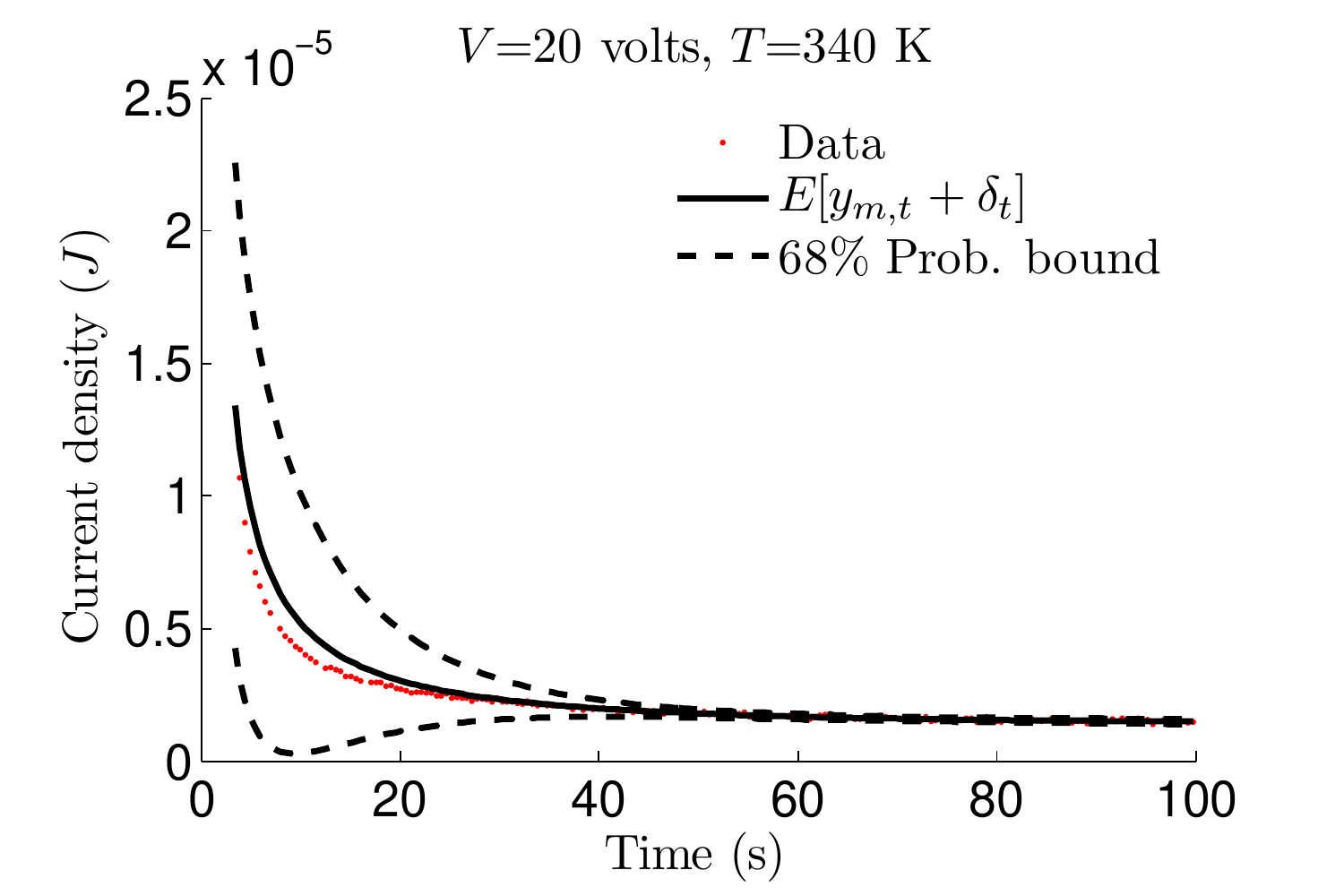}\label{fig:prediction_MAP_11}}
\subfigure[]{
\includegraphics[width=0.48\textwidth]{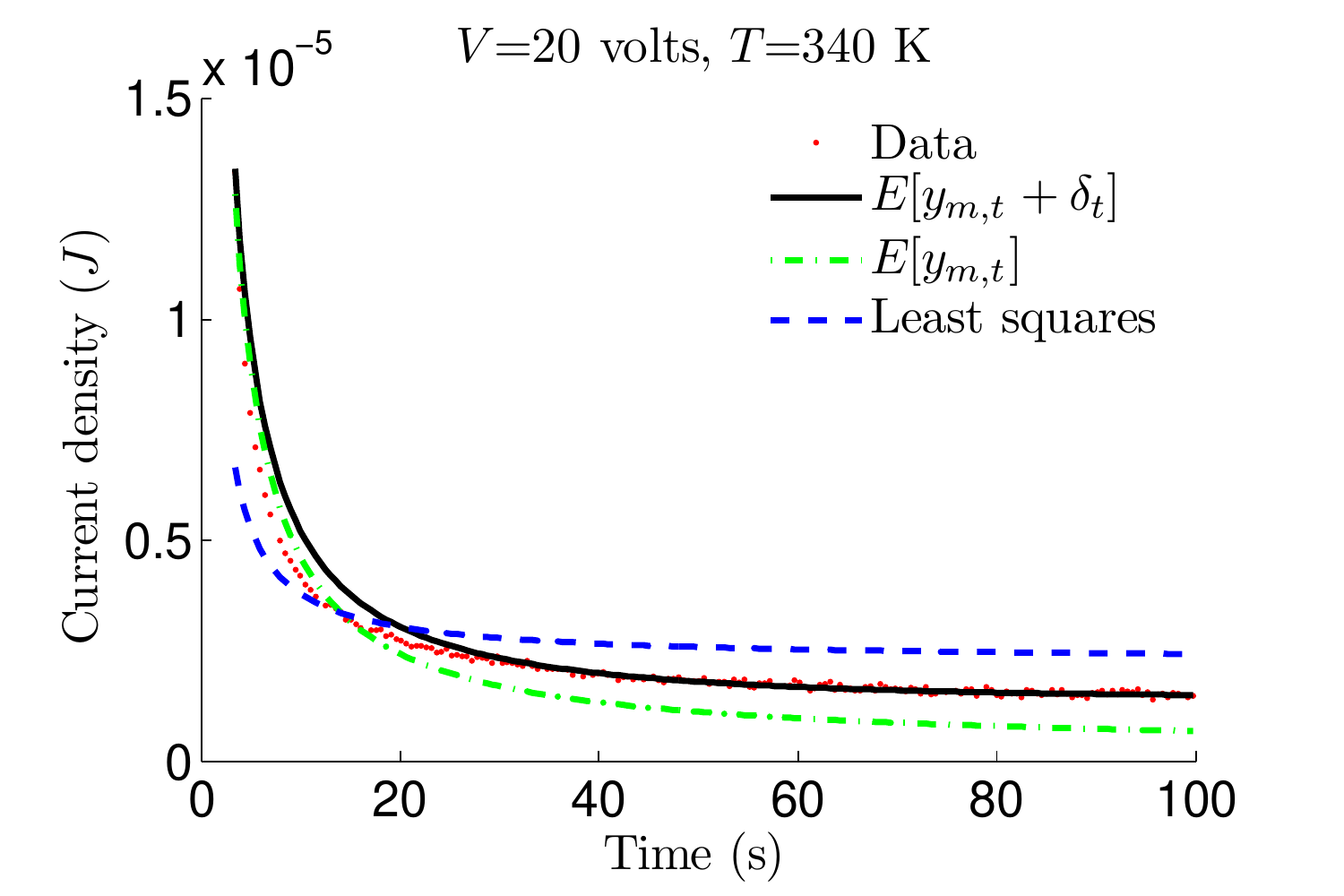}\label{fig:prediction_comparison_11}}
\end{center}
\caption{Validation of the calibrated model}
\label{fig:VALpredictionDielectric}
\end{figure}
With the calibrated model and Gaussian process discrepancy function, we can predict the current density as $J_t = y_{m,t}+\delta_{t}$. A comparison between the prediction and the unused set of data can help validate the calibrated model and discrepancy function. In this example, we compute current density using the maximum \textit{a posteriori} (MAP)~\cite{Sorenson1980} estimation of unknown parameters at the 12 combinations of inputs $V$ and $T$. An example comparison between prediction and data is shown in Fig.~\ref{fig:VALpredictionDielectric}, where $E[y_{m,t}]$ is the expected current density based on the calibrated model without discrepancy correction, $E[y_{m,t}+\delta_{t}]$ is the expected current density based on the combination of the calibrated model and discrepancy function. The blue dash line in Fig.~\ref{fig:prediction_comparison_11} is the model prediction based on least squares estimate of the parameters. 

We observe from this graphical comparison that $E[y_{m,t}+\delta_{t}]$ fits the data relatively better than $E[y_{m,t}]$ and the prediction based on least squares method, and that the one standard deviation (68\% probability) bound ($E[y_{m,t}+\delta_{t}] \pm \sqrt{V[y_{m,t}+\delta_{t}]}$, where $V[y_{m,t}+\delta_{t}]$ is the variance of prediction) fully covers the data. It should be noted that the prediction based on Bayesian calibration accounts for model uncertainty; data uncertainty can also be taken into account if the full posterior PDFs of parameters are used. However, Bayesian calibration based on MCMC sampling methods is more computationally demanding compared with calibration based on least squares analysis, since the generation of $10^6$ samples require the same amount of function evaluations. It can also be observed that there is more difference between the expected current density $E[y_{m,t}+\delta_{t}]$ and the data at the early time range ($0\sim 20$ seconds) than at the later time range, which is reflected by the wider probability bound at the early time range. This observation indicates that the combination of the physics model and the discrepancy is not sufficient to model the current density for the whole time range, and the actual model discrepancy may be a more complicated function with respect to time.

\subsection{Calibration of multi-physics models using interval and point data}\label{subsection:NumExp-2}
\begin{figure}[h!]
\begin{center}
\subfigure[Dev-1: Contacting capacitive RF MEMS switch]{
\includegraphics[trim=85mm 80mm 85mm 80mm, clip=true, width=0.48\textwidth]{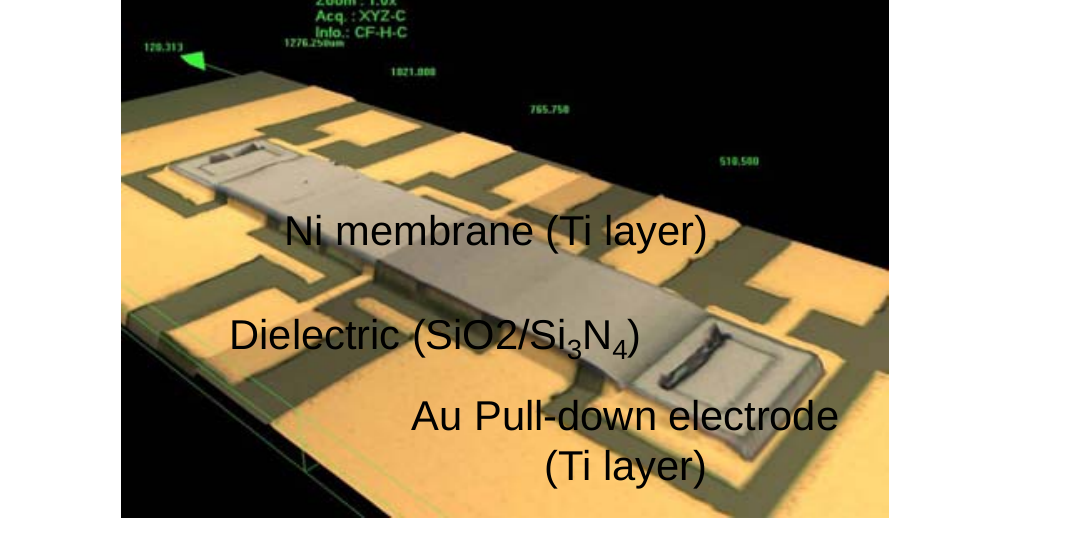}\label{fig:targetDevice-1}}
\subfigure[Dev-2: RF MEMS varactor]{
\includegraphics[trim=70mm 60mm 70mm 60mm, clip=true, width=0.48\textwidth]{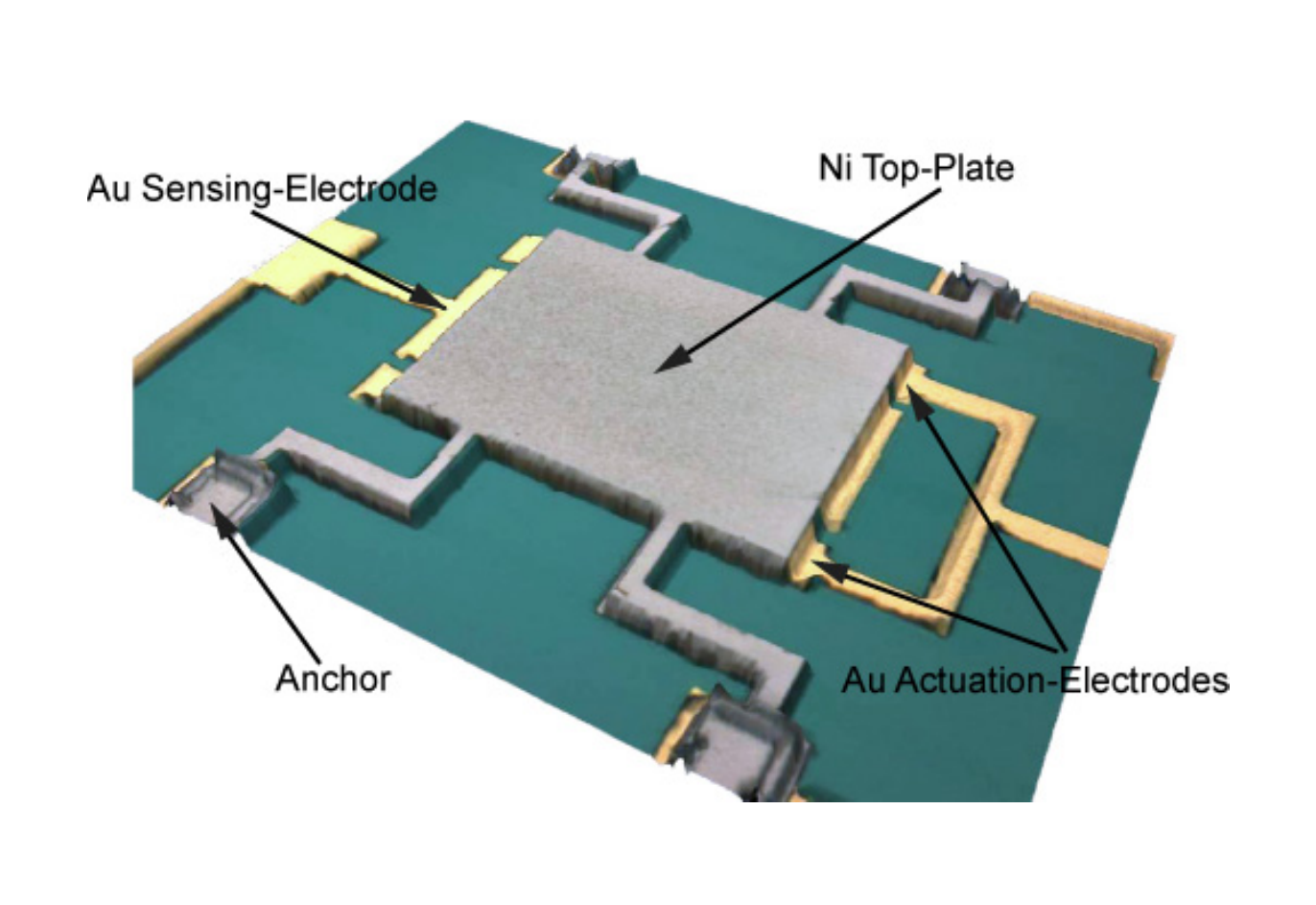}\label{fig:targetDevice-2}}
\end{center}
\caption{Example RF MEMS devices (Courtesy: Purdue PRISM center)}
\label{fig:targetDevice}
\end{figure}

The target MEMS device of this example (denoted as Dev-1) shown in Fig.~\ref{fig:targetDevice-1} is used as a switch. The membrane deflects under some applied voltage, and will contact the dielectric pad when the applied voltage exceeds a certain threshold. This threshold voltage is called pull-in voltage ($V_{pl}$), and the device will be closed when the contact occurs. Pull-in voltage is an important metric in the reliability analysis of the device after a certain period of usage. Several models are needed to calculate the pull-in voltage, namely dynamic model, electrostatic model, damping model, and creep model. A 1-D Euler-Bernoulli beam model is used to simulate the dynamic behavior of the MEMS device~\cite{Ayyaswamy2010}. The electrostatic model takes applied voltage and air gap ($g$) as inputs, and calculates electrostatic loading as output. The damping model considers the gas pressure and air gap, and the corresponding damping force is computed~\cite{Alexeenko2011}. The electrostatic loading, damping force, device geometry, material property, boundary condition, and time are the inputs of the dynamic model. The creep model calculates the plastic deformation of the device under long-term loading, and is coupled with the dynamic model. The unknown parameters include Young's modulus ($E$) and residual stress ($\sigma_{rs}$) in the dynamic model, and the creep coefficient $A_c$ in the Coble creep model~\cite{Coble1963,Hsu2011}. To predict the pull-in voltage, an iterative method is used by varying the values of applied voltage, and calculating the resulting maximum deflection of the beam. The pull-in voltage is equal to the minimum value of applied voltage that causes the beam to be in contact with the dielectric pad.

\subsubsection{Different data on two devices}
Due to the limitation of experimental resources, currently only the measurement data of pull-in voltage at an early time point is available, and the data are collected on 17 Dev-1 devices with different geometries and initial positions. Because the pull-in voltage data are obtained by keeping increasing the applied voltage by 5 volts until the switch becomes closed, the data are reported in the form of \emph{intervals}.

Study of creep modeling has been separately performed for another type of device (denoted as Dev-2, which has different boundary conditions from Dev-1 as shown in Fig.~\ref{fig:targetDevice-2}), and measurements of device deflection under constant voltage for a relative long time period ($\sim$700 hours) are available. Since these two types of devices are made of the same material, the material-related parameters $E$ and $A_c$ can be considered as the same. A polynomial chaos expansion (PCE) surrogate model is constructed based on 3-D membrane simulation for Dev-2, with $E$ and $A_c$ as inputs and the deflection at three different time points $\boldsymbol{t}=[200,400,600]$ hours as output, i.e., $g_{t2} = \text{PCE}(A_c,E)+\delta_{2}$. $\delta_{2}$ is the model discrepancy term.

\begin{figure}[h!]
\begin{center}
\includegraphics[trim=45mm 40mm 65mm 65mm, clip=true, width=0.7\textwidth]{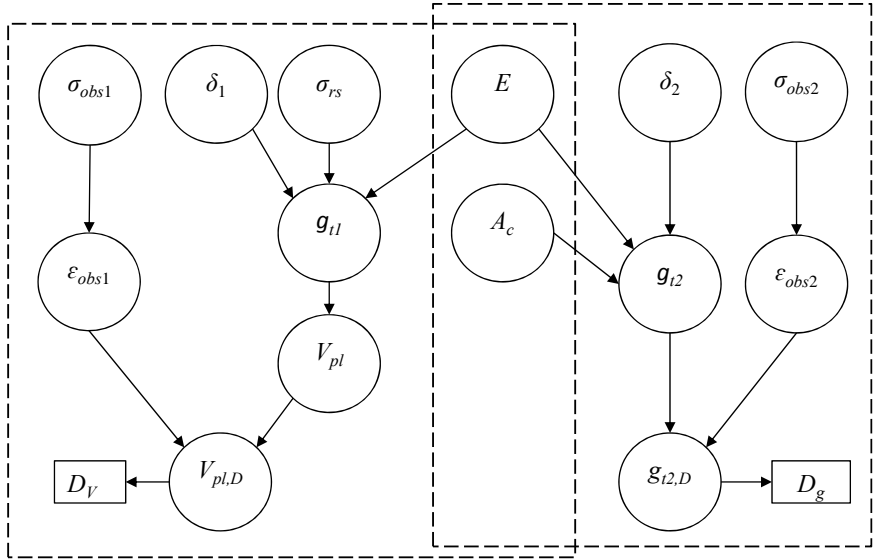}
\end{center}
\caption{Bayesian network}
\label{fig:BN_3}
\end{figure}

Based on the aforementioned models and data, we construct a Bayesian network as shown in Fig.~\ref{fig:BN_3}. Note that $A_c$ is not directly related to pull-in voltage, since the calculation of pull-in voltage at a given time point only requires dynamic simulation within microseconds, and creep is negligible in such a short time period. Therefore the only common parameter between the two physics models is $E$. The second and the third options presented in Section~\ref{section:stratBayesMP} are both implemented for the purpose of comparison in this example, although the first option is not considered due to its higher computation cost.

The identifiability of the calibration parameters in the Bayesian network given the available experimental data is checked using the first-order Taylor series expansion-based method presented in Section~\ref{section:identifiability}. Since the measurement data of pull-in voltage for 17 Dev-1 devices will be directly used to calibrate the parameters ($E$, $\sigma_{rs}$, $\delta_{1}$) in the left half of the Bayesian network in Fig.~\ref{fig:BN_3}, we obtain a $17*3$ first-order derivative matrix $\boldsymbol{A}$ with rank $r_{A}=3$, i.e., $E$, $\sigma_{rs}$, and $\delta_{1}$ are identifiable with these 17 data points of pull-in voltage. We also examine the identifiability of parameters $E$, $A_c$ and $\delta_{2}$ in the right half of the Bayesian network with the deflection data of Dev-2 at the three test time points ($200$, $400$, and $600$ hours).  In this case, the size of the matrix $\boldsymbol{A}$ is $3*3$ and the rank of $\boldsymbol{A}$ is 3, which indicates that $E$, $A_c$ and $\delta_{2}$ are all identifiable with the deflection data. Note that this method is not applicable for $\sigma_{obs1}$ and $\sigma_{obs2}$, since the standard deviations of measurement noise are the parameters of statistical models as stated in Section~\ref{section:identifiability}. 

\subsubsection{Calibration with information flowing from left to right in the Bayesian network}

\begin{figure}[h!]
\begin{center}
\subfigure[$E$]{
\includegraphics[width=0.23\textwidth]{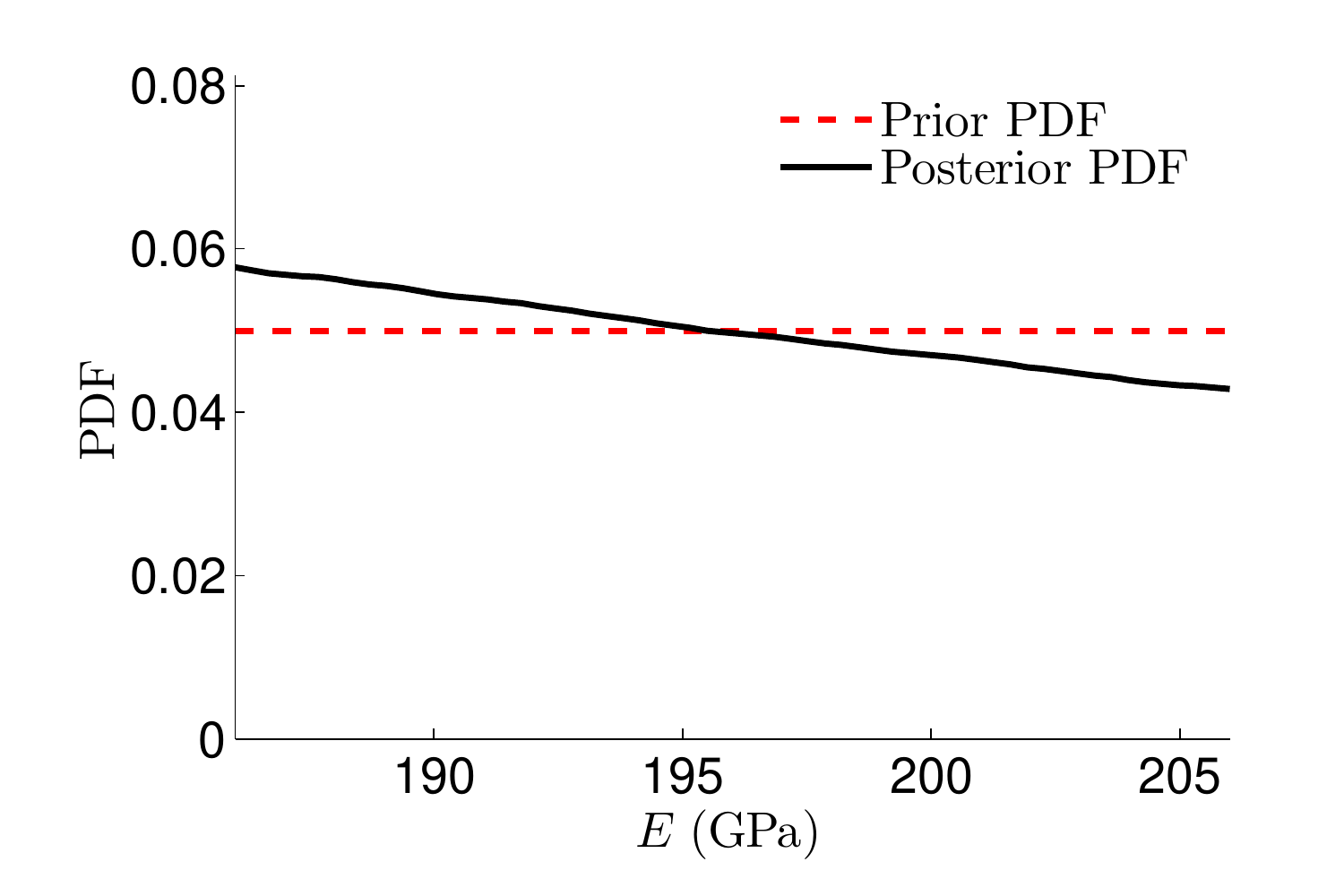}}
\subfigure[$\sigma_{rs}$]{
\includegraphics[width=0.23\textwidth]{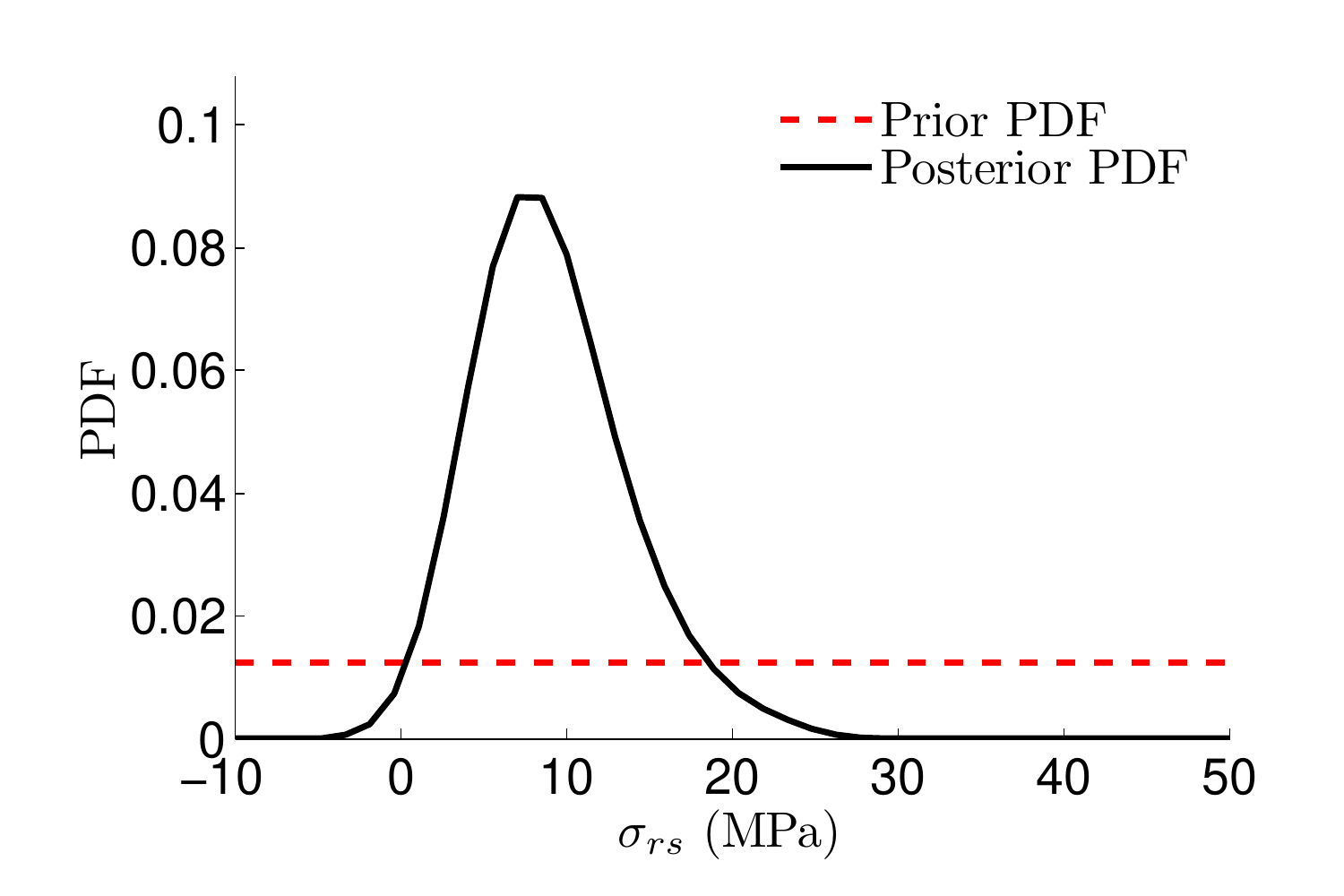}}
\subfigure[$\delta_1$]{
\includegraphics[width=0.23\textwidth]{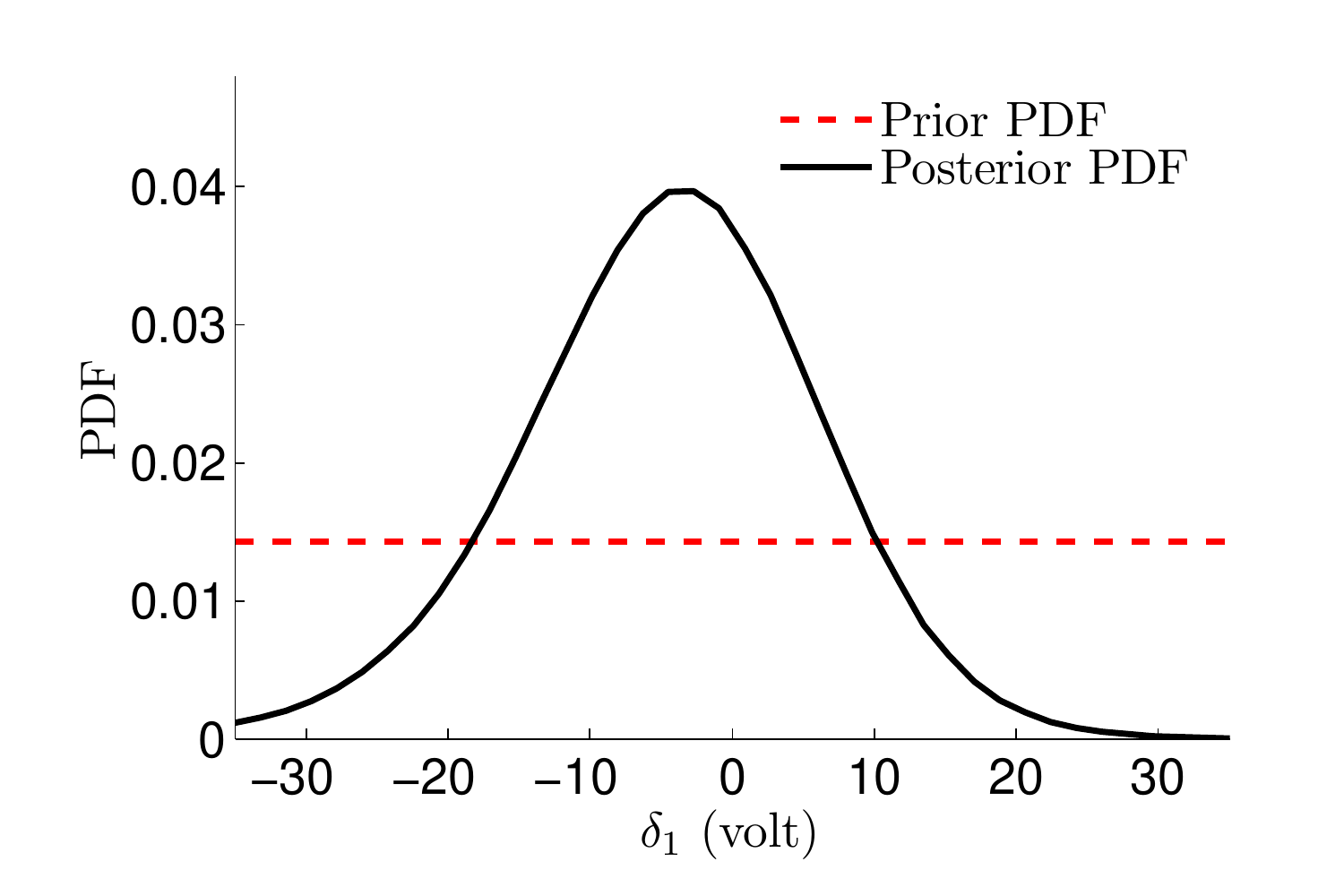}}
\subfigure[$\sigma_{obs1}$]{
\includegraphics[width=0.23\textwidth]{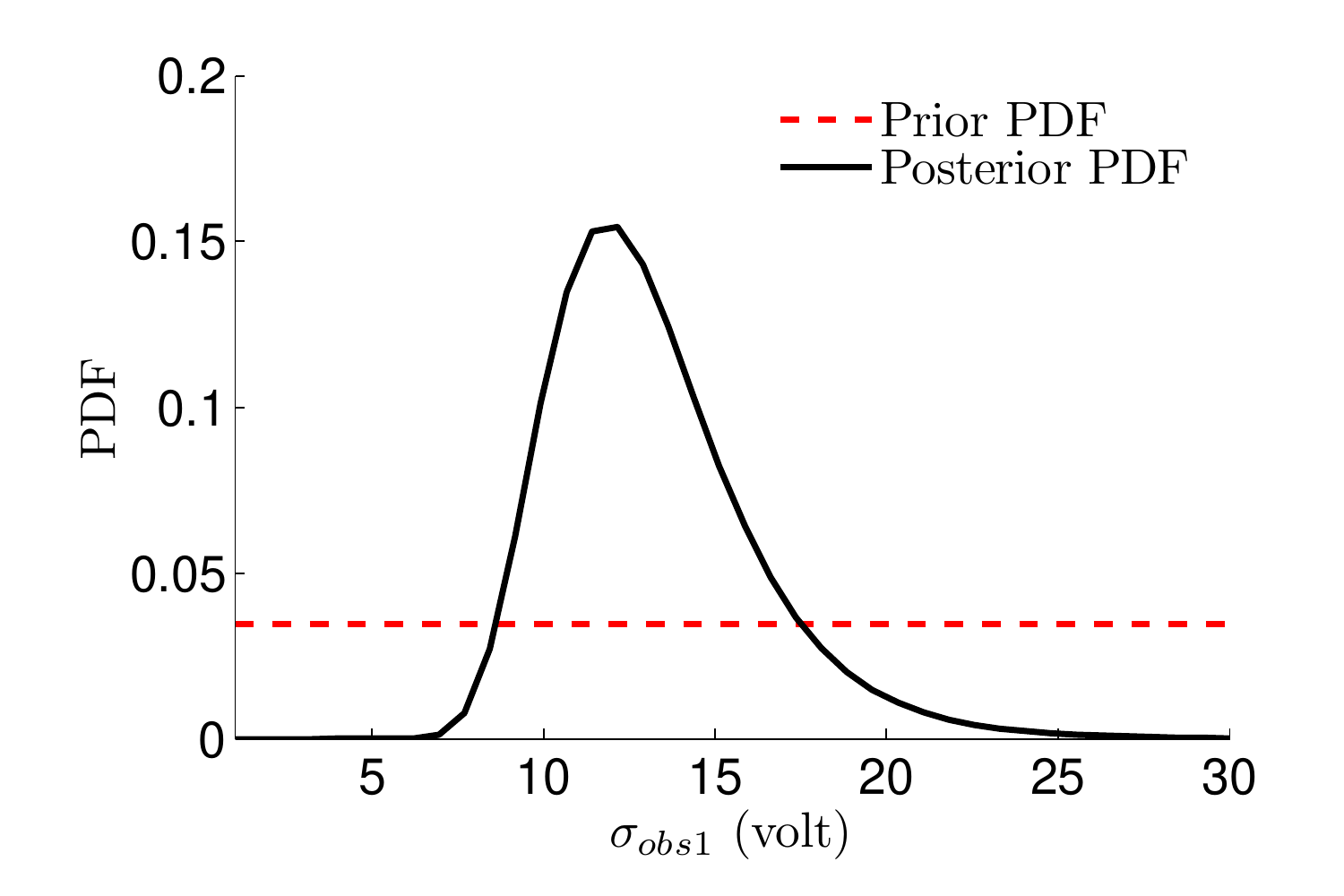}}
\end{center}
\caption{Calibration of parameters using pull-in voltage data}
\label{fig:calPullin-1}
\end{figure}

\begin{figure}[h!]
\begin{center}
\subfigure[$E$]{
\includegraphics[width=0.23\textwidth]{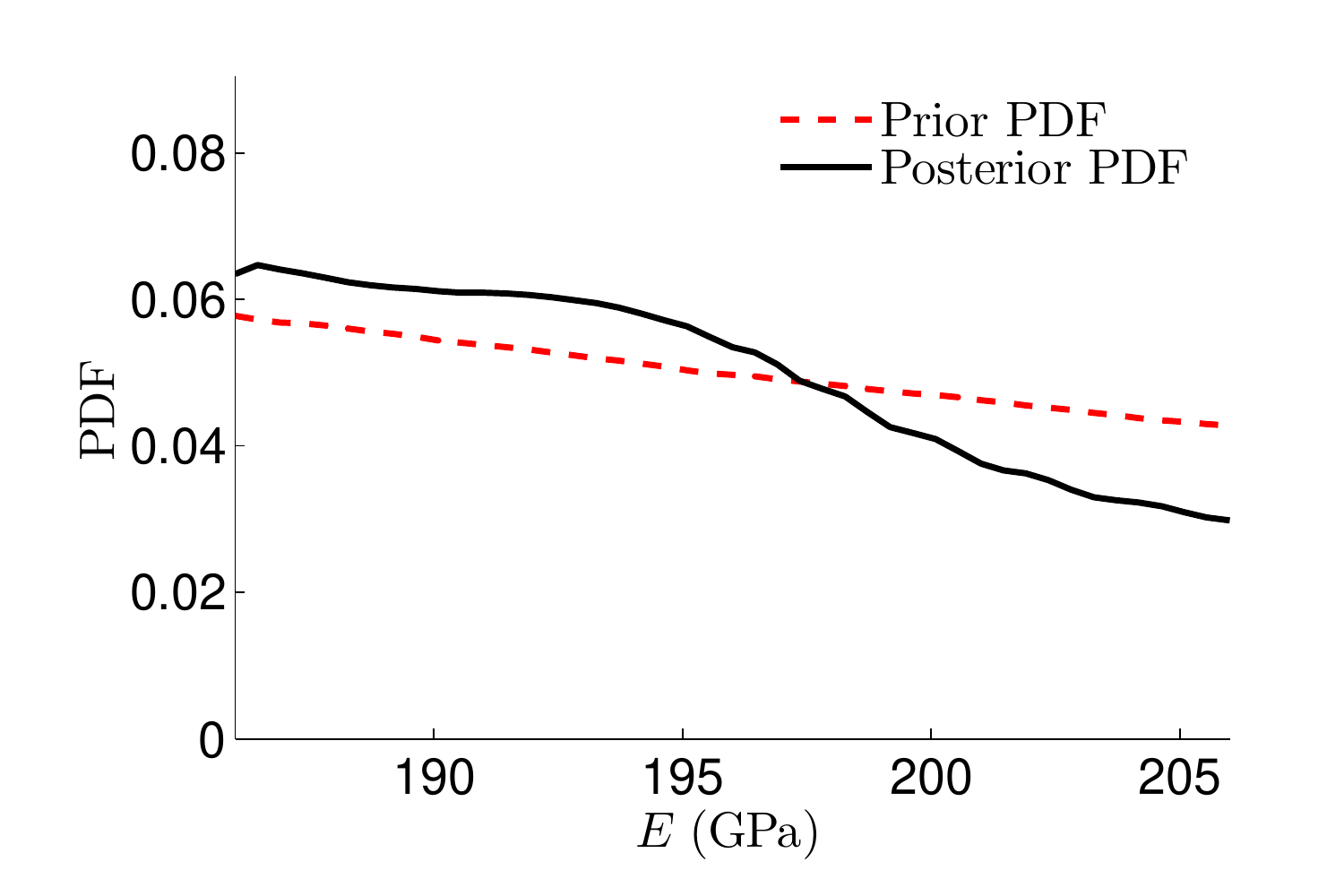}}
\subfigure[$A_c$]{
\includegraphics[width=0.23\textwidth]{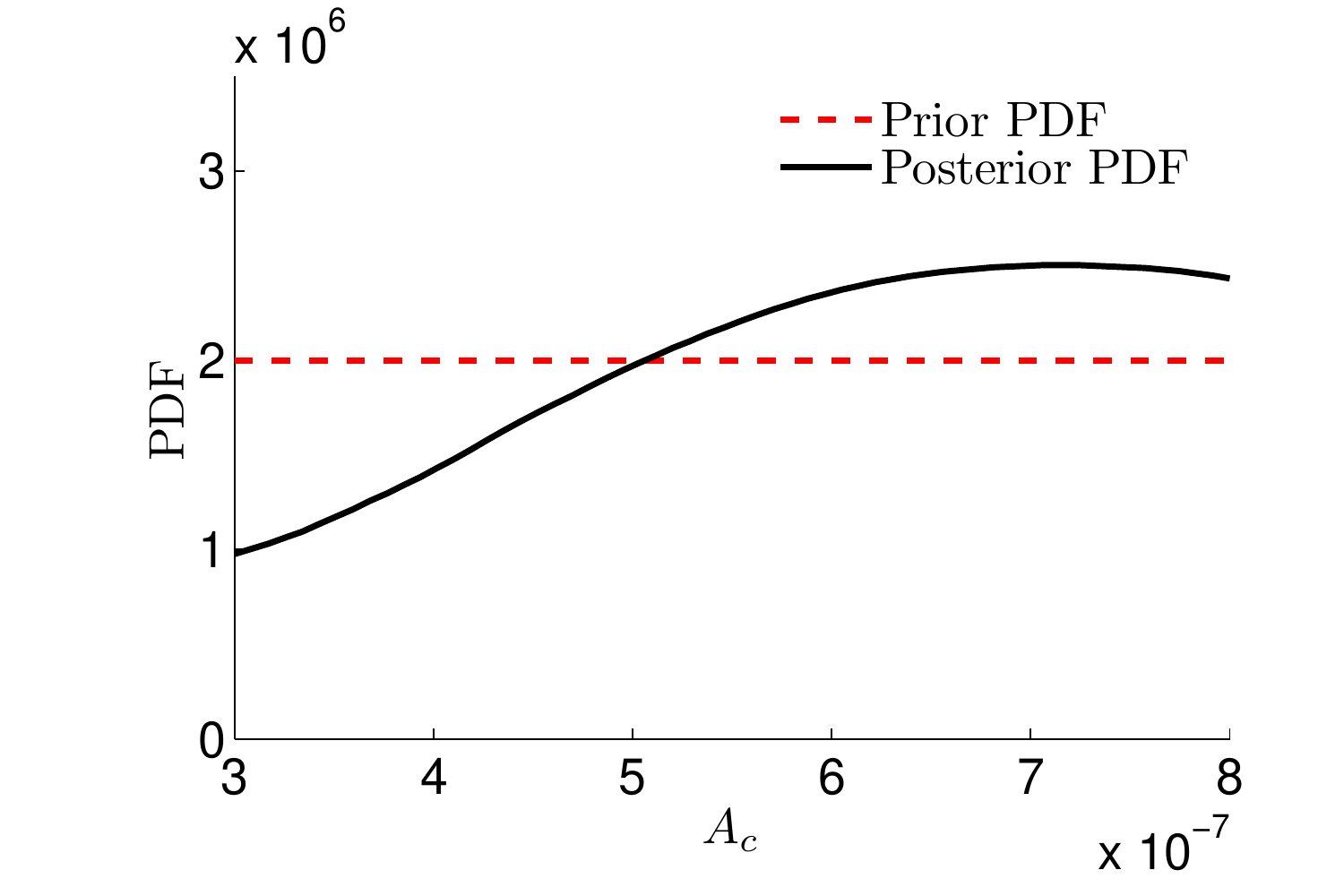}}
\subfigure[$\delta_{2}$]{
\includegraphics[width=0.23\textwidth]{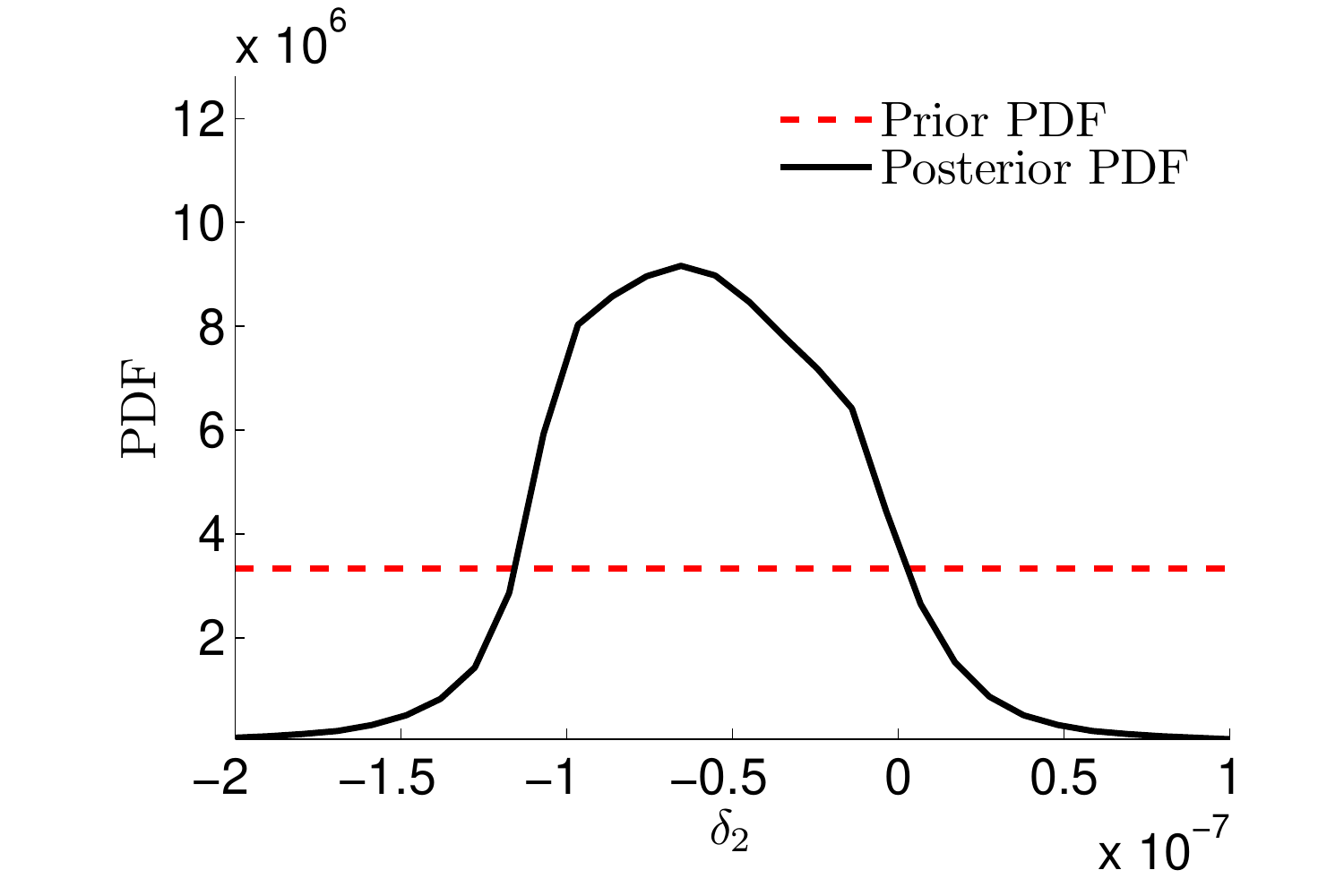}}
\subfigure[$\sigma_{obs2}$]{
\includegraphics[width=0.23\textwidth]{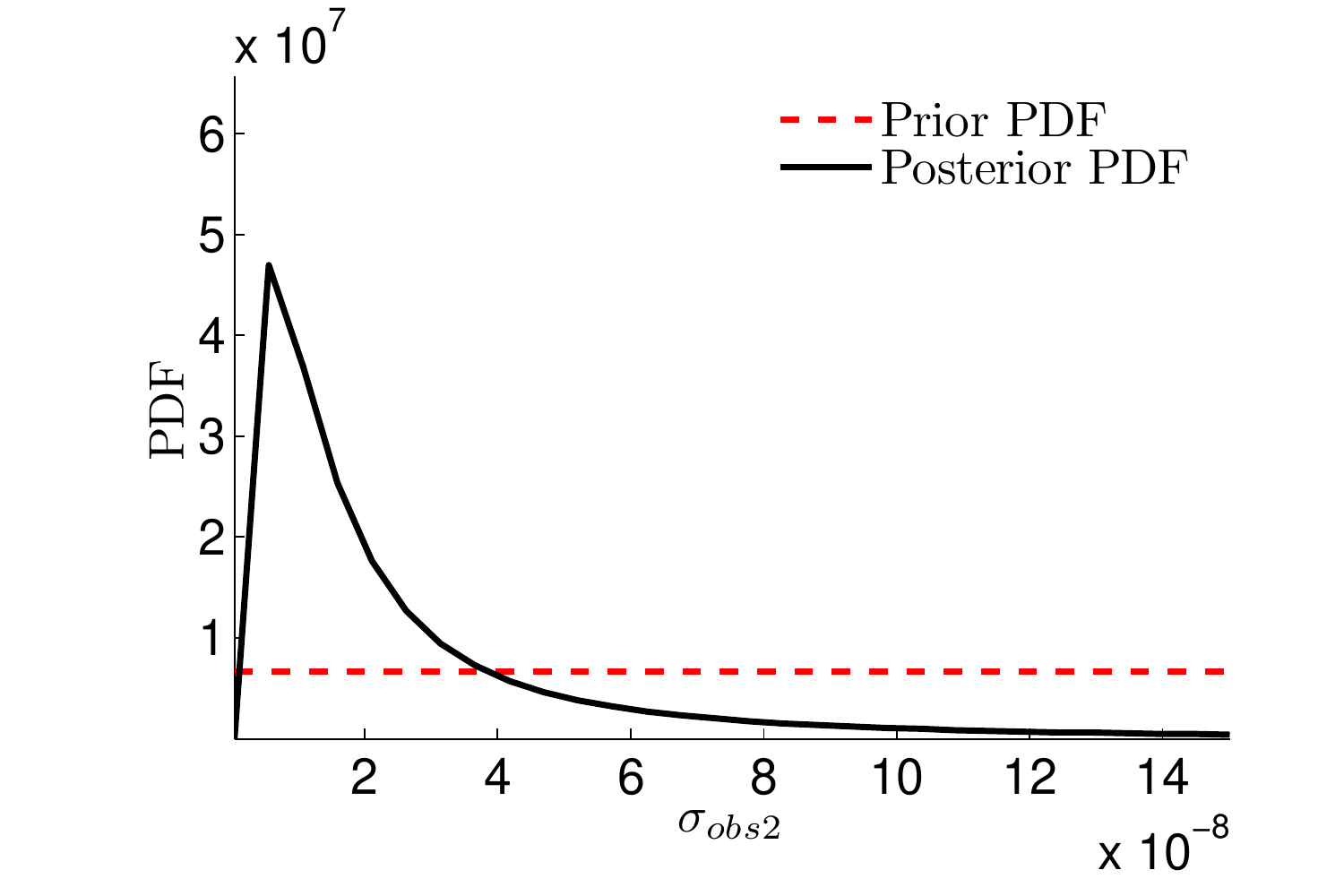}}
\end{center}
\caption{Calibration of parameters using deflection data}
\label{fig:calCreep-2}
\end{figure}

Following the second option presented in Section~\ref{section:stratBayesMP}, the left half of the Bayesian network is considered first, i.e., the parameters $E$, $\sigma_{rs}$, $\delta_{1}$, and $\sigma_{obs1}$ are calibrated using the pull-in voltage data. The prior and marginal posterior PDFs of $E$, $\sigma_{rs}$, $\delta_{1}$, and $\sigma_{obs1}$ are plotted in Fig.~\ref{fig:calPullin-1}. The prior PDFs are shown as red dash lines, whereas the posterior PDFs are shown as black solid lines (the same format applies to Figs.~\ref{fig:calCreep-2}, \ref{fig:calCreep-1}, and \ref{fig:calPullin-2}). The corresponding statistics are shown in Table~\ref{table:statDataDev-1-1}. Note that all the prior PDFs used in this example are assumed to be uniform, except for the prior PDF of the common parameter $E$ in the second step calibration, which is the posterior PDF obtained in the first step calibration.
 
\begin{table}[h!]
\begin{center}
\caption{Prior and posterior statistics of parameters (with data on Dev-1)}
\label{table:statDataDev-1-1}
\begin{tabular}{lcccc}
\toprule
& \multicolumn{2}{c}{Mean} & \multicolumn{2}{c}{Standard deviation} \\
\cmidrule(r){2-5}
& Prior & Posterior & Prior & Posterior \\
\midrule
$E$ (GPa) & 196.0 & 195.5 & 5.78  & 5.76 \\
$\sigma_{rs}$ (MPa) & 10.00 & 9.07 & 23.10 & 4.76 \\
$\delta_{1}$ (Volt) & 0 & -4.27 & 20.22 & 10.35 \\ 
$\sigma_{obs1}$ (Volt) &  15.50 & 13.16 & 8.38 & 2.98 \\  
\bottomrule
\end{tabular}
\end{center}
\end{table}

Then, the parameters in the right half of the Bayesian network are calibrated using the deflection data of Dev-2, and the posterior PDF of $E$ obtained in the first step is used as prior. Fig.~\ref{fig:calCreep-2} shows the prior and marginal posterior PDFs of $E$, $A_c$, $\delta_{2}$, and $\sigma_{obs2}$, and Table~\ref{table:statDataDev-2-2} contains the corresponding statistics.
 
\begin{table}[h!]
\begin{center}
\caption{Prior and posterior statistics of parameters (with data on Dev-2)}
\label{table:statDataDev-2-2}
\begin{tabular}{lcccc}
\toprule
& \multicolumn{2}{c}{Mean} & \multicolumn{2}{c}{Standard deviation} \\
\cmidrule(r){2-5}
& Prior & Posterior & Prior & Posterior \\
\midrule
$E$ (GPa) & 195.5& 194.7 & 5.76 & 5.51 \\
$A_c$ & 5.50e-7 & 5.85e-7 & 1.44e-7 & 1.35e-7 \\
$\delta_{2}$ ($\mu$m) & -0.050 & -0.057 & 0.087 & 0.040 \\ 
$\sigma_{obs2}$ ($\mu$m) & 0.075 & 0.026 & 0.043 & 0.027 \\  
\bottomrule
\end{tabular}
\end{center}
\end{table}

\subsubsection{Calibration with information flowing from right to left in the Bayesian network}

\begin{figure}[h!]
\begin{center}
\subfigure[$E$]{
\includegraphics[width=0.23\textwidth]{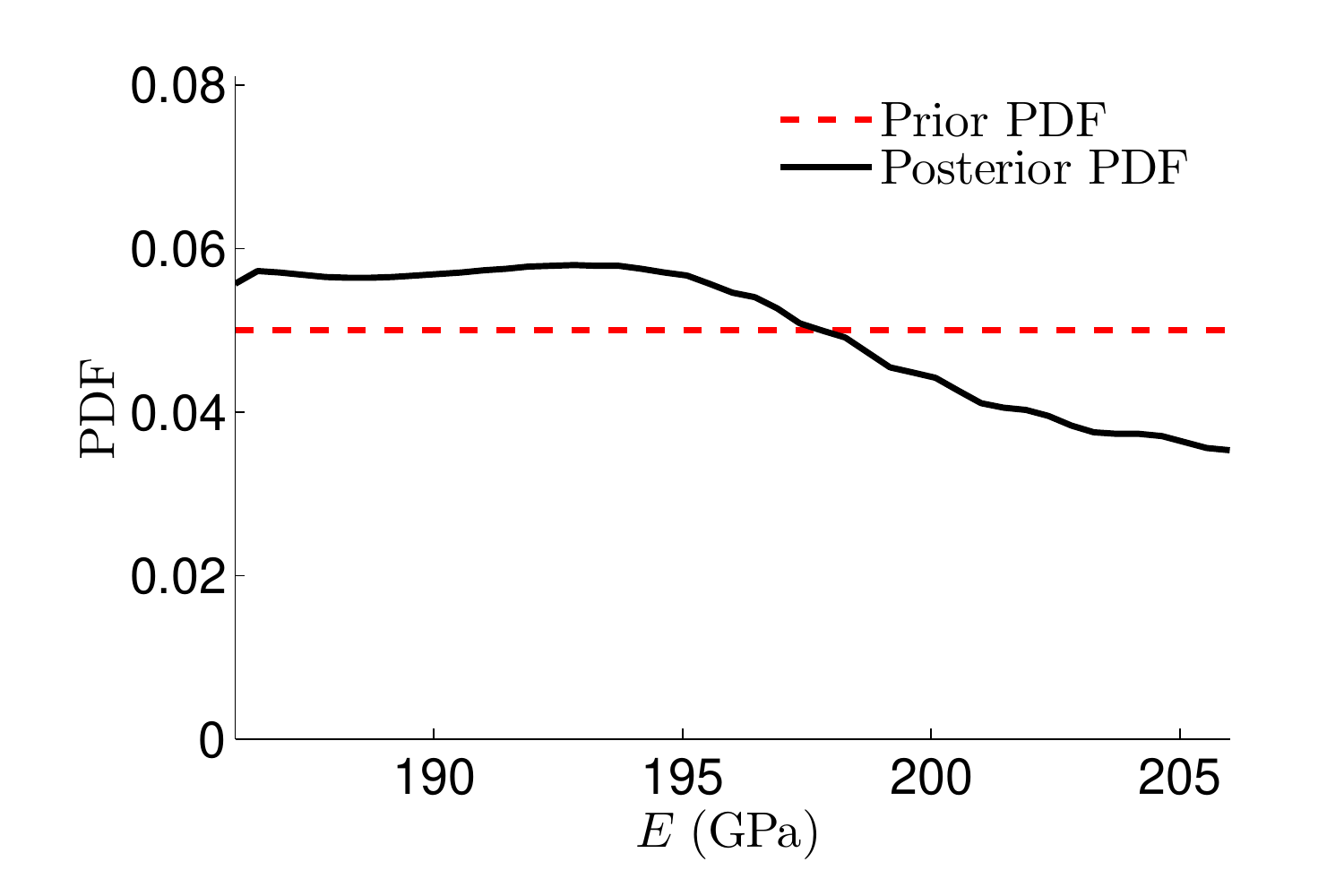}}
\subfigure[$A_c$]{
\includegraphics[width=0.23\textwidth]{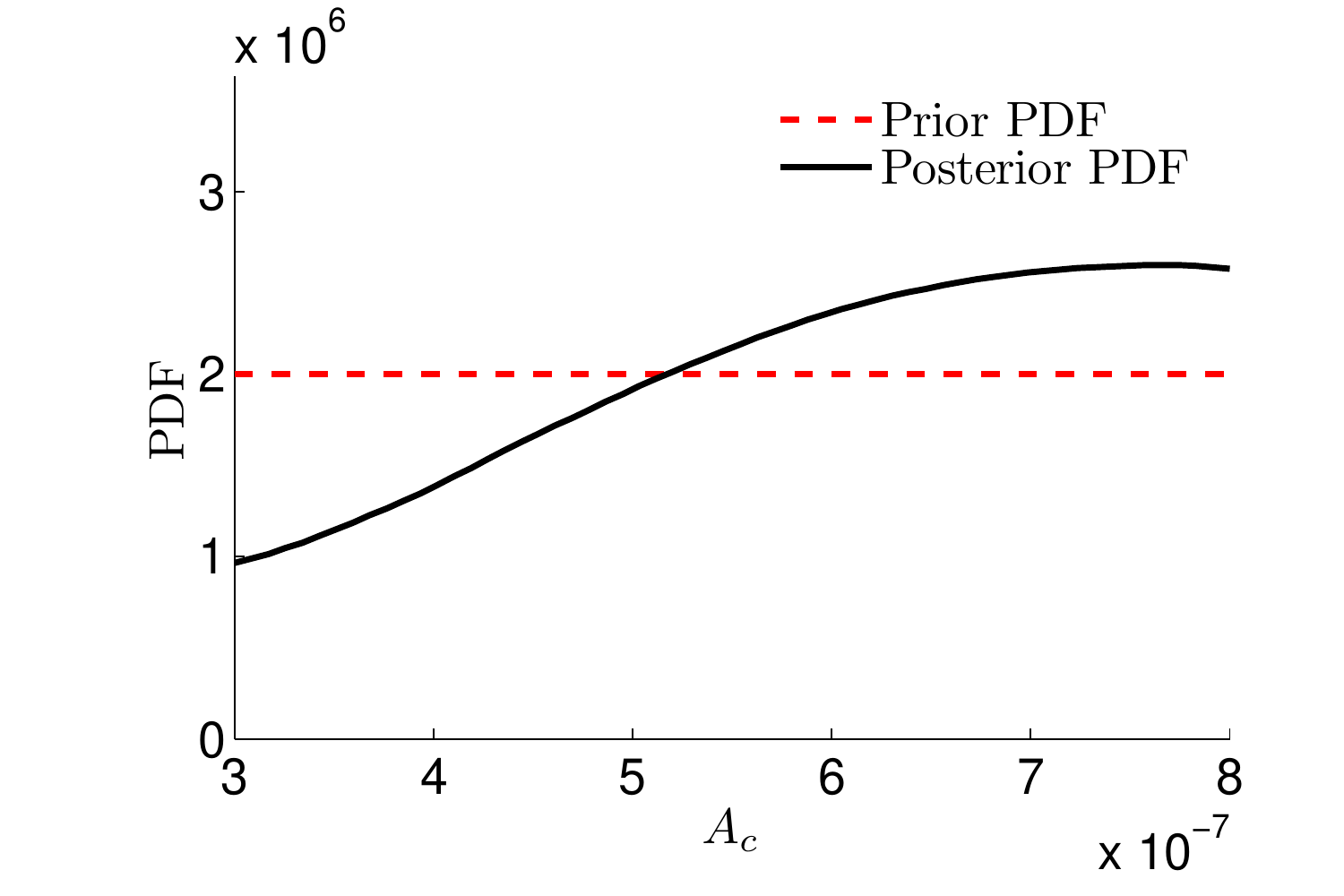}}
\subfigure[$\delta_2$]{
\includegraphics[width=0.23\textwidth]{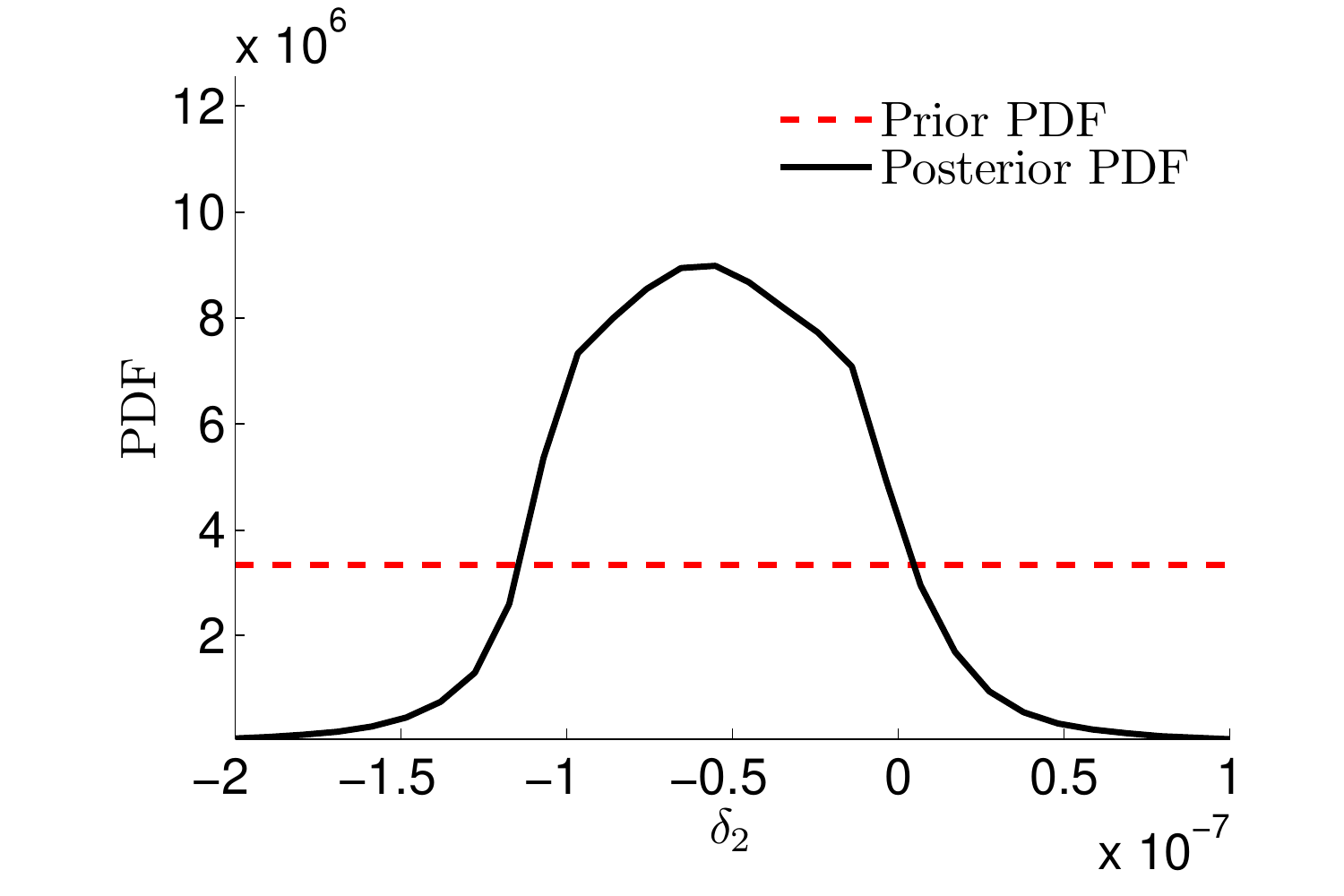}}
\subfigure[$\sigma_{obs2}$]{
\includegraphics[width=0.23\textwidth]{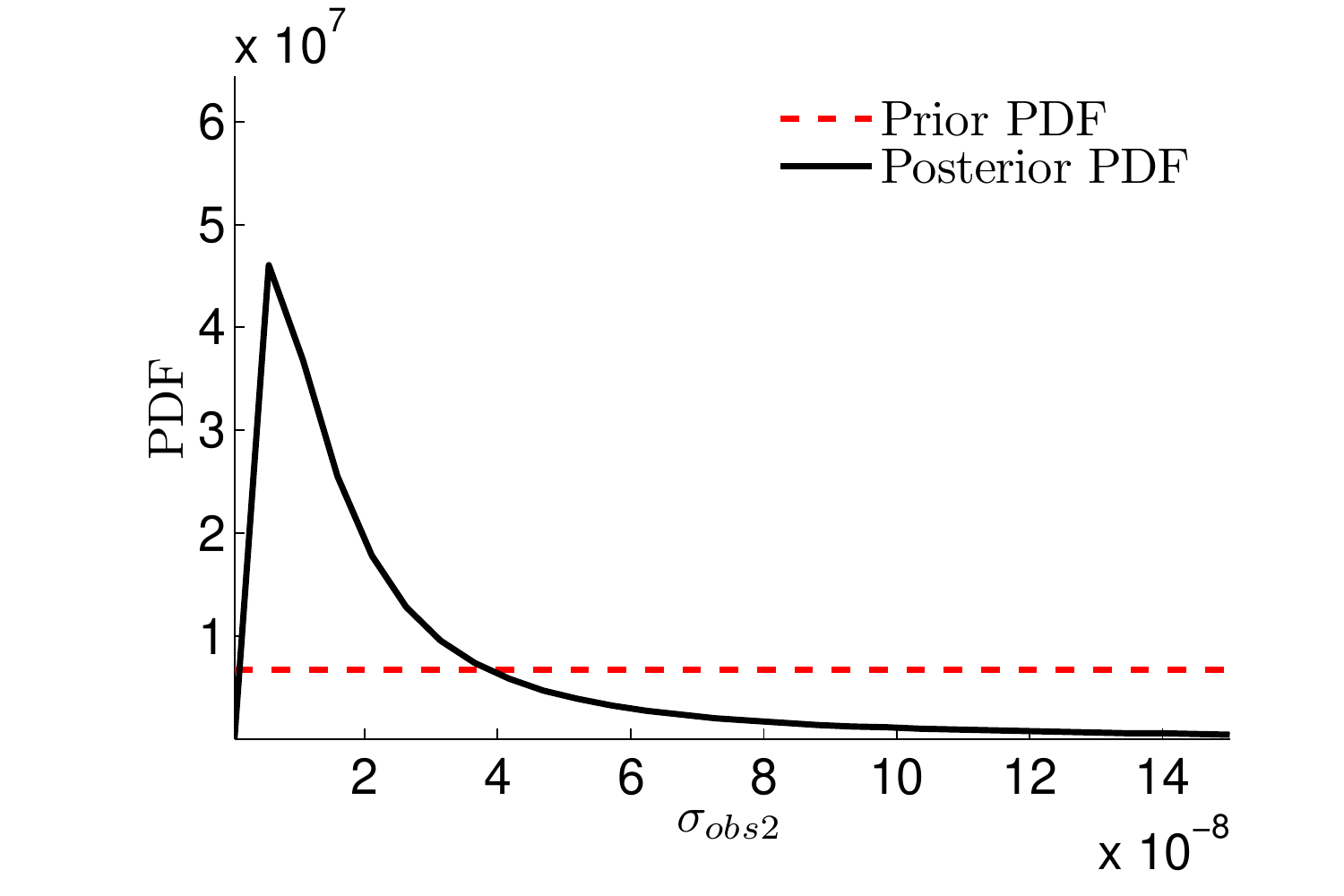}}
\end{center}
\caption{Calibration of parameters using deflection data}
\label{fig:calCreep-1}
\end{figure}

Following the third option presented in Section~\ref{section:stratBayesMP}, the sequence of calibration in the previous section is now reversed. First, the calibration parameter ($E$, $A_c$, $\delta_{2}$, and $\sigma_{obs2}$) in the right half of the Bayesian network in Fig.~\ref{fig:BN_3} are calibrated with the deflection data of Dev-2. The prior and marginal posterior PDFs, and the corresponding statistics are shown in Fig.~\ref{fig:calCreep-1} and Table~\ref{table:statDataDev-2-1}.

\begin{table}[h!]
\begin{center}
\caption{Prior and posterior statistics of parameters (with data on Dev-2)}
\label{table:statDataDev-2-1}
\begin{tabular}{lcccc}
\toprule
& \multicolumn{2}{c}{Mean} & \multicolumn{2}{c}{Standard deviation} \\
\cmidrule(r){2-5}
& Prior & Posterior & Prior & Posterior \\
\midrule
$E$ (GPa) & 196 .0& 195.1 & 5.78 & 5.57 \\
$A_c$ & 5.50e-7 & 5.88e-7 & 1.44e-7 & 1.35e-7 \\
$\delta_{2}$ ($\mu$m) & -0.050 & -0.054 & 0.087 & 0.040 \\ 
$\sigma_{obs2}$ ($\mu$m) & 0.075 & 0.025 & 0.043 & 0.027 \\  
\bottomrule
\end{tabular}
\end{center}
\end{table}

\begin{figure}[h!]
\begin{center}
\subfigure[$E$]{
\includegraphics[width=0.23\textwidth]{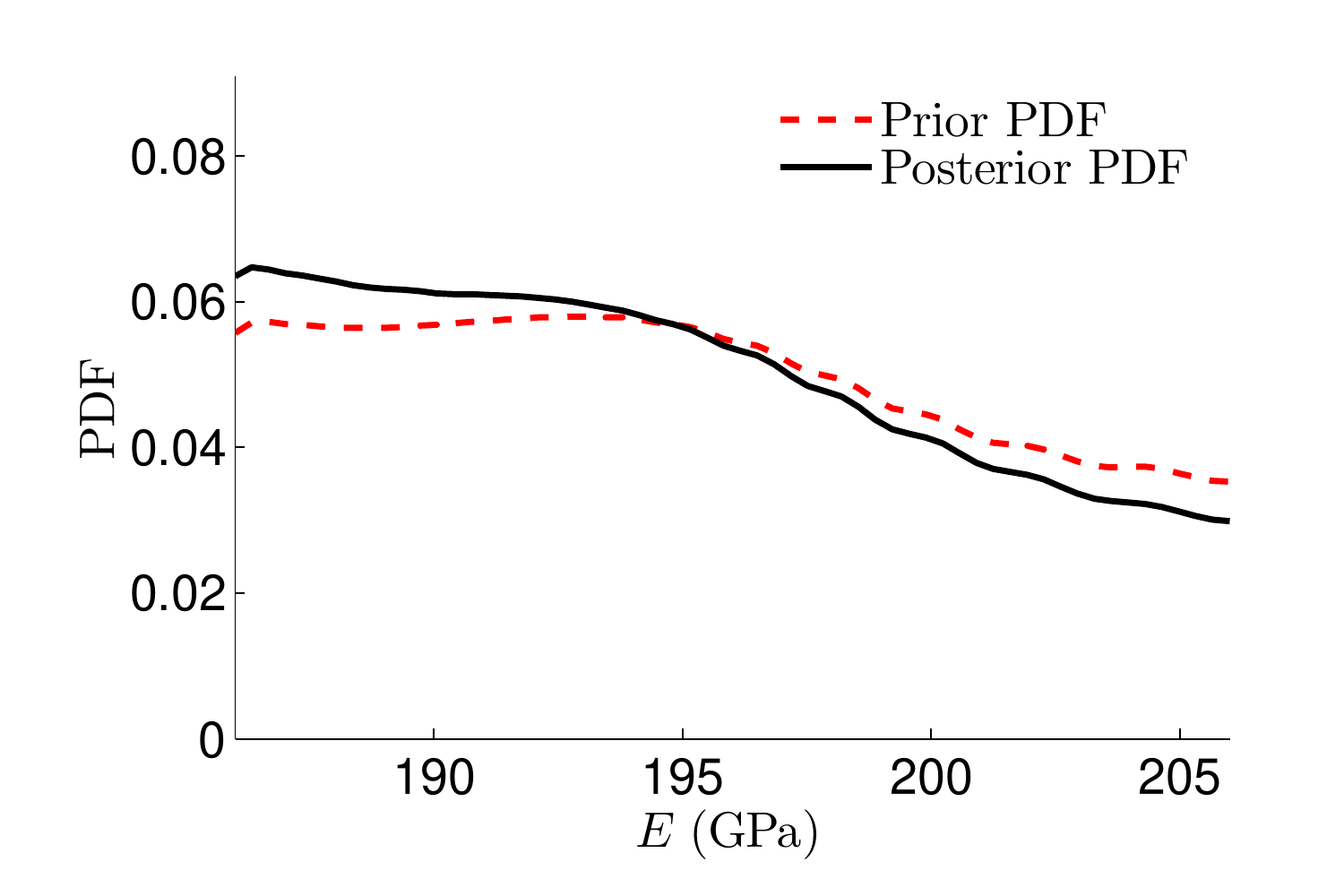}}
\subfigure[$\sigma_{rs}$]{
\includegraphics[width=0.23\textwidth]{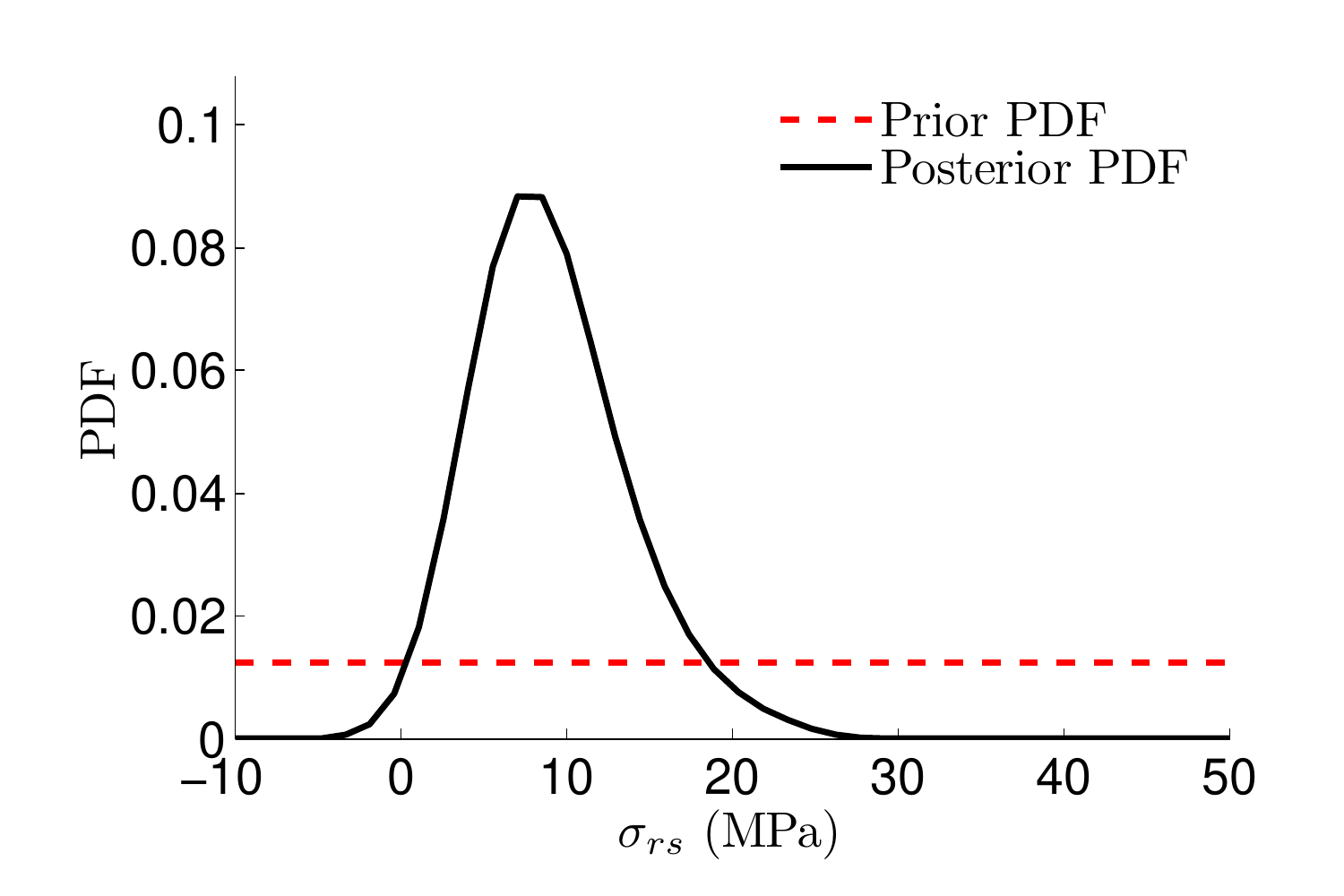}}
\subfigure[$\delta_1$]{
\includegraphics[width=0.23\textwidth]{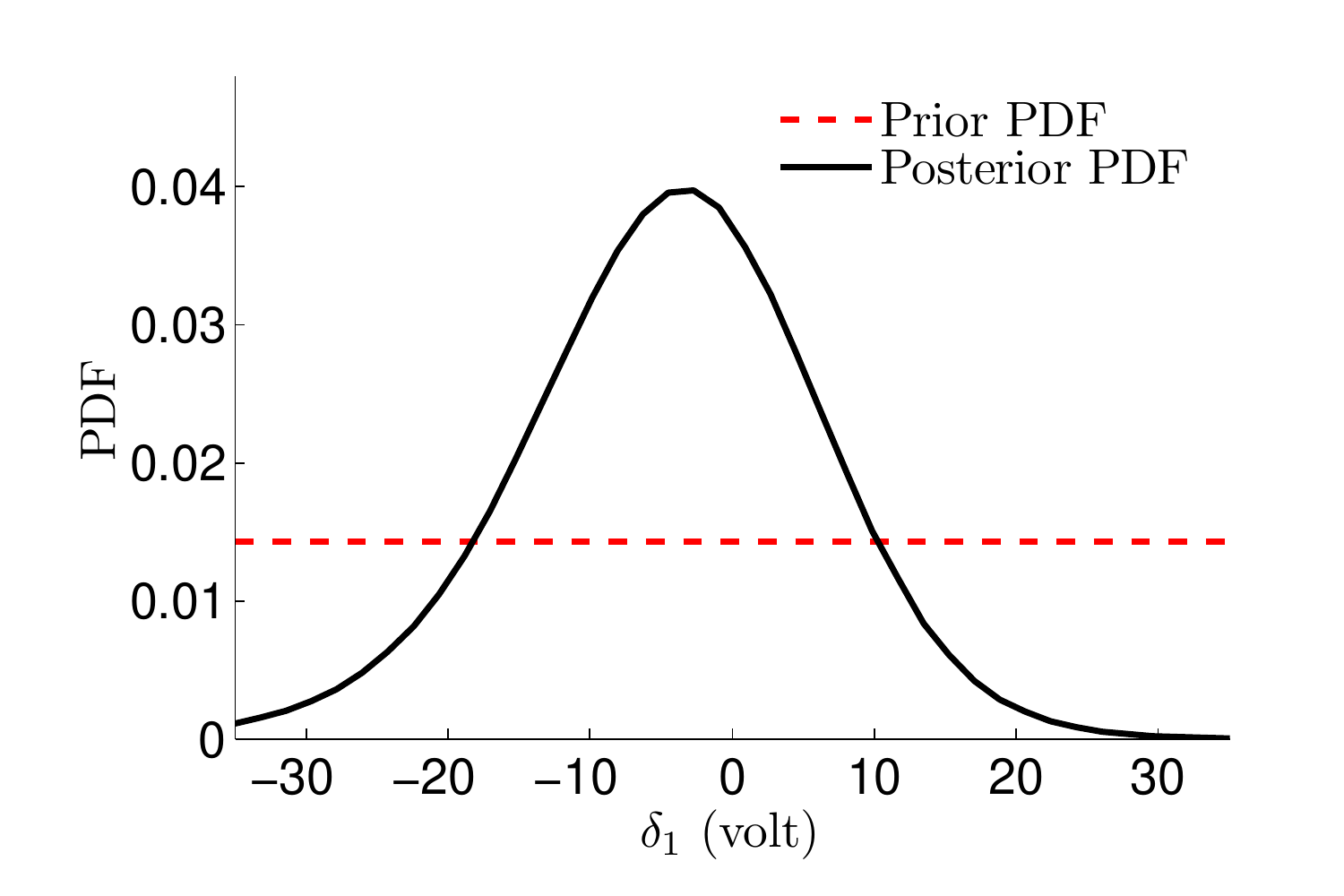}}
\subfigure[$\sigma_{obs1}$]{
\includegraphics[width=0.23\textwidth]{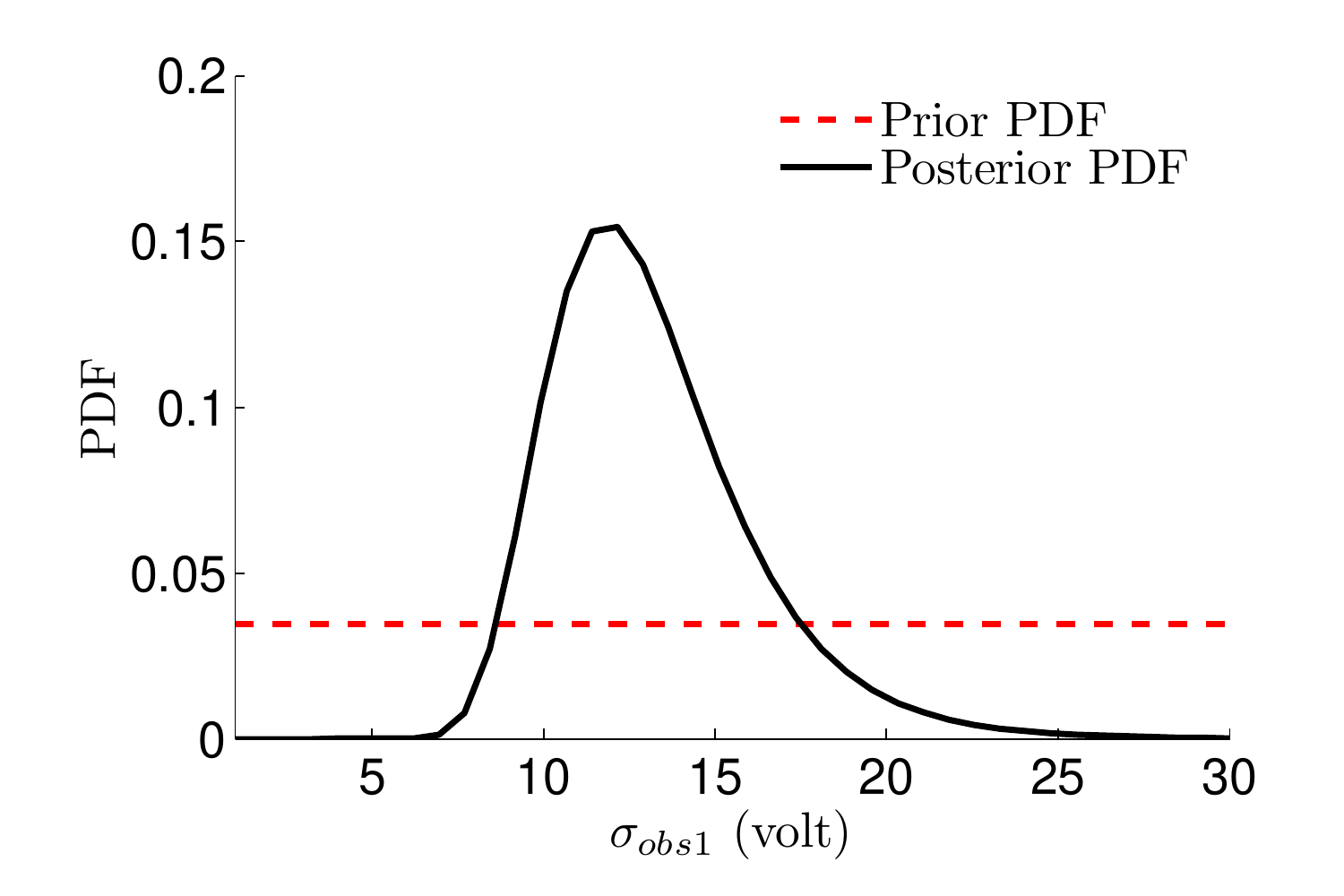}}
\end{center}
\caption{Calibration of parameters using pull-in voltage data}
\label{fig:calPullin-2}
\end{figure}

Similarly to the previous section, the posterior PDF of the common parameter $E$ obtained in the first step of calibration is used as prior, and the parameters in the left half of the Bayesian network are calibrated using the pull-in voltage data of Dev-1. The calibration results can be found in Fig.~\ref{fig:calPullin-2} and Table~\ref{table:statDataDev-1-1}.

\begin{table}[h!]
\begin{center}
\caption{Prior and posterior statistics of parameters (with data on Dev-1)}
\label{table:statDataDev-1-2}
\begin{tabular}{lcccc}
\toprule
& \multicolumn{2}{c}{Mean} & \multicolumn{2}{c}{Standard deviation} \\
\cmidrule(r){2-5}
& Prior & Posterior & Prior & Posterior \\
\midrule
$E$ (GPa) & 195.1 & 194.7 & 5.56  & 5.51 \\
$\sigma_{rs}$ (MPa) & 10.00 & 9.08 & 23.10 & 4.76 \\
$\delta_{1}$ (Volt) & 0 & -4.22 & 20.22 & 10.35 \\ 
$\sigma_{obs1}$ (Volt) &  15.50 & 13.16 & 8.38 & 2.98 \\  
\bottomrule
\end{tabular}
\end{center}
\end{table}. 
\subsubsection{Discussion}
In this example, the posterior PDFs of the parameters are computed directly using trapezoidal integration rule as only 4 parameters need to be calibrated at one time. Uniform grids are used for the numerical integration over the parameters, and the number of grid points for each parameter is selected based on the convergence of the posterior density computation. By comparing Figs.~\ref{fig:calCreep-2} and~\ref{fig:calCreep-1}, or Tables.~\ref{table:statDataDev-2-2} and~\ref{table:statDataDev-2-1}, we observe that the second and the third options give similar posterior PDFs and statistics of the calibration parameters. The same observation can be drawn from the comparison between Figs.~\ref{fig:calPullin-1} and~\ref{fig:calPullin-2}, or Tables.~\ref{table:statDataDev-1-1} and~\ref{table:statDataDev-1-2}. This is due to the fact that the posterior PDFs of the common parameter $E$ obtained in the first step of these two options are not significantly different from the uniform prior PDFs. Hence, the calibration in the second step, which uses the posterior PDF of $E$ obtained from the first step as prior, will give similar results to the case that the uniform prior PDF is used. The relatively small difference between the prior and posterior PDFs of $E$ indicates that the available experimental data are insufficient to reduce significantly the uncertainty about $E$. In addition, it can be observed from Tables.~\ref{table:statDataDev-2-2} and~\ref{table:statDataDev-1-2} that both the second and the third options give the same posterior statistics of $E$ after the two-step calibration, which is expected since in theory both options should give $\pi(E|D_1,D_2)$ as the calibrated PDF of $E$ ($D_1$ denotes the pull-in voltage data of Dev-1, and $D_2$ denotes the deflection data of Dev-2). 

\section{Conclusion}
A Bayesian network-based approach is proposed in this paper to integrate the calibration of multi-physics computational models with various sources of uncertainty and available experimental data. Several issues in Bayesian calibration for practical applications are discussed, including calibration with two different types of data - interval data and time series data, the identifiability of model parameters, efficient computation of the likelihood function, and more efficient use of available data by exchanging information on multiple physical quantities through a Bayesian network. A first-order Taylor series expansion-based method is developed to determine the identifiability of model parameters, and it is especially applicable to models with unknown expressions. This method can help to avoid the wasted effort on parameters that cannot be identified. Using a set of multi-physics models and data for two types of MEMS devices, we illustrate the Bayesian network-based approach, and the procedure of model calibration with efficient use of available information is presented. Future research efforts may include (1) applying the proposed approach to systems with more complicated structures (e.g., multi-scale, multi-level), and (2) reducing the computational effort by exploring more efficient uncertainty quantification algorithms and parallel computing.

\section*{Acknowledgment}
This paper is based upon research partly supported by the Department of Energy [National Nuclear Security Administration] under Award Number DE-FC52-08NA28617 to Purdue University (Principal Investigator: Prof. Jayathi Murthy), and subaward to Vanderbilt University. The support is gratefully acknowledged. The authors also thank the following personnel at the U. S. DOE (NNSA) PSAAP Center for Prediction of Reliability, Integrity and Survivability of Microsystems (PRISM) at Purdue University for providing the models and experimental data used in the numerical example: Professors Muhammad A. Alam, Alina Alexeenko, Marisol Koslowski, Jayathi Murthy, Dimitrios Peroulis, and Alejandro Strachan, Dr. Hao-Han Hsu, and graduate students Venkattraman Ayyaswamy,  Shankhadeep Das, Sambit Palit, Benjamin Pax, and Ravi Vedula.


\singlespacing
\bibliographystyle{elsarticle-num}
\bibliography{Research}

\begin{thebibliography}{10}
\expandafter\ifx\csname url\endcsname\relax
  \def\url#1{\texttt{#1}}\fi
\expandafter\ifx\csname urlprefix\endcsname\relax\def\urlprefix{URL }\fi
\expandafter\ifx\csname href\endcsname\relax
  \def\href#1#2{#2} \def\path#1{#1}\fi

\bibitem{Kennedy2001}
M.~Kennedy, A.~O'Hagan, {Bayesian calibration of computer models}, Journal of
  the Royal Statistical Society. Series B, Statistical Methodology 63~(3)
  (2001) 425--464.

\bibitem{Campbell2006}
K.~Campbell, {Statistical calibration of computer simulations}, Reliability
  Engineering \& System Safety 91~(10) (2006) 1358--1363.
\newblock \href {http://dx.doi.org/10.1016/j.ress.2005.11.032}
  {\path{doi:10.1016/j.ress.2005.11.032}}.

\bibitem{McFarland2008}
J.~M. Mcfarland, {Uncertainty analysis for computer simulations through
  validation and calibration}, Phd Dissertation, Vanderbilt University (2008)
  23--50.

\bibitem{Sankararaman2011}
S.~Sankararaman, Y.~Ling, C.~Shantz, S.~Mahadevan, {Inference of equivalent
  initial flaw size under multiple sources of uncertainty}, International
  Journal of Fatigue 33~(2) (2011) 75--89.
\newblock \href {http://dx.doi.org/10.1016/j.ijfatigue.2010.06.008}
  {\path{doi:10.1016/j.ijfatigue.2010.06.008}}.

\bibitem{Koutsourelakis2009}
P.~Koutsourelakis, {A multi-resolution, non-parametric, Bayesian framework for
  identification of spatially-varying model parameters}, Journal of
  Computational Physics 228~(17) (2009) 6184--6211.
\newblock \href {http://dx.doi.org/10.1016/j.jcp.2009.05.016}
  {\path{doi:10.1016/j.jcp.2009.05.016}}.

\bibitem{Bower2010}
R.~G. Bower, I.~Vernon, M.~Goldstein, a.~J. Benson, C.~G. Lacey, C.~M. Baugh,
  S.~Cole, C.~S. Frenk, {The parameter space of galaxy formation}, Monthly
  Notices of the Royal Astronomical Society 407~(4) (2010) 2017--2045.
\newblock \href {http://dx.doi.org/10.1111/j.1365-2966.2010.16991.x}
  {\path{doi:10.1111/j.1365-2966.2010.16991.x}}.

\bibitem{Sankararaman2010}
S.~Sankararaman, Y.~Ling, S.~Mahadevan, {Statistical inference of equivalent
  initial flaw size with complicated structural geometry and multi-axial
  variable amplitude loading}, International Journal of Fatigue 32~(10) (2010)
  1689--1700.
\newblock \href {http://dx.doi.org/10.1016/j.ijfatigue.2010.03.012}
  {\path{doi:10.1016/j.ijfatigue.2010.03.012}}.

\bibitem{Cheung2011}
S.~H. Cheung, T.~A. Oliver, E.~E. Prudencio, S.~Prudhomme, R.~D. Moser,
  {Bayesian uncertainty analysis with applications to turbulence modeling},
  Reliability Engineering and System Safety 96~(9) (2011) 1137--1149.
\newblock \href {http://dx.doi.org/10.1016/j.ress.2010.09.013}
  {\path{doi:10.1016/j.ress.2010.09.013}}.

\bibitem{Wan2011}
J.~Wan, N.~Zabaras, {A Bayesian approach to multiscale inverse problems using
  the sequential Monte Carlo method}, Inverse Problems 27~(10) (2011) 105004.
\newblock \href {http://dx.doi.org/10.1088/0266-5611/27/10/105004}
  {\path{doi:10.1088/0266-5611/27/10/105004}}.

\bibitem{Moselhy2011}
T.~A.~E. Moselhy, Y.~M. Marzouk, {Bayesian Inference with Optimal Maps}, Arxiv
  preprint arXiv:1109.1516\href {http://arxiv.org/abs/arXiv:1109.1516v2}
  {\path{arXiv:arXiv:1109.1516v2}}.

\bibitem{Jensen2007}
F.~Jensen, T.~Nielsen, {Bayesian networks and decision graphs (Information
  Science and Statistics)}, 2nd Edition, Springer, 2007.

\bibitem{Torres-Toledano1998}
J.~Torres-Toledano, L.~Sucar, {Bayesian networks for reliability analysis of
  complex systems}, Progress in Artificial Intelligence - IBERAMIA 98 (1998)
  465--465.

\bibitem{Langseth2007}
H.~Langseth, L.~Portinale, {Bayesian networks in reliability}, Reliability
  Engineering \& System Safety 92~(1) (2007) 92--108.
\newblock \href {http://dx.doi.org/10.1016/j.ress.2005.11.037}
  {\path{doi:10.1016/j.ress.2005.11.037}}.

\bibitem{Zabaras2008}
N.~Zabaras, B.~Ganapathysubramanian, {A scalable framework for the solution of
  stochastic inverse problems using a sparse grid collocation approach},
  Journal of Computational Physics 227~(9) (2008) 4697--4735.
\newblock \href {http://dx.doi.org/10.1016/j.jcp.2008.01.019}
  {\path{doi:10.1016/j.jcp.2008.01.019}}.

\bibitem{Guan2009}
X.~Guan, R.~Jha, Y.~Liu, {Probabilistic fatigue damage prognosis using maximum
  entropy approach}, Journal of Intelligent Manufacturing 23~(2) (2009)
  163--171.
\newblock \href {http://dx.doi.org/10.1007/s10845-009-0341-3}
  {\path{doi:10.1007/s10845-009-0341-3}}.

\bibitem{Berry2012}
R.~D. Berry, H.~N. Najm, B.~J. Debusschere, Y.~M. Marzouk, H.~Adalsteinsson,
  {Data-free inference of the joint distribution of uncertain model
  parameters}, Journal of Computational Physics 231~(5) (2012) 2180--2198.
\newblock \href {http://dx.doi.org/10.1016/j.jcp.2011.10.031}
  {\path{doi:10.1016/j.jcp.2011.10.031}}.

\bibitem{Mahadevan2001}
S.~Mahadevan, R.~Zhang, N.~Smith, {Bayesian networks for system reliability
  reassessment}, Structural Safety 23~(3) (2001) 231--251.
\newblock \href {http://dx.doi.org/10.1016/S0167-4730(01)00017-0}
  {\path{doi:10.1016/S0167-4730(01)00017-0}}.

\bibitem{Cobb2006}
B.~Cobb, P.~P. Shenoy, {Operations for inference in continuous Bayesian
  networks with linear deterministic variables}, International Journal of
  Approximate Reasoning 42~(1-2) (2006) 21--36.
\newblock \href {http://dx.doi.org/10.1016/j.ijar.2005.10.002}
  {\path{doi:10.1016/j.ijar.2005.10.002}}.

\bibitem{Cinicioglu2009}
E.~Cinicioglu, P.~P. Shenoy, {Arc reversals in hybrid Bayesian networks with
  deterministic variables}, International Journal of Approximate Reasoning
  50~(5) (2009) 763--777.
\newblock \href {http://dx.doi.org/10.1016/j.ijar.2009.02.005}
  {\path{doi:10.1016/j.ijar.2009.02.005}}.

\bibitem{Sankararaman2011a}
S.~Sankararaman, S.~Mahadevan, {Likelihood-based representation of epistemic
  uncertainty due to sparse point data and/or interval data}, Reliability
  Engineering \& System Safety 96~(7) (2011) 814--824.
\newblock \href {http://dx.doi.org/10.1016/j.ress.2011.02.003}
  {\path{doi:10.1016/j.ress.2011.02.003}}.

\bibitem{Thrun2005}
S.~Thrun, W.~Burgard, D.~Fox, {Probabilistic Robotics (Intelligent Robotics and
  Autonomous Agents series)}, The MIT Press, 2005.

\bibitem{Raue2009}
A.~Raue, C.~Kreutz, T.~Maiwald, J.~Bachmann, M.~Schilling, U.~Klingm\"{u}ller,
  J.~Timmer, {Structural and practical identifiability analysis of partially
  observed dynamical models by exploiting the profile likelihood},
  Bioinformatics 25~(15) (2009) 1923--1929.
\newblock \href {http://dx.doi.org/10.1093/bioinformatics/btp358}
  {\path{doi:10.1093/bioinformatics/btp358}}.

\bibitem{Arendt2011}
P.~D. Arendt, W.~Chen, D.~W. Apley, P.~Zhu, {Multiple Responses and Design of
  Experiments for Improving Identifiability in Model Calibration}, in: 9th
  World Congress on Structural and Multidisciplinary Optimization, Shizuoka,
  Japan, 2011.

\bibitem{Arendt2010}
P.~D. Arendt, W.~Chen, D.~W. Apley, {Updating Predictive Models: Calibration,
  Bias Correction and Identifiability}, in: Proceedings of the ASME 2010
  International Design Engineering Technical Conferences \& Computers and
  Information in Engineering Conference (IDETC/CIE2010), ASME, Montreal,
  Quebec, Canada, 2010, pp. 1089--1098.
\newblock \href {http://dx.doi.org/dx.doi.org/10.1115/DETC2010-28828}
  {\path{doi:dx.doi.org/10.1115/DETC2010-28828}}.

\bibitem{Grewal1976}
M.~Grewal, K.~Glover, {Identifiability of linear and nonlinear dynamical
  systems}, Automatic Control, IEEE Transactions on 21~(6) (1976) 833--837.

\bibitem{Walter1996}
E.~Walter, L.~Pronzato, {On the identifiability and distinguishability of
  nonlinear parametric models}, Mathematics and Computers in Simulation 42~(2)
  (1996) 125--134.

\bibitem{Jia-fan2011}
Z.~Jia-fan, Y.~Qing-hua, Z.~Tong, {Numerical Approach to Identifiability Test
  of Parametric Models in Nonlinear Mechanical Systems}, Journal of Dynamic
  Systems, Measurement, and Control 133~(5) (2011) 051002.
\newblock \href {http://dx.doi.org/10.1115/1.4004062}
  {\path{doi:10.1115/1.4004062}}.

\bibitem{Gu1994}
M.~Gu, J.~Lu, {A note on identifiability of the regression parameter and
  validity of the partial likehood approach in general relativerisk
  regression}, Biometrika 81~(4) (1994) 802--806.

\bibitem{Paulino1994}
C.~Paulino, C.~{de Bragan\c{c}a Pereira}, {On identifiability of parametric
  statistical models}, Statistical Methods and Applications 3~(1) (1994)
  125--151.
\newblock \href {http://dx.doi.org/10.1007/BF02589044}
  {\path{doi:10.1007/BF02589044}}.

\bibitem{Little2010}
M.~Little, W.~Heidenreich, G.~Li, {Parameter identifiability and redundancy:
  theoretical considerations}, PloS one 5~(1) (2010) e8915.
\newblock \href {http://dx.doi.org/10.1371/journal.pone.0008915}
  {\path{doi:10.1371/journal.pone.0008915}}.

\bibitem{Raue2011}
A.~Raue, C.~Kreutz, T.~Maiwald, U.~Klingmuller, J.~Timmer, {Addressing
  parameter identifiability by model-based experimentation}, Systems Biology,
  IET 5~(2) (2011) 120--130.
\newblock \href {http://dx.doi.org/10.1049/iet-syb.2010.0061}
  {\path{doi:10.1049/iet-syb.2010.0061}}.

\bibitem{Murphy2000}
S.~Murphy, A.~{Van der Vaart}, {On profile likelihood}, Journal of the American
  Statistical Association 95~(450) (2000) 449--465.

\bibitem{Rasmussen2006}
C.~E. Rasmussen, C.~K.~I. Williams, {Gaussian Processes for Machine Learning},
  The MIT Press, 2006.

\bibitem{Xiu2002}
D.~Xiu, G.~Karniadakis, {The Wiener-Askey polynomial chaos for stochastic
  differential equations}, SIAM Journal on Scientific Computing 24~(2) (2002)
  619--644.
\newblock \href {http://dx.doi.org/10.1137/S1064827501387826}
  {\path{doi:10.1137/S1064827501387826}}.

\bibitem{Ghanem2003}
R.~Ghanem, P.~Spanos, {Stochastic Finite Elements: A Spectral Approach}, Dover
  Pubns, 2003.

\bibitem{Marzouk2009}
Y.~M. Marzouk, H.~N. Najm, {Dimensionality reduction and polynomial chaos
  acceleration of Bayesian inference in inverse problems}, Journal of
  Computational Physics 228~(6) (2009) 1862--1902.
\newblock \href {http://dx.doi.org/10.1016/j.jcp.2008.11.024}
  {\path{doi:10.1016/j.jcp.2008.11.024}}.

\bibitem{Vapnik1999}
V.~N. Vapnik, {An overview of statistical learning theory.}, IEEE transactions
  on neural networks / a publication of the IEEE Neural Networks Council 10~(5)
  (1999) 988--99.
\newblock \href {http://dx.doi.org/10.1109/72.788640}
  {\path{doi:10.1109/72.788640}}.

\bibitem{Tipping2001}
M.~Tipping, {Sparse Bayesian learning and the relevance vector machine}, The
  Journal of Machine Learning Research 1 (2001) 211--244.

\bibitem{Ma2009}
X.~Ma, N.~Zabaras, {An efficient Bayesian inference approach to inverse
  problems based on an adaptive sparse grid collocation method}, Inverse
  Problems 25~(3) (2009) 035013 (27pp).
\newblock \href {http://dx.doi.org/10.1088/0266-5611/25/3/035013}
  {\path{doi:10.1088/0266-5611/25/3/035013}}.

\bibitem{Bliznyuk2008}
N.~Bliznyuk, D.~Ruppert, C.~Shoemaker, R.~Regis, S.~Wild, P.~Mugunthan,
  {Bayesian Calibration and Uncertainty Analysis for Computationally Expensive
  Models Using Optimization and Radial Basis Function Approximation}, Journal
  of Computational and Graphical Statistics 17~(2) (2008) 270--294.

\bibitem{Chib1995}
S.~Chib, E.~Greenberg, {Understanding the metropolis-hastings algorithm},
  American Statistician 49~(4) (1995) 327--335.

\bibitem{Green1995}
P.~J. Green, {Reversible Jump Markov Chain Monte Carlo Computation and Bayesian
  Model Determination}, Biometrika 82~(4) (1995) 711--732.
\newblock \href {http://dx.doi.org/10.2307/2337340}
  {\path{doi:10.2307/2337340}}.

\bibitem{Gelman1996}
A.~Gelman, G.~Roberts, W.~Gilks, {Efficient metropolis jumping hules}, Bayesian
  statistics 5 (1996) 599--607.

\bibitem{Haario2006}
H.~Haario, M.~Laine, A.~Mira, E.~Saksman, {DRAM: Efficient adaptive MCMC},
  Statistics and Computing 16~(4) (2006) 339--354.
\newblock \href {http://dx.doi.org/10.1007/s11222-006-9438-0}
  {\path{doi:10.1007/s11222-006-9438-0}}.

\bibitem{Zuev2012}
K.~M. Zuev, J.~L. Beck, S.-K. Au, L.~S. Katafygiotis, {Bayesian post-processor
  and other enhancements of Subset Simulation for estimating failure
  probabilities in high dimensions}, Computers \& Structures 92-93 (2012)
  283--296.
\newblock \href {http://dx.doi.org/10.1016/j.compstruc.2011.10.017}
  {\path{doi:10.1016/j.compstruc.2011.10.017}}.

\bibitem{Casella1992}
G.~Casella, E.~George, {Explaining the Gibbs sampler}, The American
  Statistician 46~(3) (1992) 167--174.
\newblock \href {http://dx.doi.org/10.2307/2685208}
  {\path{doi:10.2307/2685208}}.

\bibitem{Neal2003}
R.~Neal, {Slice sampling}, The annals of statistics 31~(3) (2003) 705--741.
\newblock \href {http://dx.doi.org/10.1214/aos/1056562461}
  {\path{doi:10.1214/aos/1056562461}}.

\bibitem{Palit2012}
S.~Palit, M.~Alam, {Theory of charging and charge transport in
  “intermediate” thickness dielectrics and its implications for
  characterization and reliability}, Journal of Applied Physics 111~(5) (2012)
  054112.
\newblock \href {http://dx.doi.org/10.1063/1.3691962}
  {\path{doi:10.1063/1.3691962}}.

\bibitem{Jain2011}
A.~Jain, S.~Palit, M.~A. Alam, {A Physics-Based Predictive Modeling Framework
  for Dielectric Charging and Creep in RF MEMS Capacitive Switches and
  Varactors}, Microelectromechanical Systems, Journal of~(99) (2011) 1--11.

\bibitem{Hartzell2011}
A.~L. Hartzell, M.~G. da~Silva, H.~R. Shea, {MEMS reliability (MEMS reference
  shelf)}, 1st Edition, Springer, 2011.

\bibitem{Quinonero-Candela2005}
J.~Quinonero-Candela, C.~E. Rasmussen, {A unifying view of sparse approximate
  Gaussian process regression}, Journal of Machine Learning Research 6 (2005)
  1939--1959.

\bibitem{Sorenson1980}
H.~Sorenson, {Parameter estimation: principles and problems}, Vol.~9, Marcel
  Dekker, Inc, 1980.

\bibitem{Ayyaswamy2010}
V.~Ayyaswamy, A.~Alexeenko,
  \href{http://memshub.org/resources/prismcg}{{Coarse-grained Model for RF MEMS
  Device}} (2010).
\newline\urlprefix\url{http://memshub.org/resources/prismcg}

\bibitem{Alexeenko2011}
A.~Alexeenko, S.~Chigullapalli, J.~Zeng, X.~Guo, A.~Kovacs, D.~Peroulis,
  {Uncertainty in microscale gas damping: Implications on dynamics of
  capacitive MEMS switches}, Reliability Engineering and System Safety 96~(9)
  (2011) 1171--1183.
\newblock \href {http://dx.doi.org/10.1016/j.ress.2011.01.002}
  {\path{doi:10.1016/j.ress.2011.01.002}}.

\bibitem{Coble1963}
R.~L. Coble, {A Model for Boundary Diffusion Controlled Creep in
  Polycrystalline Materials}, Journal of Applied Physics 34~(6) (1963)
  1679--1682.
\newblock \href {http://dx.doi.org/10.1063/1.1702656}
  {\path{doi:10.1063/1.1702656}}.

\bibitem{Hsu2011}
H.~Hsu, M.~Koslowski, D.~Peroulis, {An Experimental and Theoretical
  Investigation of Creep in Ultrafine Crystalline Nickel RF-MEMS Devices},
  Microwave Theory and Techniques, IEEE Transactions on 59~(10) (2011)
  2655--2664.

\end{thebibliography}

\end{document}